\documentclass[11pt]{article}
\pdfoutput=1 
\usepackage{./jheppub}

\usepackage[T1]{fontenc}

\usepackage{graphicx}
\usepackage{amsfonts,amsmath,amssymb}
\usepackage{verbatim}
\usepackage{array}
\usepackage{mathrsfs}
\usepackage{mathrsfs}
\usepackage{yfonts}
\usepackage{dsfont}
\usepackage{bbm}
\usepackage{colonequals}
\usepackage{amscd}

\usepackage{subcaption}

\usepackage{wrapfig}

\usepackage{suffix,mathtools,cancel,bbm}

\usepackage{multirow}

\usepackage{enumerate}

	\usepackage{tikz, tikz-3dplot, pgfplots}
	\usepackage{tkz-graph}
	\usetikzlibrary[positioning,patterns] 
	\usepackage{genyoungtabtikz}

\newcommand{\bn}{\begin{enumerate}}
\newcommand{\en}{\end{enumerate}}

\newcommand{\beq}{\begin{equation}}
\newcommand{\eeq}{\end{equation}}

\newcommand\nn{\nonumber}

\newcommand{\cA}{\mathcal{A}}

\newcommand{\cC}{\mathcal{C}}

\newcommand{\cE}{\mathcal{E}}
\newcommand{\cF}{\mathcal{F}}

\newcommand{\cH}{\mathcal{H}}
\newcommand{\cI}{\mathcal{I}}

\newcommand{\cM}{\mathcal{M}}
\newcommand{\cN}{\mathcal{N}}
\newcommand{\cO}{\mathcal{O}}

\newcommand{\cR}{\mathcal{R}}
\newcommand{\cS}{\mathcal{S}}

\newcommand{\cU}{\mathcal{U}}

\numberwithin{equation}{section}

\def\bea{\begin{eqnarray}}
\def\eea{\end{eqnarray}}

\DeclarePairedDelimiterX\MeijerM[3]{\lparen}{\rparen}%
{\begin{smallmatrix}#1 \\ #2\end{smallmatrix}\delimsize\vert\,#3}

\newcommand\MeijerG[8][]{%
  G^{\,#2,#3}_{#4,#5}\MeijerM[#1]{#6}{#7}{#8}}

\WithSuffix\newcommand\MeijerG*[7]{%
  G^{\,#1,#2}_{#3,#4}\MeijerM*{#5}{#6}{#7}}

\def\cH{\mathcal{H}}

\def\cN{\mathcal{N}}

\def \beg#1{\begin{#1}} 

\def \bea{\beg{eqnarray}}
\def \eea{\end{eqnarray}}
\def \ee{\end{equation}}

\def \restr#1#2{{\left.\kern-\nulldelimiterspace#1\vphantom{\big|}\right|_{#2}}}

\def \nn{\nonumber}

\usepackage{colortbl}
\definecolor{mygray}{gray}{0.93}

\title{\boldmath Anomaly Inflow Methods for SCFT Constructions in Type IIB}

\author[a]{Ibrahima Bah,}
\author[a]{Federico Bonetti,}
\author[b,c]{Ruben Minasian,}
\author[a,b]{and Peter Weck} 
\affiliation[a]{Department of Physics and Astronomy, Johns Hopkins University, 3400 North Charles Street, Baltimore, MD 21218, USA}
\affiliation[b]{Institut de Physique Th\'{e}orique, Universit\'{e} Paris Saclay, CNRS, CEA, F-91191, Gif-sur-Yvette, France}
\affiliation[c]{School of Physics, Korea Institute for Advanced Study, Seoul 02455, Korea}

\emailAdd{iboubah@jhu.edu, fbonett3@jhu.edu, ruben.minasian@ipht.fr, pweck1@jhu.edu}

\abstract
{
We extend the anomaly inflow methods developed in M-theory to SCFTs
engineered via D3-branes in type IIB. We show that the 't Hooft anomalies of such SCFTs can be computed systematically from their geometric definition. 
Our procedure is tested
in several 4d examples and   applied   to 2d theories
obtained by wrapping D3-branes on a Riemann surface.
In particular, we show how to analyze
half-BPS regular punctures for 4d $\mathcal N = 4$ SYM
on a Riemann surface.
We   discuss generalizations of this formalism
to type IIB configurations with $F_3$, $H_3$ fluxes,
as well as to F-theory setups.
}

\usepackage{etoolbox}
\appto\appendix{\addtocontents{toc}{\protect\setcounter{tocdepth}{1}}}

\appto\listoffigures{\addtocontents{lof}{\protect\setcounter{tocdepth}{1}}}
\appto\listoftables{\addtocontents{lot}{\protect\setcounter{tocdepth}{1}}}

\begin{document} 

\maketitle
\flushbottom



\section{Introduction}

Geometric engineering is a powerful tool in the construction and analysis of
quantum field theories (QFTs) in various dimensions.
In many situations,
geometric methods in string/M-/F-theory allow one to study strongly coupled QFTs 
for which a Lagrangian description is not available.  
A prototypical example
is furnished by 4d QFTs obtained by wrapping M5-branes on a Riemann surface with punctures,
preserving   $\cN = 2$ \cite{Gaiotto:2009we,Gaiotto:2009hg} or $\cN = 1$ 
 \cite{Maruyoshi:2009uk,Benini:2009mz,Bah:2011je,Bah:2011vv,Bah:2012dg} supersymmetry.

't Hooft anomalies are among the most interesting quantities to compute
in a geometrically engineered theory. Since 't Hooft anomalies are invariant under RG flow,
they are particularly robust observables and can be used to constrain
the phases of theories in a non-perturbative way. 
In this work, we focus on 't Hooft anomalies for
continuous 0-form symmetries. These anomalies 
only occur for QFT in even $d$ spacetime dimensions,
and are conveniently summarized in the anomaly
polynomial, which is a $(d+2)$-form characteristic class constructed
with the curvatures of the background fields that couple to the global
symmetries.

In \cite{Bah:2018gwc, Bah:2018jrv, Bah:2019jts, Bah:2019rgq, Bah:2019vmq} systematic
tools have been developed 
to compute the anomaly polynomial 
for even-dimensional QFTs obtained from wrapped M5-branes.
The methods of \cite{Bah:2018gwc, Bah:2018jrv, Bah:2019jts, Bah:2019rgq, Bah:2019vmq} are based on the 
anomaly inflow mechanism for M5-branes, first studied  
in \cite{Duff:1995wd,Witten:1996hc,Freed:1998tg,Harvey:1998bx}. 
The main strategy underlying these inflow-based tools is to shift
the focus from the worldvolume theory on the M5-branes
to the supergravity fields in 11d ambient space that surrounds the branes.
In the presence of the M5-brane stack, the supergravity fields
acquire non-trivial boundary conditions, which in turn generate
a classical anomalous variation of the 11d effective action.
This classical variation counterbalances the quantum anomalies
of the worldvolume theory on the M5-branes
(including anomalies  of modes that decouple in the deep IR).

Having a systematic toolkit for the computation of anomalies in
this class of theories is beneficial in various ways.  For instance,
the analysis of \cite{Bah:2018gwc, Bah:2018jrv, Bah:2019jts} shows that   the 
``bulk'' and ``puncture'' contributions to 't Hooft anomalies
in a 4d $\cN = 2$ theory of class $\cS$ with regular punctures can be
treated on an equal footing.
Indeed,   both can   be analyzed by studying the boundary
conditions for the $G_4$-flux configuration near the M5-branes.
Furthermore, the inflow perspective can be applied to holographic
setups, where it has the potential of  yielding finite terms in $N$ without resorting
to AdS loop computations \cite{Bah:2019rgq}. As exemplified in \cite{Bah:2019vmq},
a careful treatment of the boundary conditions for the 11d supergravity
fields can be efficiently used as a proxy to track non-trivial
dynamics on the worldvolume of the branes, including
the emergence of accidental symmetries and spontaneous symmetry
breaking.

Given the success of anomaly inflow methods in M-theory, it is natural
to ask whether similar tools can be developed for other 
string constructions. The main objective of this work is
to formulate a proposal for inflow tools in type IIB string theory.
In the M-theory case, an essential role is played
by a formal 12-form $\cI_{12}$, constructed with
a 4-form $E_4$ that encodes the boundary condition
for $G_4$ near the M5-branes.
In the type IIB context, we have a formal 11-form $\cI_{11}$,
which is constructed with a 5-form $E_5$, and 3-forms $\cF_3$, $\cH_3$, which
capture the boundary conditions for $F_5$, $F_3$, $H_3$,
respectively. The structure of $\cI_{11}$ is expected to be considerably richer
if we upgrade from type IIB to F-theory
(\emph{i.e.}~type IIB backgrounds
in which the axio-dilaton profile
has non-trivial monodromies around singular loci).
 We also
comment about such extension,
making contact with the constructions of \cite{Lawrie:2018jut}.

To begin with, we   consider setups
with D3-branes in the absence of $F_3$, $H_3$ fluxes
and with constant dilaton profile. These type IIB constructions
have been studied intensively over the years. Typically, the worldvolume
theory on the D3-branes is a well-understood Lagrangian theory.
We may then use these setups to test our inflow methods.
In particular, in these examples we have full 
control over the modes that decouple in the IR and we can
explicitly verify that, keeping decoupling into account, anomaly inflow gives results
that are exact in $N$ (the number of D3-branes in the stack).
It is worth pointing out  that recent work \cite{Fazzi:2019gvt}
demonstrates that there are still interesting aspects
of the dynamics of D3-branes at the tip of a Calabi-Yau cone
that are not fully understood and deserve further
investigation. We propose that our inflow tools should
be applicable to these less-understood cases, as well.

Next, we apply our inflow proposal to some 2d theories.
In particular, we exploit the intuition developed in \cite{Bah:2018gwc, Bah:2018jrv, Bah:2019jts}
to compute the anomaly of
4d $\cN = 4$ SYM compactified on a Riemann surface
with   half-BPS punctures.

The rest of this paper is organized as follows.
In section \ref{sec_generalstory} we formulate our proposal for the computation
of the inflow anomaly polynomial, introducing the 
main objects $E_5$, $\cF_3$, $\cH_3$, and $\cI_{11}$.
Section \ref{sec_4dexamples} is devoted to a careful study of several examples
of 4d QFTs engineered using D3-branes in type IIB, which provide various
checks of our proposal. In section \ref{sec_2dstory} we consider 
some 2d examples, obtained from reduction from four dimensions
on a Riemann surface without punctures, or with half-BPS
punctures. Section \ref{sec_ftheory}  is dedicated to a preliminary
investigation of the extension of $\cI_{11}$ to F-theory setups.
We conclude with a brief discussion. Some derivations
and other   technical material are collected in the appendices.



\section{Inflow anomaly polynomial for type IIB} \label{sec_generalstory}

In this section we describe our proposal for the computation
of the inflow anomaly polynomial for type IIB setups.
A crucial role is played by a formal 11-form
  class $\cI_{11}$,
which captures the anomalous variation of the type IIB action
in the presence of a boundary.
Our method is inspired by the tools of \cite{Bah:2018gwc, Bah:2018jrv, Bah:2019jts, Bah:2019rgq, Bah:2019vmq} 
for the analysis of anomaly inflow for M5-branes in M-theory.
We therefore start with a quick review of M5-brane
inflow before discussing our proposal for type IIB setups.

\subsection{Review: anomaly inflow for M5-branes}

Let us consider a stack of $N$ M5-branes
with worldvolume $W_6$
inside the 11d spacetime $M_{11}$.
We suppose that $W_6$ is of the form
\beq
W_6 =W_d \times \cS_{6-d} \ ,
\eeq
where
$W_d$ is external $d$-dimensional spacetime
and
 $\cS_{6-d}$ is a smooth compact 
internal space.
The low-energy dynamics of the degrees of freedom
localized at the stack furnishes a QFT in the $d$ external dimensions.
We focus on the case of $d$ even.
Our task is it exploit anomaly inflow from the 11d ambient space
of M-theory to compute the 't Hooft anomalies
for global continuous symmetries 
of the   QFT on $W_d$.

In order to specify the M5-brane configuration, we need both the geometry
of the internal space $\cS_{6-d}$, and 
information about the normal bundle $NW_6$ to the branes in the ambient 11d space.
A convenient way to encode these data is to introduce
a compact $(10-d)$-dimensional space $M_{10-d}$,
which is an $S^4$ fibration over $\cS_{6-d}$,
\beq \label{M10minusd}
S^4 \hookrightarrow M_{10-d} \rightarrow \cS_{6-d} \ .
\eeq
We think of the $S^4$ fiber as
the unit sphere in the fibers of $NW_6$,
or equivalently as
 the $S^4$ that surrounds the M5-brane
stack in its five transverse directions.
The fibering of $S^4$ over $\cS_{6-d}$
encodes the partial topological twist 
of the 6d theory on $W_6$   compactified 
on $\cS_{6-d}$.

The key observation for anomaly inflow in this setup
is that the M5-brane stack acts as a singular magnetic
source the M-theory flux $G_4$.
Following \cite{Freed:1998tg,Harvey:1998bx},
the singularity is removed by 
 excising 
a small tubular neighborhood of the 
M5-brane stack. 
As a result, the 11d spacetime $M_{11}$ acquires
a boundary $\partial M_{11} = M_{10}$. If $r$ denotes the radial coordinate
away from the M5-brane stack,
 $M_{10}$ is located at $r = \epsilon$,
 where $\epsilon$ is a small positive constant.
The space $M_{10}$ is a fibration of $M_{10-d}$
over $W_d$, 
\beq \label{M10_fibration}
M_{10-d} \hookrightarrow M_{10} \rightarrow W_d \ .
\eeq
The   fibering of the internal space $M_{10-d}$
over the external spacetime $W_d$
is due to the fact that we are turning on
background gauge connections
for the continuous global symmetries of the QFT
living on $W_d$.

The magnetic source for $G_4$ is modeled by
imposing suitable boundary conditions
for $G_4$ near 
   $r = \epsilon$. More precisely, we have \cite{Freed:1998tg,Harvey:1998bx}
\beq \label{G4_boundary}
\frac{G_4}{2\pi} = - \rho \, E_4 + \dots \ .
\eeq   
In the above expression, $\rho$ is a bump function depending on $r$ only,
which interpolates smoothly between $\rho = -1$ at $r = \epsilon$
and $\rho = 0$ at large $r$. The ellipsis stands for terms
with $dr$ legs and/or subleading terms in the limit $r \rightarrow \epsilon$.
The quantity $E_4$ is a closed and globally-defined 4-form on 
$M_{10}$. (Thus, by definition, $E_4$ has no legs along $r$.)
In order to be globally defined, $E_4$ must be gauge invariant
under a gauge transformation of the external background
gauge fields. Furthermore, the integral of $E_4$
along the $S^4$ fibers of $\cM_{10-d}$, see \eqref{M10minusd},
counts the total number of M5-branes in the stack,
\beq
\int_{S^4} E_4 = N \ .
\eeq
In the simplest situation in which we consider six uncompactified 
directions, the 4-form $E_4$ is given by
\beq \label{simple_E4}
E_4 = N \, e_4    \qquad \text{(for $d=6$)}\ ,
\eeq
where $e_4$ is the global angular form of $SO(5)$,
normalized to integrate to 1 along $S^4$.
The form $e_4$ is the $SO(5)$ invariant and closed completion
of the volume form on $S^4$.
Its expression can be found \emph{e.g.}~in \cite{Harvey:1998bx}.

The boundary condition \eqref{G4_boundary} for $G_4$
is used to build a formal 12-form characteristic class
\beq
\cI_{12} = - \frac 16 \, E_4^3 - E_4 \, X_8  \ ,
\eeq
where we suppressed wedge products for brevity, and we introduced
the 8-form
\beq \label{X8def}
X_8 = \frac{1}{192} \bigg[ p_1(TM_{11})^2
- 4 \, p_2(TM_{11}) \bigg] \ ,
\eeq
where the quantities $p_{1,2}(TM_{11})$
are the first and second Pontryagin
classes of the 11d tangent bundle,
implicitly pulled back to the boundary at $r = \epsilon$.
The importance of the 12-form $\cI_{12}$ stems
from the fact that it encodes the variation of the M-theory
effective action in the presence of the boundary $M_{10}$.
More precisely, the two terms in $\cI_{12}$ originate from the
Chern-Simons $C_3 \, G_4 \, G_4$ coupling and 
$C_3 \, X_8$ coupling, respectively.
The variation of the 11d action is related to $\cI_{12}$
via standard
descent procedure,
\beq
\cI_{12} = d\cI_{11}^{(0)} \ , \qquad
\delta \cI_{11}^{(0)} = d\cI_{10}^{(1)} \ , \qquad
\delta S_{\rm 11d} = 2\pi \, \int_{M_{10}} \cI_{10}^{(1)} \ .
\eeq
As a result, the inflow anomaly polynomial for the worldvolume theory on $W_d$
is obtained by integrating $\cI_{12}$ along the $M_{10-d}$ fibers
of $M_{10}$, see \eqref{M10_fibration},
\beq \label{Iinflow}
I_{d+2}^{\rm inflow} = \int_{M_{10-d}} \cI_{12} \ .
\eeq
This quantity cancels
  against the 't Hooft anomalies
of all degrees of the freedom on $W_d$.
We are mainly interested 
in the situation in which, at low energies, 
the worldvolume theory  consists of 
an interacting CFT,
together with decoupled fields.
We may then write
\beq
I_{d+2}^{\rm inflow} + I_{d+2}^{\rm CFT} + I_{d+2}^{\rm decoupl} = 0 \ .
\eeq
Uusually, one is interested in deriving $I_{d+2}^{\rm CFT}$.
In this case, the quantity $I_{d+2}^{\rm decoupl}$ has to be identified
and subtracted by hand from $I_{d+2}^{\rm inflow}$.

\subsection{Inflow tools for D3-branes} \label{sec_D3branes}

We would like to develop a formalism analogous to the
one of the previous section that can be applied to type IIB setups.
For definiteness, we first consider a stack of $N$ D3-branes
with worldvolume $W_4$ inside the spacetime $M_{10}$.

The stack supports localized degrees of freedom
that yield a non-trivial QFT coupled to the 10d bulk.
In the IR, it consists of $\cN = 4$ super Yang-Mills (SYM) with gauge
group $SU(N)$, together with a free 4d $\cN = 4$ vector multiplet.
The local Lorentz symmetry $SO(1,9)$ of type IIB
is broken to $SO(1,3) \times SO(6)$,
with $SO(1,3)$ identified with the local Lorentz symmetry
of the worldvolume theory, and $SO(6)$ identified
with its   R-symmetry. 
(More precisely, the R-symmetry is $\text{Spin}(6) \cong SU(4)$.)
The 4d worldvolume theory
contains chiral degrees of freedom that induce a cubic 't Hooft
anomaly for the $SU(4)$ R-symmetry.
Both the interacting SCFT and the decoupling modes
admit a Lagrangian description, and the anomaly can be computed 
with standard methods. One has
\beq \label{worldvolume_anomaly}
I_6^{\rm CFT}  =   \frac 12  \, (N^2-1) \, c_3(SU(4)) \ , \qquad
I_6^{\rm decoupl} = \frac 12   \, c_3(SU(4)) \ , \qquad
\eeq
where $c_3$ denotes the third Chern class.

We expect that the anomaly \eqref{worldvolume_anomaly} is counterbalanced by
inflow from the type IIB bulk.
This was indeed verified in \cite{Kim:2012wc} by using the 
coupling $\int_{W_4} i^* C_4$ on the D3-brane worldvolume, where $C_4$ is the type IIB
4-form potential and $i^*$ is pullback along the embedding of $W_4$ inside $M_{10}$.
Our strategy, however, is different. 
In analogy with our M-theory analysis, we 
aim at performing anomaly inflow by removing 
a small neighborhood of the D3-brane stack.
Instead of using the coupling 
$\int_{W_4} i^* C_4$, our goal is to describe the variation
of the action for the 10d bulk of type IIB supergravity
in the presence of a boundary.
In the rest of this subsection we describe
a prescription to do so
when only D3-brane sources are activated.
 At the moment we do not have a direct first-principle
derivation of our formulae, also due to the fact that 
the self-duality of $F_5$ flux in type IIB makes it harder to
write down an action. We nonetheless offer a motivation
for our method. Moreover, we test it thoroughly in
several examples in the following sections.

Let us remove a small tubular neighborhood
of the D3-brane stack.
The 10d spacetime $M_{10}$ acquires a boundary $M_9$,
located at $r = \epsilon$, where $r$ is the radial coordinate
away from the branes, and $\epsilon$ is a small
positive constant.
The space $M_9$   is an $S^5$ fibration over $W_4$,
\beq \label{M9_bundle}
S^5 \hookrightarrow M_9 \rightarrow W_4 \ .
\eeq
We think of the $S^5$ fiber as
the unit sphere in the fibers of normal bundle $NW_4$
to $W_4$,
or equivalently as
 the $S^5$ that surrounds the D3-brane
stack in its six transverse directions.

To proceed we must give an appropriate boundary condition
for the $F_5$ flux of type IIB supergravity
in the vicinity of $r = \epsilon$. In analogy
with \eqref{G4_boundary}, we write
\beq \label{F5_boundary}
\frac{F_5}{2\pi} = (1 + *_{10}) \, \Big[ - \rho \, E_5 + \dots \Big] \ .
\eeq
In the previous expression, $*_{10}$ denotes the
Hodge star with respect to the 10d metric,
so that $F_5$ is manifestly self-dual.
Inside the bracket, the bump function $\rho = \rho(r)$ is as above,
and the ellipsis denote terms with $dr$ legs and/or subleading
terms in the limit $r \rightarrow \epsilon$.
The quantity $E_5$ is a globally-defined 5-form on $M_9$.

In analogy with \eqref{simple_E4}, the natural guess for $E_5$  is
\beq
E_5 = N  \, e_5 \ ,
\eeq
where  $e_5$ is the global angular form of $SO(6)$.
The latter is globally defined on $M_9$ and integrates to 1
along the $S^5$ fibers of $M_9$.
The explicit expression of $e_5$ is as follows,
\begin{align}
e_5  &= \frac{1}{\pi^3} \, \bigg[
 \frac{1}{5!} \, \epsilon_{ABCDEF} \, y^A \, Dy^B \, Dy^C \, Dy^D \, Dy^E \, Dy^F
- \frac{1}{48} \, \epsilon_{ABCDEF} \, F^{AB} \, y^C \, Dy^D \, Dy^E \, Dy^F
\nn \\
& \phantom{= \frac{1}{\pi^3} \, \bigg[} {}+ \frac{1}{64} \, \epsilon_{ABCDEF} \, F^{AB}  \, F^{CD} \, y^E \, Dy^F
\bigg] \ , \qquad Dy^A := dy^A - A^{AB} \, y_B \ .  \label{full_e5} 
\end{align}
The quantities $y^A$, $A = 1, \dots,6$ are 
constrained coordinates on $S^5$, satisfying $y^A \, y_A =1$,
with $SO(6)$ indices raised and lowered with $\delta$.
The 1-forms $A^{AB}$ are the components of the $SO(6)$ connection,
and $F^{AB}$ denote the corresponding field strengths.
In contrast 
with $e_4$,   the 5-form $e_5$ is not closed.
More precisely, the 6-form $de_5$ has only legs
along the external $W_4$ directions, and is given by
\beq \label{full_de5}
de_5   = \frac{1}{48} \, \frac{1}{(2\pi)^3} \, \epsilon_{ABCDEF} \, F^{AB } \, F^{CD} \, F^{DE}  \ .
\eeq
An equivalent, more compact way of expressing \eqref{full_de5} is 
\beq \label{de5}
de_5 = - \pi^* \, \Big[ \chi_6(SO(6)) \Big] =  - \pi^* \, \Big[ c_3(SU(4)) \Big] \ .
\eeq
In the above expression, the 6-form
$\chi_6(SO(6))$ is the Euler class of the normal bundle to the D3-brane stack.
Under $SO(6) \cong SU(4)$, it
yields the third Chern class $ c_3(SU(4))$.
The map $\pi:M_9 \rightarrow W_4$
is the projection map of the bundle \eqref{M9_bundle},
and $\pi^*$ in the pullback from the base $W_4$ to
the total space $W_9$. In what follows,
 for the sake of notational simplicity,
we omit $\pi^*$ from formulae like \eqref{de5}.

The next step in our analysis is to use the boundary
condition $E_5$ to build a suitable 11-form    
$\cI_{11}$, which is going to be the type IIB analog of $\cI_{12}$
in M-theory.
The class $\cI_{11}$ must be such that
the inflow anomaly polynomial $I_6^{\rm inflow}$
is given by integrating $\cI_{11}$ along the $S^5$
fibers of $M_9$. The quantity $I_6^{\rm inflow}$ 
should counterbalance the 't Hooft anomalies
of interacting and decoupling modes on the D3-branes,
\beq
I_6^{\rm inflow} = \int_{S^5} \cI_{11} \ , \qquad
I_{6}^{\rm inflow} + I_{6}^{\rm CFT} + I_{6}^{\rm decoupl} = 0 \ ,
\eeq
with $I_{6}^{\rm CFT}$, $I_{6}^{\rm decoupl}$ given in \eqref{worldvolume_anomaly}. 
The relation \eqref{de5} suggests a simple
definition of $\cI_{11}$,
\beq \label{cI11_simple}
\cI_{11} =  \frac 12 \, E_5 \, dE_5 \ .
\eeq
Indeed, we have
\begin{align}
I_6^{\rm inflow} = - \frac 12 \, N^2 \int_{S^5} \, e_5 \, c_3(SU(4))
= - \frac 12 \, N^2 \, c_3(SU(4)) \ ,
\end{align}
where in the last step we used the fact that  in our conventions   $e_5$
integrates to $1$ on the $S^5$ fibers.
We see that our definition of $\cI_{11}$ reproduces
the anomalies of $\cN = 4$ SYM with gauge group $SU(N)$,
plus one free vector multiplet.
Notice that \eqref{cI11_simple}
does not originate from a Chern-Simons coupling in the
type IIB effective action. Indeed, we argue below that its
origin is the kinetic term for $F_5$, due to self-duality
of the latter.

The fact that \eqref{cI11_simple}   reproduces the anomalies
of 4d $\cN = 4$ SYM is  non-trivial.
In section \ref{sec_4dexamples} 
we test our definition \eqref{cI11_simple}
in several other examples, including  D3-branes at a tip
of a Calabi-Yau cone. We find that 
\eqref{cI11_simple} correctly captures  the
inflow anomaly polynomial for all these 4d theories.

\subsection{The class $\cI_{11}$ } \label{cI11_in_IIB}

Before proceeding with tests of  \eqref{cI11_simple},
we would like to discuss its generalization
to more general type IIB setups.
More precisely, we aim at including 
the boundary conditions for the $F_3$ and $H_3$
fluxes of type IIB inside $\cI_{11}$.
For the time being, we do not include terms
involving derivatives of the axion $C_0$ or the dilaton
$\phi$.
We comment on such terms in section \ref{sec_ftheory}.

Since we are focusing on backgrounds with $dC_0 = 0$,
the Bianchi identities of $F_3$ and $H_3$ are standard,
$dF_3 = 0$, $dH_3 = 0$.
In analogy with \eqref{G4_boundary},
we write
\beq
\frac{F_3}{2\pi} = - \rho \, \cF_3 + \dots \ , \qquad
\frac{H_3}{2\pi} = - \rho \, \cH_3 + \dots \ ,
\eeq
where $\cF_3$ and $\cH_3$ are closed and globally
defined 3-forms on $M_9$.
We then argue that \eqref{cI11_simple} generalizes to
\beq \label{cI11_medium}
\cI_{11} =  \frac 12 \, E_5 \, dE_5 + E_5 \, \cF_3 \, \cH_3 \ .
\eeq
The new term in $\cI_{11}$  is consistent with the $SL(2,\mathbb Z)$ symmetry
of type IIB. Indeed $F_5$ (and hence $E_5$)
is an $SL(2,\mathbb Z)$ singlet,
while 
$F_3$ and $H_3$ (and hence $\cF_3$ and $\cH_3$) transform as a doublet.
As a result, the 6-form $\cF_3 \, \cH_3$ is an $SL(2,\mathbb Z)$
singlet.

The term $E_5 \, \cF_3 \, \cH_3$ in $\cI_{11}$
originates from the   Chern-Simons coupling
$C_4 \, F_3\, H_3$ in the type IIB effective action.
In contrast, the term $E_5 \, dE_5$ is intuitively related 
to the kinetic term for $F_5$. Notice that, due to the self-duality constraint on
$F_5$, the na\"ive kinetic term in the type IIB pseudo-action
vanishes. In order to clarify the relation between
$E_5 \, dE_5$ and the kinetic term for $F_5$
we can consider the reduction of type IIB on a circle to nine dimensions.
This is discussed in appendix \ref{app_circle},
where we provide indirect evidence for the relative
weight of the two terms in \eqref{cI11_medium}.

As a final remark, we point out that no 
higher-derivative
corrections
to \eqref{cI11_medium} are allowed, under the assumption
that $dC_0 = 0 = d\phi$. More precisely, we cannot include
any  terms involving the curvature 2-form of the 10d metric.
A priori, the 11-form $\cI_{11}$ might contain 
the terms
\beq
p_1(TM_{10}) \, \omega_7 \ , \qquad
p_1(TM_{10})^2 \, \omega_3 \ , \qquad
p_2(TM_{10}) \, \omega_3' \ , \qquad
\chi_{10}(TM_{10}) \, \omega_1 \ ,
\eeq
where $\chi_{10}(TM_{10})$
is the Euler class of $TM_{10}$.
The forms $\omega_7$, $\omega_3$, $\omega_3'$, $\omega_1$
must be built with $E_5$, $\cF_3$, $\cH_3$ and be
$SL(2,\mathbb Z)$ invariant.
It is easy to see, however, that such forms cannot be constructed.
The structure of $\cI_{11}$ is much richer
if we allow terms built with gradients of $C_0$, $\phi$,
as we   discuss in section \ref{sec_ftheory}.



\section{Four-dimensional examples} \label{sec_4dexamples}

In this section we verify that the 11-form $\cI_{11}$
given in \eqref{cI11_simple} correctly captures the inflow anomaly
polynomial of the worldvolume theory of a stack
of D3-branes at the tip of a Calabi-Yau cone. 
Since we consider setups that only have D3-brane charge,
the dilaton profile is constant and the fluxes $F_3$ and $H_3$ play no role.

After discussing some general properties of the 5-form $E_5$,
we compute the inflow anomaly polynomial for the case
of a Calabi-Yau which is a cone over a smooth Sasaki-Einstein 
manifold. We compare the inflow result to the known 4d worldvolume theory,
which  consists of an interacting $\cN = 1$ SCFT plus decoupling modes.
We show that the inflow anomaly polynomial computed from   
\eqref{cI11_simple} cancels exactly against the anomalies
of the SCFT and of the decoupling modes, up to terms involving
accidental symmetries that emerge in the IR.

As another example, we consider D3-branes probing a
$\mathbb C^2/\Gamma$ singularity, with $\Gamma$ an ADE
subgroup of $SU(2)$. 
The worldvolume theory
is an $\cN = 2$ SCFT, plus decoupling modes.
We check the inflow anomaly polynomial against the
total anomalies of the worldvolume theory,
and we get a match.

\subsection{General form of  $E_5$}  \label{sec_general_properties}

We consider a stack of D3-branes extended 
along an uncompactified worldvolume $W_4$.
In the six transverse directions, the branes
sit at the tip of a Calabi-Yau cone $Y_3$.
The latter is a metric cone over a compact smooth
Sasaki-Einstein space ${\rm SE}_5$,
\beq \label{good_cone}
ds^2(Y_3) = dr^2 + r^2 \, ds^2({\rm SE}_5) \ .
\eeq
The metric $g_{mn}$ on ${\rm SE_5}$
satisfies the Einstein condition $R_{mn} = 4 \, g_{mn}$.
The worldvolume theory in the IR
consists of an interacting 4d $\cN = 1$ SCFT,
together with decoupling modes.

Let us consider the 5d supergravity theory that is obtained
from compactification of type IIB supergravity on $\rm SE_5$.
This supergravity theory contains massless gauge fields.
They correspond to global continuous  symmetries of the
worldvolume theory.
In the 5d supergravity, massless gauge fields originate from two sources:
\begin{enumerate}
\item Isometries of $\rm SE_5$: the 5d massless vectors are off-diagonal components of the 10d metric along the direction of Killing vectors of $\rm SE_5$.
\item Harmonic 3-forms on $\rm SE_5$: the 5d massless vector are obtained expanding
the $C_4$ potential of type IIB supergravity onto a basis of harmonic 3-forms.
\end{enumerate}
After we remove a small tubular neighborhood
of the D3-branes, the boundary $M_9$ of 10d spacetime takes the form
\beq \label{SE_is_fibered}
{\rm SE_5} \hookrightarrow M_9 \rightarrow W_4 \ .
\eeq
The   fibering of $\rm SE_5$ over $W_4$ is due to the
background connections for symmetries associated to isometries of $\rm SE_5$.

Our task is the construction of the 5-form $E_5$
that enters the boundary condition for $F_5$ on $M_9$,
as in \eqref{F5_boundary}.
The form $E_5$ contains the external connections listed
in points 1. and 2.~above. Moreover, 
there are two natural requirements on $E_5$:
\begin{enumerate}[(i)]
\item The form $E_5$ is globally defined on $M_9$, and in particular
it is invariant under gauge transformations of the background connections
associated to isometries of $\rm SE_5$.
\item If all external connections are turned off, the form $E_5$
reduces to $N \, V_5$, where $N$ is the number of D3-branes in the 
stack, and $V_5$ is the volume form on $\rm SE_5$.  
\end{enumerate}
In our conventions, $V_5$ is normalized as
\beq
\int_{\rm SE_5} V_5 =  1 \ .
\eeq
In order to discuss efficiently the fibration \eqref{SE_is_fibered},
it is convenient to introduce some notation
for isometries of ${\rm SE_5}$.

We denote the Killing vector of $\rm SE_5$
as $k_I^m$, where $m = 1, \dots, 5$ is a curved tangent on index $\rm SE_5$
and $I$ labels the generators of the isometry group
of $\rm SE_5$. The Lie algebra of Killing vectors reads
\beq
[k_I , k_J]^m = f_{IJ}{}^K \, k_K^m \ ,
\eeq
where $f_{IJ}{}^K$ are the structure constants.
Let $\xi^m$ be local coordinates on $\rm SE_5$,
and let $\Lambda$ be a $p$-form on $\rm SE_5$,
$\Lambda = \frac{1}{p!} \, \Lambda_{m_1 \dots m_p} \, d\xi^{m_1} \dots d\xi^{m_p}$.
The form $\Lambda$ is not  invariant under a gauge transformation
of the background connections. We can 
remedy this problem by introducing a ``gauged'' version of $\Lambda$.
It is denoted $\Lambda^{\rm g}$ and it is defined by
\beq \label{gauging_def}
\Lambda^{\rm g} = \frac{1}{p!} \, \Lambda_{m_1 \dots m_p} \, D\xi^{m_1} \dots
D\xi^{m_p} \ , \qquad
D\xi^m = d\xi^m + k^m_I \, A^I \ ,
\eeq
where $A^I$ is the background connection for the symmetry
associated  to the  $I$-th isometry generator of $\rm SE_5$.
The field strength of $A^I$ reads
\beq
F^I = dA^I - \frac 12 \, f_{JK}{}^I \, A^K \, A^K \ .
\eeq
A useful identity to compute derivatives of $\Lambda^{\rm g}$ is
\beq \label{isom_identity}
d(\Lambda^{\rm g}) + A^I \, (\pounds_I \Lambda)^{\rm g} = (d\Lambda)^{\rm g} + F^I \,
(\iota_I \Lambda)^{\rm g} \ ,
\eeq
where $\pounds_I$ is the Lie derivative
along $k_I^m$, and $\iota_I$ denotes the interior product
of the vector $k_I^m$ with a $p$-form.

After these preliminaries we are in a position to present $E_5$.
It is given by
\beq \label{simplest_E5}
E_5 = N \, \bigg( V_5^{\rm g} +  \frac{F^I}{2\pi} \,  \omega_I^{\rm g} 
+ \frac{F^\alpha}{2\pi} \, \omega_\alpha^{\rm g} \bigg) \ .
\eeq
In the above expressions, $\omega_\alpha$ is a basis of harmonic 3-forms 
on $\rm SE_5$. The external 2-forms $F^\alpha = dA^\alpha$ are the field strengths of the connections
associated to the harmonic 3-forms, as per point 2.~above.
We notice that a harmonic 3-form is automatically invariant
under Lie derivative along all isometry directions,\footnote{From
$d\omega_\alpha=0$ we derive
$\pounds _I \omega_\alpha = d(\iota_I \omega_\alpha)$.
Making use of $\nabla_{(m} k_{I|n)}  = 0$ and  $\nabla^m \omega_{\alpha mnp} = 0$,
we verify
$(\pounds_I \omega_\alpha)_{mnp} = \nabla^q (k_I \wedge \omega_\alpha)_{qmnp}$.
We have thus established that the 3-form $\pounds_I \omega_\alpha$
is both exact and co-exact.
It follows that 
 $\int_{{\rm SE_5}} (\pounds_I \omega_\alpha) *  (\pounds_I \omega_\alpha) = 0$ (no sum over $\alpha$, $I$), which in turn guarantees
$\pounds_I \omega_\alpha = 0$.
} 
\beq
\pounds_I \omega_\alpha = 0 \ .
\eeq
This condition ensures that the term $F^\alpha \, \omega_\alpha^{\rm g}$ in $E_5$
is invariant under gauge transformations of the  external connections $A^I$. 
We stress that, while $d\omega_\alpha = 0$,
we have $d(\omega_\alpha^{\rm g}) = F^I \, (\iota_I \omega_\alpha)^{\rm g}$
by virtue of \eqref{isom_identity}.

The quantities $\omega_I$ in \eqref{simplest_E5} are 3-forms on $\rm SE_5$,
determined as follows.
The volume form $V_5$ is closed and invariant under the action
of the isometries of $\rm SE_5$, $dV_5  = 0$, $\pounds_I V_5 = 0$.
It follows that, for each $I$, $\iota_I V_5$ is a closed 4-form on $\rm SE_5$.
A Sasaki-Einstein space, however, has no harmonic 4-forms,\footnote{Its first Betti number is zero because the first Betti number of any compact and orientable 
Riemannian manifold of positive definite Ricci curvature is zero,
see {\rm e.g.}~\cite{goldberg2011curvature} theorem 3.2.1 page 87.
}
and thus $\iota_I V_5$ is   exact.
The 3-form $\omega_I$ is then defined by the relation
\beq \label{domegaI}
d\omega_I + 2 \pi \, \iota_I V_5 = 0  \ .
\eeq
We notice that,
in order to ensure that $E_5$ is invariant under gauge transformations of 
the connections $A^I$, the 3-forms $\omega_I$ must satisfy
\beq \label{omega_gauge_inv}
\pounds_I \omega_J = f_{IJ}{}^K \, \omega_K \ .
\eeq
This relation is compatible with \eqref{domegaI}.\footnote{Indeed, 
using \eqref{domegaI} 
and the identities $\pounds_I \iota_J - \iota_J \pounds_I = f_{IJ}{}^K \, \iota_K$,
$\pounds_I V_5 = 0$, we derive $d\pounds_I \omega_J = f_{IJ}{}^K \, d\omega_K$.
By modifying $\omega_I$ by a  exact piece if necessary, we can achieve
 \eqref{omega_gauge_inv}.}

The form $E_5$ in \emph{not} closed.
Indeed, with the help of \eqref{isom_identity} and the Bianchi identities
for $F^I$, $F^\alpha$, we find
\begin{align} \label{simplest_dE5}
dE_5 & = N \, F^I \, F^J \, \frac{(\iota_I \omega_J)^{\rm g}}{2\pi}
+ N \, F^I \, F^\alpha \, \frac{(\iota_I \omega_\alpha)^{\rm g}}{2\pi} \ .
\end{align}
Crucially, by virtue
of \eqref{domegaI} there
is a cancellation between $d(V_5^{\rm g})$ and $F^I d(\omega_I^{\rm g})$,
in such a way that all terms in $dE_5$ have two external field strengths.

\subsubsection*{Comments}

The expressions \eqref{simplest_E5}, \eqref{simplest_dE5} deserve some comments.

Firstly, we point out that $E_5$ contains terms associated to an expansion
onto harmonic 3-forms, but does not contain terms associated to expansion
onto the dual harmonic 2-forms. Including such terms would be redundant,
since they are generated by $*_{10} E_5$ when we construct
$F_5 = E_5 + *_{10} \, E_5$.

Secondly, we notice that a non-zero $dE_5$ is not in contradiction
with the Bianchi identity for $F_5   = E_5 + *_{10} \, E_5$.
The latter is the boundary condition for the physical 5-form field of type IIB, which (in the absence of $F_3$, $H_3$)
must be closed and self-dual on shell.
In appendix \ref{sec_dF5app} we show that our expression  \eqref{simplest_E5} for $E_5$
is indeed compatible with $dF_5 = 0$. Moreover, we show that $dF_5 = 0$
is the origin of the condition \eqref{domegaI} on the 3-forms $\omega_I$.

Next, there seems to be a tension between \eqref{simplest_dE5},
which holds for a general Sasaki-Einstein manifold,
and \eqref{full_de5}, which holds for the global angular form $e_5$ associated to a round $S^5$
and shows that $de_5$ is purely horizontal.
We also notice that $e_5$ in \eqref{full_e5}
contains terms quadratic in $F^{AB}$, which are 
crucial in guaranteeing \eqref{full_de5} but are
absent 
from the parametrization \eqref{simplest_E5}.
The key observation to reconcile \eqref{full_e5} and \eqref{simplest_E5}
is that we can modify $e_5$ into a different $e_5'$ without affecting the 
inflow anomaly polynomial,
\beq
\int_{S^5} e_5 \, de_5 = \int_{S^5} e_5' \, de_5' \ ,
\eeq
where the new form $e_5'$ is obtained from $e_5$ by omitting
the term quadratic in $F^{AB}$,
\begin{align}
e_5'  &= \frac{1}{\pi^3} \, \bigg[
 \frac{1}{5!} \, \epsilon_{ABCDEF} \, y^A \, Dy^B \, Dy^C \, Dy^D \, Dy^E \, Dy^F
- \frac{1}{48} \, \epsilon_{ABCDEF} \, F^{AB} \, y^C \, Dy^D \, Dy^E \, Dy^F
\bigg] \ , \nn   \\
de_5' & = -  \frac{1}{8} \, \frac{1}{(2\pi)^3} \, \epsilon_{ABCDEF} \, F^{AB } \, F^{CD} \, Dy^D \, Dy^E  \label{new_de5}  \ .
\end{align}
As expected, $de_5'$ in \eqref{new_de5} is no longer purely external,
but rather has the structure \eqref{simplest_dE5}.

The equivalence between $e_5$ and $e_5'$ for the purposes
of anomaly inflow is a specific example of a more general property
of $E_5$, demonstrated in appendix \ref{sec_dF5app}:
as soon as \eqref{domegaI} holds, we are free to add arbitrary ``non-minimal'' $FF\lambda$
terms to $E_5$ (where $\lambda$ is a 1-form on $\rm SE_5$)
without modifying the value of the integral $\int_{\rm SE_5} E_5 \, dE_5$.

The example of $S^5$ shows that non-minimal terms can be tuned in such a way
as to ensure that $de_5$ is purely horizontal.
It is natural to wonder if this holds true for a generic Sasaki-Einstein space.
We show in appendix \ref{sec_dF5app} that, as soon as $\rm SE_5$ admits
harmonic 3-forms, there is an obstruction to making $dE_5$ purely horizontal:
there is no choice of non-minimal terms such that $dE_5$ is the pullback of a 6-form
in external spacetime. Thus, in the presence of harmonic 3-forms,
a relation of the form \eqref{simplest_dE5} is the ``most horizontal possible''
for $dE_5$.

Finally, we would like to point out  that the 3-forms $\omega_I$
are not uniquely determined by \eqref{domegaI}, since they can be shifted
by an arbitrary closed 3-form. We argue in appendix \ref{sec_dF5app}
that this ambiguity has no effect on the inflow anomaly polynomial.

\subsubsection*{Collective notation}

In what follows, it is   convenient to introduce a shorthand
  notation for describing all external connections  collectively.
We introduce the new index $X = (I,\alpha)$ and we write
\beq \label{collective_notation}
F^X = ( F^I \ , \  F^\alpha  ) \ , \qquad
\omega_X = (\omega_I \ , \ \omega_\alpha)  
 \ .
\eeq
As a result, we may rewrite \eqref{simplest_E5} and \eqref{simplest_dE5} as
\beq \label{very_compact}
E_5 = N \, \bigg( V_5^{\rm g} + \frac{F^X}{2\pi} \, \omega_{X}^{\rm g} \bigg) \ , \qquad
dE_5 = 2\pi N \,  \frac{F^X}{2\pi} \, \frac{F^Y}{2\pi} \, (\iota_X \omega_Y)^{\rm g} \ ,
\eeq
with the understanding that the operation $\iota_X$
is defined to be $\iota_I$ if $X = I$ and is defined to be
zero if $X = \alpha$.
We also notice that
the closure property $d\omega_\alpha = 0$ for the harmonic 3-forms
can be combined with \eqref{domegaI} into a single relation,
\beq \label{domegaX}
d\omega_X + 2\pi\, \iota_X V_5 = 0 \ .
\eeq

\subsection{Inflow analysis for smooth $\rm SE_5$}  \label{sec_smoothSE5}

In this subsection we compute the inflow anomaly polynomial
in the case in which $\rm SE_5$ is a smooth manifold
admitting a possibly non-Abelian isometry group
and an arbitrary number of harmonic 3-forms. 

\subsubsection*{Computation of the inflow anomaly polynomial}
Making use of \eqref{very_compact} it is immediate to verify that
\beq
\int_{\rm SE_5} E_5 \, dE_5 = 2\pi N^2 \, \frac{F^X}{2\pi} \, \frac{F^Y}{2\pi} \, 
\frac{F^Z}{2\pi} \, \int_{\rm SE_5} \omega_X \, \iota_Y \omega_Z \ .
\eeq
As a result, the inflow anomaly polynomial obtained from \eqref{cI11_simple}
can be written as 
\beq \label{cone_inflow}
I_6^{\rm inflow} = 
\frac 16 \, c_{XYZ} \, \frac{F^X}{2\pi} \, \frac{F^Y}{2\pi} \, \frac{F^Z}{2\pi} \ , \qquad
\frac 16 \, c_{XYZ} = \frac 12 \, N^2 \cdot 2\pi \,
\int_{\rm SE_5}  \omega_{(X}  \, \iota_Y \omega_{Z)} \ ,
\eeq
where the total symmetrization $(XYZ)$ is performed
with weight 1, \emph{i.e.}~with the combinatorial prefactor $1/6$.

Our expression for $I_6^{\rm inflow}$  agrees with the
results of   \cite{Benvenuti:2006xg},
where the anomalies of the interacting SCFT on the D3-brane
were derived at leading order in $N$ from the 5d supergravity effective action.\footnote{The collective index $X$ here corresponds to the index $I$ in \cite{Benvenuti:2006xg}. The
normalization of the 3-forms $\omega$ here and in \cite{Benvenuti:2006xg} is the same,
as can be seen from (2.15) in that paper,
taking into account that ${\rm vol}^\circ$ there is the same as $V_5$ here,
and that the quantity $k_I$ there contains
a factor $2\pi$, as stated above their (2.15).
By a similar token, our expression for the $c$ coefficients
agrees with (2.20) in \cite{Benvenuti:2006xg}, taking into account that 
they have
 an implicit $2\pi$ factor in the interior product $\iota$.
In our expression, this $2\pi$ factor is explicit.}
Notice in particular that $I_6^{\rm inflow}$ is proportional to $N^2$,
without subleading terms. This is due to the fact that
we have included a prefactor $N$ in front of the harmonic
3-forms $\omega_\alpha$ in $E_5$.
As explained in \cite{Benvenuti:2006xg}, this is the correct prescription
to reproduce the charge of D3-brane states that are charged
under the baryonic $U(1)$ symmetries associated to the
harmonic 3-forms $\omega_\alpha$.

Anomaly inflow should yield results that are exact in $N$,
and not just the leading order part in the large $N$ limit.
To verify this claim, we must take into account the 
whole worldvolume theory, including decoupled sectors.
We address this analysis in a class of examples in the next subsection.

\subsubsection*{Comparison with worldvolume theory}

For the sake of concreteness, in this subsection we focus on the
case of a toric Calabi-Yau cone with smooth Sasaki-Einstein base.
We expect, however, that the   picture 
we describe should hold for general Calabi-Yau cones.

The worldvolume theory on a stack of 
D3-branes at the tip of a toric Calabi-Yau cone
is an 
 $\cN = 1$ quiver gauge theory
with  bifundamental and adjoint matter chiral superfields,
and a superpotential.
The quiver and the superpotential are extracted from
the toric diagram of the Calabi-Yau cone \cite{Franco:2005sm}.
Let the label $i$ enumerate the nodes of the quiver.
At the node $i$ we have a gauge
group $U(N_i)$. In the toric phase, $N_i = N$ for each $i$,
but for the sake of generality we consider possibly distinct $N_i$'s  in what follows.

The quiver gauge theory with $U(N_i)$ gauge groups is not conformal.
In the IR, the $U(1)$ factor inside each $U(N_i)$ decouples.
We are then left with a quiver with $SU(N_i)$ gauge groups,
and one free vector multiplet for each node in the quiver.
Moreover, each chiral field in the adjoint representation of $U(N_i)$,
of dimension $N_i^2$, splits into
a chiral field in the adjoint representation of $SU(N_i)$,
of dimension $N_i^2-1$,
plus one free massless chiral field.
In contrast, the bifundamental representation of $U(N_i) \times U(N_j)$,
of dimension $N_i \, N_j$, simply
becomes the bifundamental representation of $SU(N_i) \times SU(N_j)$,
of the same dimension, without any free chiral field decoupling.

For $i \neq j$, let $m_{ij}$ be the number of 
chiral superfields in
the bifundamental  of $SU(N_i) \times SU(N_j)$.
We denote these fields as $X_{ij,\alpha}$,
with $\alpha = 1, \dots, m_{ij}$.
In a similar way, if there are $m_{ii}$ chiral superfields
in the adjoint of $SU(N_i)$, we denote them as $X_{ii,\alpha}$
with $\alpha = 1, \dots, m_{ii}$.
From the discussion of the previous paragraph,
we know that each $X_{ii,\alpha}$ comes accompanied by a free
chiral superfield, which we denote $Y_{ii,\alpha}$.

The interacting CFT defined by the quiver with $SU(N_i)$ gauge
groups admits 
global symmetries. We choose a basis in which
$R_0$ is a 
 reference $U(1)$ R-symmetry,
while all other global symmetries are flavor
symmetries. We ignore
non-Abelian
flavor symmetries, if present,
and we denote the generators of $U(1)$
flavor symmetries as $T_\cI$.

The generators $R_0$ and $T_\cI$
must be free of ABJ anomalies
with the generators of each gauge group $SU(N_i)$. This requirement gives
\begin{align} \label{noABJ}
0 & = N_i  + \sum_{\alpha = 1}^{m_{ii}} N_i \, \big( R_0[X_{ii,\alpha}] -1 \big)
+ \frac 12 \, \sum_{j \neq i} \, \sum_{\alpha = 1}^{m_{ij}} \, N_j \,
\big( R_0[X_{ij,\alpha}] -1 \big)
+ \frac 12 \, \sum_{j \neq i} \, \sum_{\alpha = 1}^{m_{ji}} \, N_j \,
\big( R_0[X_{ji,\alpha}] -1 \big) \ , \nn \\
0 & = \sum_{\alpha = 1}^{m_{ii}} N_i \, T_\cI [X_{ii,\alpha}]
+ \frac 12 \, \sum_{j \neq i} \, \sum_{\alpha = 1}^{m_{ij}} \, N_j \,
 T_\cI [X_{ij,\alpha}]  
+ \frac 12 \, \sum_{j \neq i} \, \sum_{\alpha = 1}^{m_{ji}} \, N_j \,
T_\cI [X_{ji,\alpha}]   \ .
\end{align}
The symbol $R_0[X_{ii,\alpha}]$
denotes the charge of the scalar $X_{ii,\alpha}$
under the generator $R_0$, and similarly
for other scalars and generators.
The $R_0$ and $T_\cI$ charges of the free chiral superfields
$Y_{ii,\alpha}$ are not constrained by ABJ anomalies,
because the fields $Y_{ii,\alpha}$ are gauge singlets.
Given the common origin of $Y_{ii,\alpha}$ and $X_{ii,\alpha}$
from the adjoint representation of $U(N_i)$,
the natural charge assignments for $Y_{ii,\alpha}$ are
\beq \label{Y_charges}
R_0[Y_{ii,\alpha}] = R_0[X_{ii,\alpha}] \ ,\qquad
T_\cI[Y_{ii,\alpha}] = T_\cI[X_{ii,\alpha}] \ .
\eeq
It follows that, if we consider the interacting CFT together
with the free chiral fields $Y_{ii,\alpha}$, and one free vector multiplet
for each node in the quiver, we have
\begin{align} \label{no_linear_traces}
{\rm Tr}_\text{CFT \!+\! free} \, R_0  = 0  \ , \qquad
{\rm Tr}_\text{CFT \!+\! free} \, T_\cI  = 0   \ .
\end{align} 
This is   derived by multiplying the 
conditions \eqref{noABJ} for the $i$th node by $N_i$,
and summing over $i$, as in \cite{Benvenuti:2004dw}.

Let us  now consider the quantity ${\rm Tr}_\text{CFT \!+\! free} \, (abc)$,
where $a,b,c\in\{R_0, T_\cI\}$ not necessarily distinct.
If a bifundamental field $X_{ij,\alpha}$ contributes to
${\rm Tr}_\text{CFT \!+\! free} \, (abc)$, it does so with a prefactor
$N_i \, N_j$,  because this is the dimension of the gauge representation
in which $X_{ij,\alpha}$ transforms.
By a similar token, if $X_{ii,\alpha}$ contributes,
it does so with a prefactor $N_i^2 -1$.
Because of the charge assignments \eqref{Y_charges},
the contribution of $Y_{ii,\alpha}$ is identical to that of
$X_{ii,\alpha}$. As a result, 
$X_{ii,\alpha}$ and $Y_{ii,\alpha}$ give together a contribution
with a prefactor 
$N_i^2$.
From these considerations, it follows that 
${\rm Tr}_\text{CFT \!+\! free} \, (abc)$ is 
an order $N^2$ quantity,
without any $\cO(1)$ terms.
More precisely, 
let $N$ be the greatest common divisor of the $N_i$'s,
so that we may write $N_i = N \, n_i$ with coprime $n_i$'s.
Then all dependence of 
${\rm Tr}_\text{CFT \!+\! free} \, (abc)$ on $N$
is via an overall factor $N^2$.

It should be noted that each free chiral field $Y_{ii,\alpha}$
comes together with an additional $U(1)$ factor in
the global symmetry group of the theory,
with generator $\widehat T_{ii,\alpha}$.
These  
are accidental symmetries of the IR theory.
All fields in the system have charge zero
under $\widehat T_{ii,\alpha}$, except the 
free chiral field $Y_{ii,\alpha}$, which by convention
has charge $1$.

The superconformal R-symmetry of the total system
comprised of the interacting CFT and the free fields is of the form
\beq
R_{\cN = 1} = R_0 + \sum_\cI \, s_\cI \, T_\cI + \sum_{i} \sum_{\alpha = 1}^{m_{ii}} s_{ii,\alpha} \, \widehat T_{ii,\alpha} \ ,
\eeq
for suitable values of the parameters $s_I$, $s_{ii,\alpha}$.
We notice that, if we did not include the $\widehat T_{ii,\alpha}$
generators, then the interacting field $X_{ii,\alpha}$
and the free field $Y_{ii,\alpha}$ would have had the same
charge under $R_{\cN = 1}$, because they have the same
charges under $R_0$ and $T_\cI$.
Clearly this would be in tension with the fact that
$X_{ii,\alpha}$ has a non-trivial anomalous dimension,
while $Y_{ii,\alpha}$ has dimension 1.
This puzzle is resolved by the terms
with $\widehat T_{ii,\alpha} $ in $R_{\cN = 1}$.
The parameter $s_{ii,\alpha}$ can always be tuned in such a way that
$R_{\cN = 1}[Y_{ii,\alpha}] = 2/3$, as appropriate for a free chiral field.

Let $c_1^0$ be the first Chern class of the background
connection for the $R_0$ symmetry,
$c_1^\cI$ the first Chern class for the symmetry $T_\cI$,
and $c_1^{ii,\alpha}$ for the accidental symmetry $\widehat T_{ii,\alpha}$.
The anomaly polynomial of the  CFT together with 
 the free fields takes the form
\begin{align}
I_6^\text{CFT} + I_6^{\rm decoupl} & = I_6^{N^2}(c_1^0, c_1^\cI) + I_6^{\rm accidental}(
c_1^0, c_1^\cI, c_1^{ii,\alpha} ,  p_1(TW_4)
) \ .
\end{align}
We have collected all terms containing 
$c_1^{ii,\alpha}$ in $I_6^{\rm accidental}$,
while 
the remaining terms without any $c_1^{ii,\alpha}$ factor
are gathered in $ I_6^{N^2}$.
Notice that $ I_6^{N^2}$ does not contain $p_1(TW_4)$
by virtue of \eqref{no_linear_traces}.
Moreover,  $ I_6^{N^2}$ has an overall $N^2$ factor.
In contrast, $I_6^{\rm accidental}$ is independent of $N$.
In fact, $I_6^{\rm accidental}$ only receives contributions
from the free chiral fields $Y_{ii,\alpha}$.
The total number of such fields is determined
by the quiver to be $\sum_i m_{ii}$, but it does not scale
with the ranks of the gauge groups at the nodes of the quiver.

We notice that the quantity $I_6^{N^2}(c_1^0, c_1^\cI)$
has an  equivalent interpretation:
it is the leading large-$N$ part of the anomaly polynomial
of the interacting CFT without free fields.
In \cite{Benvenuti:2006xg} it is demonstrated that 
the formula \eqref{cone_inflow} for the inflow anomaly
coefficients agrees with the large-$N$ anomaly coefficients
on the field theory side for any toric Calabi-Yau cone.
This means that we can write
\beq
I_6^{\rm inflow} = - I_6^{N^2}(c_1^0, c_1^\cI) \ , \qquad
I_6^\text{CFT} + I_6^{\rm decoupl} 
+ I_6^{\rm inflow}  =  I_6^{\rm accidental}(
c_1^0, c_1^\cI, c_1^{ii,\alpha} , p_1(TW_4)
) \ .
\eeq
In conclusion, the inflow anomaly polynomial
matches exactly the anomalies of the worldvolume
theory on the D3-branes, up to accidental symmetries
that emerge in the IR from the decoupling of free chiral multiplets.
Our expectation is that this conclusion should hold
for any Calabi-Yau cone. It would be interesting to
explore the relations between this proposal
and the theories discussed in \cite{Fazzi:2019gvt}.

\subsection{D3-branes probing a $\mathbb C^2  / \Gamma$ singularity}

In this subsection we consider a class of examples  that yield
4d $\cN = 2$ SCFTs. The background geometry probed by the D3-branes
is $Y_3 = (\mathbb C^2/\Gamma)\times \mathbb C$, where $\Gamma$ is
an ADE subgroup of $SU(2)$. While $Y_3$ is a Calabi-Yau cone, the associated Sasaki-Einstein
base is $S^5/\Gamma$ and has orbifold singularities.
To compute the inflow anomaly polynomial we resolve
these singularities by blow-up, in the spirit of \cite{Klebanov:1998hh}.

\subsubsection*{Anomaly inflow computation}

Let us consider the type IIB background $\mathbb R^{1,3} \times (\mathbb C^2/\Gamma) \times \mathbb C$,
where $\Gamma$ is an ADE subgroup
of $SU(2)$.  We insert  a stack of D3-branes extended along
$\mathbb R^{1,3}$ and located at the origin of $(\mathbb C^2/\Gamma) \times \mathbb C$. This setup preserves 4d $\cN = 2$ supersymmetry and 
has been studied in \cite{Douglas:1996sw,Johnson:1996py}.
We introduce coordinates $z_1 = y_1 + i \, y_2$, $z_2 = y_3 + i \, y_4$
for the $\mathbb C^2$ factor acted upon by $\Gamma$,
while we use $z_3 = y_5 + i \, y_6$ for the other $\mathbb C$ factor. 
The isometry group $SO(6)$ of $\mathbb C^3 \cong \mathbb R^6$
is reduced by the action of $\Gamma$ according to
\beq \label{preserved}
SO(6) \rightarrow G_L \times SU(2)_R \times U(1)_\phi \ .
\eeq
The factors $ G_L \times SU(2)_R$ are the subgroup
of the $SO(4) \cong SU(2)_L \times SU(2)_R$ rotating $y_1$, $y_2$, $y_3$, $y_4$ that commutes
with the action of $\Gamma$,
\beq \label{GL_def}
G_L = \left\{
\begin{array}{ll}
SU(2)_L & \quad \text{for $\Gamma = \mathbb Z_2$}  \ , \\
U(1)_L & \quad \text{for $\Gamma = \mathbb Z_k$, $k \ge 3$} \  , \\
\text{trivial}& \quad \text{for $\Gamma$ of D, E type} \ .
\end{array}
\right.
\eeq
The group $U(1)_\phi$ is identified with rotations in the $y_5 y_6$ plane,
with $\phi$ defined to be the polar angle in the usual way, $z_3 = |z_3| \, e^{i\phi}$.
The isometries $SU(2)_R \times U(1)_\psi$ are identified 
with the R-symmetry of the worldvolume theory on the D3 branes.

All points on the $y_5 y_6$ plane, with $y_1 = \dots = y_4 = 0$,
are fixed points of the action of $\Gamma$.
The unit sphere $S^5 \subset \mathbb R^6$ intersects
the set of fixed points along the circle $y_5^2 + y_6^2 = 1$ in the $y_5 y_6$ plane,
which we denote as $S^1_\phi$. As a result, the quotient
$S^5/\Gamma$ has a circle of orbifold singularities
located along $S^1_\phi$.

If we consider $\mathbb C^2/\Gamma$ in isolation, the orbifold singularity at the origin
can be resolved in a canonical way, introducing a set of resolution 
2-cycles. Each resolution cycles is a copy of $\mathbb C \mathbb P^1$.
We have $\text{rank}(\mathfrak g_\Gamma)$ resolution cycles, where $\text{rank}(\mathfrak g_\Gamma)$ is the rank
of the ADE Lie algebra $\mathfrak g_\Gamma$ associated to $\Gamma$.
We use the notation $\mathbb C \mathbb P^1_\alpha$,
$\alpha = 1, \dots, \text{rank}(\mathfrak g_\Gamma)$.
The intersection pairing of the resolution 2-cycles reproduces the Cartan matrix of
$\mathfrak g_\Gamma$. To each resolution 2-cycle in $\mathbb C^2/\Gamma$
we can associate a Poincar\'e dual harmonic 2-form. We denote these   harmonic 2-forms as $\widetilde \omega_\alpha$.
We have
\beq \label{Cartan_integral}
\int_{\mathbb C^2/\Gamma} \widetilde \omega_\alpha \, \widetilde \omega_\beta
= - \cC_{\alpha \beta} \ ,
\eeq
where $\cC_{\alpha \beta}$ is the Cartan matrix of $\mathfrak g_\Gamma$.

If we now turn to $S^5/\Gamma$, if we blow up the orbifold singularities
along $S^1_\phi$ we introduce a set of $\text{rank}(\mathfrak g_\Gamma)$ 3-cycles,
of the form $\mathbb C \mathbb P^1_\alpha \times S^1_\phi$.
The blow-up can be performed while preserving the $U(1)_\phi$ isometry
of $S^5/\Gamma$. The 3-cycles $\mathbb C \mathbb P^1_\alpha \times S^1_\phi$
in the blow-up of $S^5/\Gamma$ are dual to a set of harmonic 3-forms,
denoted $\omega_\alpha$. We can write
\beq
\omega_\alpha = \widetilde \omega_\alpha \, \frac{d\phi}{2\pi} \ .
\eeq
The 2-forms $\widetilde \omega_\alpha$ were previously defined
on $\mathbb C^2/\Gamma$. We can extend them to $S^5/\Gamma$;
by abuse of notation, we use the same symbol $\widetilde \omega_\alpha$.
After the extension, these 2-forms are supported along the $S^1_\phi$
circle at $y_1 = \dots = y_4= 0$. They do not depend on the coordinate
$\phi$, and they do not have any $d\phi$ leg.

Let us now discuss $E_5$ for the setup under consideration.
It takes the form
\beq \label{GammaE5}
E_5 = N \, |\Gamma| \, e_5^{S^5} + \frac{F^\alpha}{2\pi} \, \bigg[
(\omega_\alpha)^{\rm g}
+ F^{AB}  \, (\lambda_{AB\alpha})^{\rm g}
\bigg] \ .
\eeq
In the previous expression, $e_5^{S^5}$ 
can be taken to be the global angular form of $SO(6)$.
Its expression is recorded in appendix \ref{app_Gamma}.
The quantities $F^{AB}$ are the components of the  curvature for the
background $SO(6)$ connection.
As stated in \eqref{preserved}, only a subgroup of $SO(6)$
is preserved by the action of $\Gamma$.
It is therefore implicitly understood that 
the only non-zero components of $F^{AB}$
are those along the generators of the subgroup
$G_L \times SU(2)_R \times U(1)_\phi$.
The 2-forms $F^\alpha$ in \eqref{quotient_F5} are external
field strengths for the $U(1)^{\text{rank}(\mathfrak g_\Gamma)}$ global symmetry
originating from the 3-cycles in the blow-up of $S^5/\Gamma$.
Moreover, we can write
\beq \label{gauged_omega_alpha}
(\omega_\alpha)^{\rm g} = \widetilde \omega_\alpha \,  \frac{D\phi}{2\pi} \ , \qquad
D\phi = d\phi - A_\phi   \ ,
\eeq
where $A_\phi$ is the background connection for $U(1)_\phi$.
Notice that the gauging does not affect the 2-forms 
$\widetilde \omega_\alpha$. This is because they are localized
at $y_1 = \dots =y_4 = 0$, they do not depend on $\phi$,
and they do not have any $d\phi$ leg.
The 1-forms $\lambda_{AB\alpha}$ can be left arbitrary, since
we check   that the anomaly does not depend on them.

The computation of the inflow anomaly polynomial
from $E_5$ in \eqref{GammaE5}
is recorded in appendix \ref{app_Gamma}. The result reads
\beq \label{Gamma_inflow}
I_6^{\rm inflow} = \frac 12 \, \int_{S^5/\Gamma} E_5 \, dE_5 
= N^2 \, |\Gamma| \, c_1^R \, \Big[ c_2(SU(2)_R) - c_2(G_L) \Big]
+ \cC_{\alpha \beta} \, c_1^R \, c_1^\alpha \, c_1^\beta \ .
\eeq
In writing the above expressions,
we have identified the field strengths $F_\phi$, $F^\alpha$
with 4d Chern classes according to
\beq \label{4d_identifications}
\frac{F_\phi}{2\pi} = 2 \, c_1(U(1)_{R_{\cN = 2}}) \equiv   2\, c_1^R  \  , \qquad
\frac{F^\alpha}{2\pi} = c_1(U(1)_\alpha) \equiv c_1^\alpha \ .
\eeq
Moreover, we have introduced the shorthand notation
\beq \label{c2GL}
c_2( G_L ) = \left\{
\begin{array}{ll}
c_2(SU(2)_L) & \quad \text{for $\Gamma = \mathbb Z_2$}  \ , \\
- c_1(U(1)_L)^2 & \quad \text{for $\Gamma = \mathbb Z_k$, $k \ge 3$} \  , \\
0 & \quad \text{for $\Gamma$ of D, E type} \ .
\end{array}
\right.
\eeq

\subsubsection*{Comparison with worldvolume theory}

The worldvolume theory on a stack of D3-branes probing the 
$\mathbb C^2/\Gamma$ singularity is a  4d $\cN = 2$ quiver gauge theory 
\cite{Douglas:1996sw,Johnson:1996py}. The quiver has the shape of the affine Dynkin diagram
of the Lie algebra $\mathfrak g_\Gamma$ associated to $\Gamma$.
The total gauge group is of the form
\beq
G_{\rm gauge} = \prod_i U(N \, n_i) \ ,
\eeq
where the product is over nodes of the affine Dynkin
diagram, and the quantities $n_i$  are integers associated to each node.
In table \ref{table_quivers} we depict the quivers
with their $n_i$ assignments. 
Each link in the quiver represents a 
bifundamental hypermultiplet. In the IR,
the $U(1)$ factors in the gauge group
decouple. We are left with a quiver with $SU$ 
gauge groups, which describes an interacting $\cN = 2$ SCFT,
together with a number of free $\cN = 2$ vector multiplets,
equal to the number of nodes in the quiver,
which is ${\rm rank}(\mathfrak g_\Gamma) + 1$.

According to the general anomaly inflow paradigm,
$I_6^{\rm inflow}$ should balance against the contributions
of all degrees of freedom on the worldvolume theory of the branes.
We should then have
\beq \label{quiver_claim}
I_6^{\rm inflow} + I_6^{\rm worldvol} = 0 \ , \qquad
  I_6^{\rm worldvol} = I_6^\text{$SU$ quiver} + I_6^\text{free vec.~multiplets}  \ .
\eeq
Since the worldvolume theory is a Lagrangian theory,
we can readily compute   $I_6^{\rm worldvol}$
and use it as a check of 
 $I_6^{\rm inflow}$ given in \eqref{Gamma_inflow}.

\begingroup
\renewcommand{\arraystretch}{1.4}

\begin{table}
\centering
 \begin{tabular}{| c  | c | c | c | c }
\hline  $\mathfrak g_\Gamma$   & $\text{rank}(\mathfrak g_\Gamma)$ & $|\Gamma|$ & quiver   \\ \hline \hline
\noindent\parbox[c]{1cm}{\centering $\mathfrak {su}(k)$} & 
\noindent\parbox[c]{1cm}{\centering $k-1$} & 
\noindent\parbox[c]{1cm}{\centering $k$} & 
\noindent\parbox[c]{3.5cm}{\centering
\includegraphics[height=1.75cm]{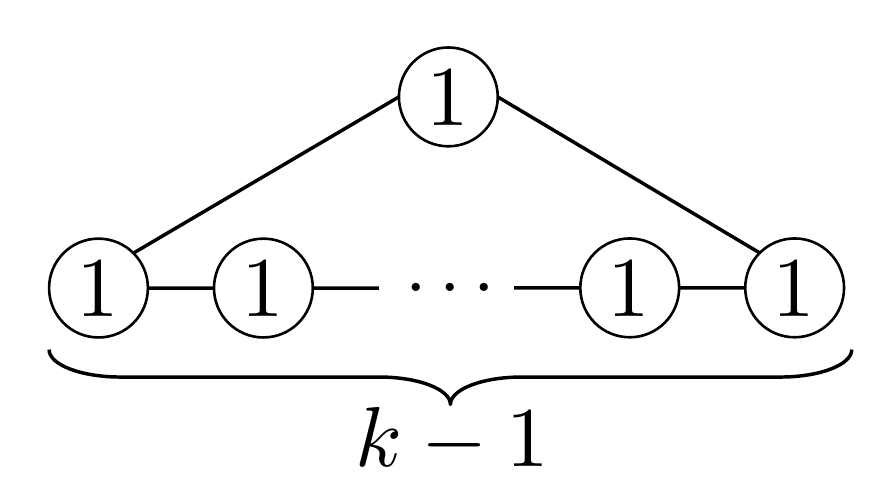}
}
\\ \hline
\noindent\parbox[c]{1cm}{\centering $\mathfrak {so}(2k)$} & 
\noindent\parbox[c]{1cm}{\centering $k$} & 
\noindent\parbox[c]{1cm}{\centering $4k-8$} & 
\noindent\parbox[c]{5.cm}{\centering
\includegraphics[height=1.6cm]{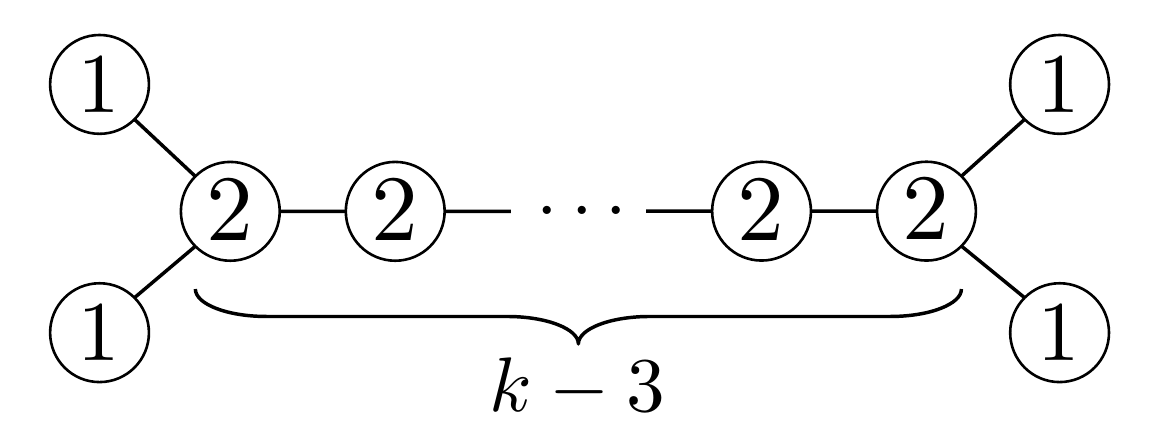}
}
\\ \hline
\noindent\parbox[c]{1cm}{\centering $\mathfrak {e}_6$} & 
\noindent\parbox[c]{1cm}{\centering $6$} & 
\noindent\parbox[c]{1cm}{\centering $24$} & 
\noindent\parbox[c]{5.cm}{\centering
\includegraphics[height=1.8cm]{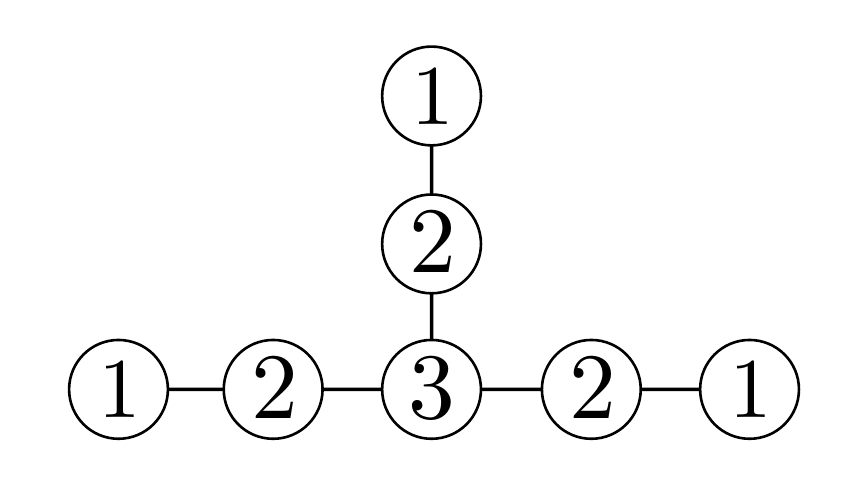}
}
\\ \hline
\noindent\parbox[c]{1cm}{\centering $\mathfrak {e}_7$} & 
\noindent\parbox[c]{1cm}{\centering $7$} & 
\noindent\parbox[c]{1cm}{\centering $48$} & 
\noindent\parbox[c]{5.cm}{\centering
\includegraphics[height=1.5cm]{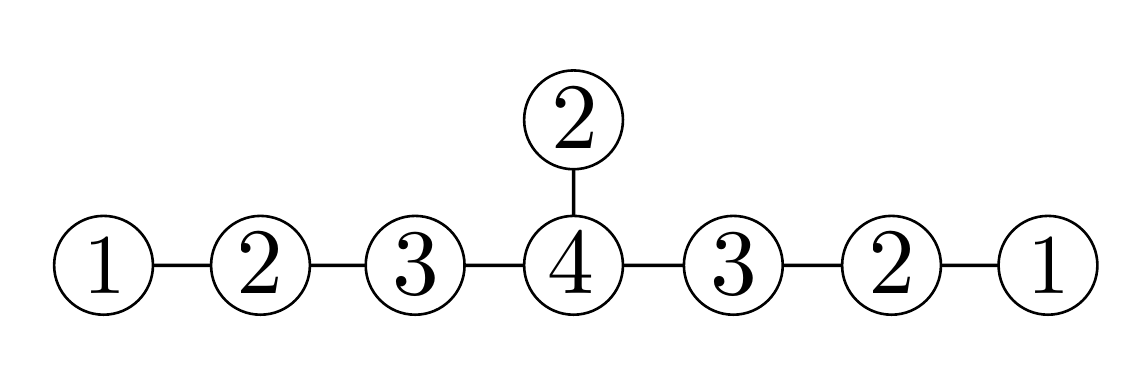}
}
\\ \hline
\noindent\parbox[c]{1cm}{\centering $\mathfrak {e}_8$} & 
\noindent\parbox[c]{1cm}{\centering $8$} & 
\noindent\parbox[c]{1cm}{\centering $120$} & 
\noindent\parbox[c]{5.cm}{\centering
\includegraphics[height=1.5cm]{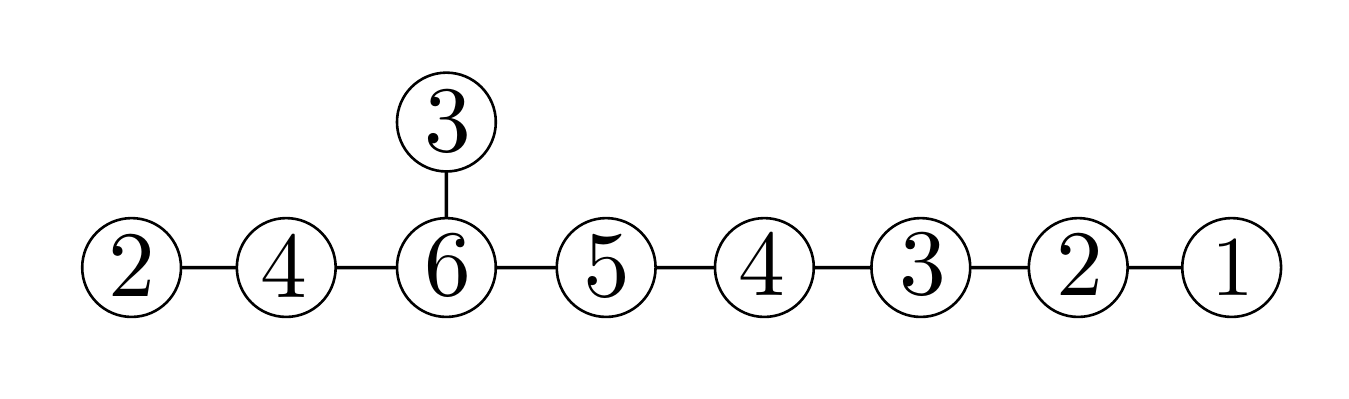}
}
\\ \hline
\end{tabular}
\caption{
For each ADE subgroup $\Gamma$  of $SU(2)$,
we give the associated Lie algebra $\mathfrak g_\Gamma$,
its rank, the order $\Gamma$ of the group,
and the quiver that describes the worldvolume theory
of D3-branes probing $\mathbb C^2/\Gamma$.
The number associated to each node is denoted $n_i$
in the text. A node with label $n_i$ corresponds to a gauge
group $U(N \, n_i)$.
} 
\label{table_quivers}
\end{table}

\endgroup

As a first check, let us verify that the symmetries
visible in the inflow computation correspond to
the global symmetries of the worldvolume theory.
The inflow geometry $S^5/\Gamma$ has isometry
group $G_L \times SU(2)_R \times U(1)_\phi$,
with $G_L$ given in \eqref{GL_def}.
Moreover, the resolution 3-cycles of $S^5/\Gamma$
provide an additional $U(1)^{\text{rank}(\mathfrak g_\Gamma)}$
global symmetry. 
On the field theory side
we have an 
 $SU(2)_R \times U(1)_{R_{\cN = 2}}$ R-symmetry,
which is identified with the isometries  $SU(2)_R \times U(1)_\phi$.
Moreover, each hypermultiplet gives a $U(1)$ global symmetry.
The case $\Gamma = \mathbb Z_2$ is special, since 
the quiver has two nodes connected by two links.
As a result, the hypermultiplets 
contribute a factor $U(2) \cong SU(2) \times U(1)$ to the global symmetry of the theory.
In summary, the flavor symmetry of the D3-brane worldvolume theory for each
$\Gamma$ is
\begin{align}
\mathfrak g_\Gamma &= \mathfrak{su}(2) \; :  & 
 G_{\rm flavor} &=  SU(2)_L \times U(1) \ , \nn \\
\mathfrak g_\Gamma &= \mathfrak{su}(k)  \ , \; k \ge 3\; : &    G_{\rm flavor} &= U(1)_L \times U(1)^{k-1} \ , \nn \\
\mathfrak g_\Gamma &= \mathfrak{so}(2k) \; : &    G_{\rm flavor} &=  U(1)^{k-1} \ , \nn \\
\mathfrak g_\Gamma &= \mathfrak e_{6,7,8}  \; : &    G_{\rm flavor} &=  U(1)^{6,7,8} \ .
\end{align}
These global symmetries correspond to those
visible in the inflow computation. The factors with a subscript
$L$ in the A series are identified with the $G_L$ isometry 
of $S^5/\mathbb Z_k$.
The other factors are $U(1)$'s and their number is equal
to the number of resolution 3-cycles in the blow-up
of $S^5/\Gamma$.

Let us now discuss $I_6^{\rm worldvol}$. 
We compute\footnote{In
our conventions,
${\rm Tr} \, R_{\cN = 2} \, I^a \, I^b = {\rm Tr} \, R_{\cN = 2} \, (I^3)^2 \, \delta^{ab}$,
$\delta^{ab} \, \frac{F_a}{2\pi} \,
\frac{F_b}{2\pi} =  p_1(SO(3)_R)  =  - 4 \, c_2(SU(2)_R)$,
where $I^a$ are the generators of $SU(2)_R$.
}
\begin{align} \label{quiver_anomaly}
I_6^\text{worldvol} = 
-   N^2 \,  c_1^R \, c_2(SU(2)_R) \sum_i n_i^2
- \sum_x M_x \, c_1^R \, (c_1^x)^2 \ .
\end{align}
In the above expression, $i$ labels the nodes of the quiver,
while $x$ labels the links. The quantity $c_1^x$ is the
first Chern class of the $U(1)_x$ flavor symmetry of the
hypermultiplet at the link $x$. The integer $M_x$
is the product of the ranks of the two $U$ gauge groups
connected by the link~$x$.
If we describe the hypermultiplet living at the link $x$
as the pair $(Q_x, \widetilde Q_x)$ of $\cN = 1$ chiral multiplets,
then in our conventions $Q_x$ has charge $+1$ and $\widetilde Q_x$
has charge $-1$ under the flavor symmetry $U(1)_x$.
The expression \eqref{quiver_anomaly} holds for $\Gamma \neq \mathbb Z_2$.
For $\Gamma = \mathbb Z_2$, we have
\begin{align} \label{quiver_anomalyZ2}
I_\text{$6$, $\Gamma = \mathbb Z_2$}^\text{worldvol} = 
-  2\, N^2 \,  c_1^R \, c_2(SU(2)_R) 
+ 2 \, N^2 \, c_1^R \, c_2(SU(2)_L)
-2\,  N^2 \, c_1^R \, c_1(U(1))^2
 \ .
\end{align}
We have recalled that the flavor symmetry associated 
to the double link is $SU(2)_L \times U(1)$.
The chiral multiplet $Q$ is in the fundamental of $SU(2)_L$
and has charge $+1$ under $U(1)$,
while $\widetilde Q$ is in the antifundamental of $SU(2)_L$
and has charge $-1$ under $U(1)$.

We can now compare \eqref{quiver_anomaly} and \eqref{Gamma_inflow}
to verify \eqref{quiver_claim}.
Let us first check the case $\Gamma = \mathbb Z_2$.
The inflow result \eqref{Gamma_inflow} reads in this case
\beq \label{Gamma_inflowZ2}
I_6^{\rm inflow} 
= 2\, N^2 \,c_1^R \, \Big[ c_2(SU(2)_R) - c_2(SU(2)_L) \Big]
+ 2  \, c_1^R \, (c_1^{\alpha=1})^2  \ ,
\eeq
where $c_1^{\alpha = 1}$ denotes the first Chern 
class associated to the unique resolution 3-cycle
in the blow up of $S^5/\mathbb Z_2$.
We match \eqref{quiver_anomalyZ2}
with the identification $c_1^{\alpha = 1} = N \, c_1(U(1))$.

Next, let us consider the case $\Gamma = \mathbb Z_k$,
or $\mathfrak g_\Gamma  = \mathfrak{su}(k)$.
The quiver gauge theory result \eqref{quiver_anomaly} becomes
\begin{align} \label{quiver_anomalyZk}
I_6^\text{worldvol} = 
-   N^2 \, k\, c_1^R \, c_2(SU(2)_R)  
- N^2 \, \sum_{i = 1}^{k}  c_1^R \, (c_1^{(i,i+1)})^2 \ .
\end{align}
For quivers of A type, it is convenient to trade the link label
$x$ for a pair $(i,i+1)$, with the understanding that
the link $(i,i+1)$ connects the $i$-th and $(i+1)$-th nodes in the quiver.
(The $i$ index is understood modulo $k$, so that the $(k+1)$-th node is
by definition the first node.)
Let us consider the following redefinition of
the external curvatures,
\begin{align}
\begin{array}{l c lllll}
N \, c_1^{(1,2)} & = & N \, c_1(U(1)_L) &   &   & + & c_1^{\alpha = 1} \ , \\
N \, c_1^{(2,3)} & = & N \, c_1(U(1)_L) & - & c_1^{\alpha = 1} & + & c_1^{\alpha = 2} \ , \\
& \vdots\\
N \, c_1^{(k-1,k)} & = & N \, c_1(U(1)_L) & - & c_1^{\alpha = k-2} & + & c_1^{\alpha = k-1} \ , \\
N \, c_1^{(k,1)} & = & N \, c_1(U(1)_L) & - & c_1^{\alpha = k-1}  \ .&   &  
\end{array}
\end{align}
The anomaly polynomial of the worldvolume theory
takes the form
\begin{align} \label{quiver_anomalyZkBIS}
I_6^\text{worldvol} = 
-   N^2 \, k\, c_1^R \, c_2(SU(2)_R)  
- N^2 \, k \, c_1^R \, c_1(U(1)_L)^2
-   \sum_{\alpha, \beta = 1}^{k-1}  \cC_{\alpha \beta} \, c_1^R \, 
c_1^\alpha \, c_1^\beta
\ ,
\end{align}
where $\cC_{\alpha \beta}$ is the standard Cartan matrix of 
$\mathfrak{su}(k)$, with 2's on the diagonal entries and $-1$'s
on the subdiagonal and superdiagonal entries.
The expression \eqref{quiver_anomalyZkBIS}
shows that $-I_6^\text{worldvol}$
is exactly equal to $I_6^{\rm inflow}$ in 
\eqref{Gamma_inflow}.

Finally, let us briefly discuss the D and E cases.
Let us focus first on the mixed 't Hooft anomaly between
$U(1)_{R_{\cN = 2}}$ and $SU(2)_R$.
The relation \eqref{quiver_claim} holds for this part of the anomaly
polynomial by virtue of the relation
\beq
\sum_i n_i^2 = |\Gamma| \ ,
\eeq
which is valid for every choice of $\Gamma$,
see table \ref{table_quivers}.
If $\mathfrak g_\Gamma$ is of D or E type,
 the number of links in the quiver is equal
to the rank of $\mathfrak g_\Gamma$.
As a result, the labels $\alpha$ and $x$ both have range
$1$ to $\text{rank}(\mathfrak g_\Gamma)$.
By a suitable change of basis, we can obtain
\beq
\sum_{\alpha, \beta =1}^{\text{rank} (\mathfrak g_\Gamma)}  \cC_{\alpha \beta} \, c_1^\alpha \, c_1^\beta = 
\sum_{x,y=1}^{\text{rank} (\mathfrak g_\Gamma)} \, M_x \, \delta_{x,y} \, c_1^x \, c_1^y \ .
\eeq
Notice that $M_x$ is proportional to $N^2$. As a result 
there is a factor $N$
in the 
change of basis relating $c_1^\alpha$ to $c_1^x$,
as in the case of the A series discussed above.

For the sake of completeness, let us give the anomaly polynomial
of the free vector multiplets that decouple in the IR,
\beq
I_6^\text{free vec.~multiplets}  = \big[
{\rm rank}(\mathfrak g_\Gamma) + 1
 \big] \bigg[
\frac 13 \, (c_1^R)^3 - \frac{1}{12} \, c_1^R \, p_1(TW_4)
- c_1^R \, c_2(SU(2)_R)
 \bigg] \ .
\eeq
Let us also notice that the central charges of the total
worldvolume theory are
\beq
 a^{\rm worldvol} = c^{\rm worldvol} =    \frac 14 \, N^2 \, |\Gamma|  \ ,
 \eeq
while the decoupling    vector multiplets contribute
\beq
(a,c)^\text{free vec.~multiplets} = \left( 
\frac{5}{24} \ ,
 \frac{1}{6} 
 \right) \, \big[
{\rm rank}(\mathfrak g_\Gamma) + 1
 \big] \ .
\eeq



\section{Two-dimensional examples} \label{sec_2dstory}

In this section we use the 11-form $\cI_{11}$ to compute the inflow
anomaly polynomial for setups with D3-branes wrapping a Riemann surface.
We first discuss a setup with D3-branes at the tip of a generic
Calabi-Yau cone, with worldvolume compactified on a Riemann surface
without punctures.
Compactifications of D3-brane theories on Riemann surfaces
have been intensively investigated \cite{Vafa:1994tf,Bershadsky:1995vm,Maldacena:2000mw,Benini:2012cz,Benini:2013cda,Bobev:2014jva,Benini:2015bwz}.
Next,
we focus on 4d $\cN = 4$ SYM   on a Riemann
surface with half-BPS punctures.

\subsection{$\rm SE_5$ fibrations over a smooth Riemann surface} \label{sec_smoothSigma}

In this section, our starting point is the 4d SCFT living on
a stack of D3-branes probing a given Calabi-Yau cone,
with base $\rm SE_5$. This 4d SCFT is compactified to two dimensions
on a genus-$g$ Riemann surface without punctures.
We focus on the case $g\neq 1$.
In order to preserve supersymmetry,  we perform the appropriate 
twist of R-symmetry over the Riemann surface. We also allow
for twists of $U(1)$ flavor symmetries of the SCFT associated 
to isometries of~$\rm SE_5$.

As expected on the grounds of anomaly matching
across dimensions, the inflow anomaly polynomial $I_4^{\rm inflow}$
for the 2d theory is closely related
to the inflow anomaly polynomial of the parent 4d theory $I_6^{\rm inflow}$.
Our analysis demonstrates how to correctly
identify 4d and 2d background curvatures
in the integration of $I_6^{\rm inflow}$ over $\Sigma_g$.

\subsection*{Some preliminaries}
The relevant internal geometry for anomaly inflow is the 7d space
\beq \label{M7_def}
{\rm SE_5} \hookrightarrow M_7 \rightarrow \Sigma_g \ .
\eeq
The fibering of $\rm SE_5$ over $\Sigma_g$ encodes the partial
topological twist of the parent 4d theory in the compactification
to two dimensions.
Throughout this section, we use a bar to distinguish objects and 
labels associated to the ${\rm SE_5}$ fibers of $M_7$.
For example, the normalized volume form on $\rm SE_5$
is denoted $\overline V_5$. The isometries of $\rm SE_5$
are labelled by the indices $\bar I$, $\bar J$, and so on.

The fibration \eqref{M7_def} can be described 
by assigning   background  fluxes for the connections associated to the isometries
of $\rm SE_5$.
We may parametrize such background fluxes by writing
\beq \label{Sigma_background}
F^{\bar I}_\Sigma = p^{\bar I} \, V_\Sigma \ , \qquad
\int_{\Sigma_g} V_\Sigma = 2\pi \ ,
\eeq
where the integer parameters $p^{\bar I}$ specify which generators of the (Cartan
subalgebra   of) isometries of $\rm SE_5$ are twisted over the Riemann surface.
For any given choice of parameters $p^{\bar I}$, the residual isometry
group of $\rm SE_5$ that is preserved by the twist is comprised
by those linear combination of generators that commute
with the background flux.
We use the index $I$ to label the generators
of the preserved subgroup. We may then write
\beq \label{embedding_isom}
t_I = s_I{}^{\bar I} \, t_{\bar I} \ ,
\eeq
where $t_{\bar I}$ are all generators of the isometry group
of $\rm SE_5$, $t_I$ are the generators of the preserved subgroup,
and $s_I{}^{\bar I}$ are suitable constants. The latter satisfy
\beq \label{subgroup_condition}
s_I{}^{\bar I} \, p^{\bar J} \, f_{\bar I \bar J}{}^{\bar K } = 0 \ ,
\eeq
where $f_{\bar I \bar J}{}^{\bar K } $ are the structure constants
of the full isometry group of $\rm SE_5$.
The condition \eqref{subgroup_condition} is simply encoding the 
fact that the generators $t_I$ commute with the background flux.

In this work we only consider twists that preserve (0,2) supersymmetry
in two dimensions.  
Let us fix a reference R-symmetry generator
$R_0$ in the  4d SCFT, and suppose $R_0$ is
given in terms of the isometry generators of $\rm SE_5$ as
\beq
R_{0} =    s_{R_0}{}^{\bar I} \, t_{\bar I} \ ,
\eeq
for suitable constants $ s_{R_0}{}^{\bar I}$.  We may then write
\beq \label{susy_condition}
p^{\bar I} = p^{R_0} \, s_{R_0}{}^{\bar I} + p^{\bar I}_{\rm flavor} \qquad
\text{with}\qquad p^{R_0} = - \chi  \ .
\eeq
The condition $p^{R_0} = - \chi$ is needed to cancel the curvature
of $T\Sigma_g$. The term $p^{\bar I}_{\rm flavor}$ 
describes any further twisting along isometry generators that are
not R-symmetries (\emph{i.e.}~such that all Killing spinors 
of the Calabi-Yau cone are neutral under them).

Finally,
recall from section \ref{sec_general_properties} that, for each $\bar I$, the 4-form $\iota_{\bar I} \overline V_5$
is exact, \emph{i.e.}~there exists a 3-form $\overline \omega_{\bar I}$ 
on $\rm SE_5$
such that
\beq \label{iotaV5bar}
d\overline \omega_{\bar I} + 2\pi \, \iota_{\bar I} \, \overline V_5 = 0 \ .
\eeq
We use the notation $\overline \omega_{\bar \alpha}$ for the harmonic
3-forms on $\rm SE_5$, with index $\bar \alpha = 1, \dots, b^3(\rm SE_5)$.

\subsection*{Results of the anomaly inflow computation}

In \eqref{embedding_isom} we have parametrized the generators
of the isometries of the $\rm SE_5$ fiber that are compatible with the
fibration, and hence give isometries of the total space $M_7$.
These isometries correspond to global symmetries of the 2d theory.
The space $M_7$, however, might have additional isometries.
For instance, if the Riemann surface is a sphere we have an additional
$SO(3)$ isometry. Moreover, the space $M_7$ generically has
harmonic 3-forms, which correspond to additional $U(1)$ global symmetries
of the 2d field theory.
For the sake of simplicity, in this work we only discuss the 't Hooft anomalies
for the 2d symmetries associated to the isometries of $M_7$
that originate from the $\rm SE_5$ fiber.
We refer the reader to appendix \ref{app_no_punctures} for the derivation of the results stated 
below.

 The inflow anomaly polynomial $I_4^{\rm inflow}$ for the 2d theory
is conveniently expressed in terms of the inflow anomaly polynomial
$I_6^{\rm inflow}$ of the parent theory. As derived in section \ref{sec_smoothSE5},
the latter is given by \eqref{cone_inflow} and therefore takes the form
\beq \label{parent_inflow}
I_6^{\rm inflow} = \frac 16 \, c_{\bar I \bar J \bar K}  \, \frac{F^{\bar I}_{\rm 4d}}{2\pi}
\, \frac{F^{\bar J}_{\rm 4d}}{2\pi} \, \frac{F^{\bar K}_{\rm 4d}}{2\pi}
+ \frac 12 \, c_{\bar I \bar J \bar \alpha}  \, \frac{F^{\bar I}_{\rm 4d}}{2\pi}
\, \frac{F^{\bar J}_{\rm 4d}}{2\pi} \, \frac{F^{\bar \alpha}_{\rm 4d}}{2\pi}
+ \frac 12 \, c_{\bar I \bar \alpha \bar \beta}  \, \frac{F^{\bar I}_{\rm 4d}}{2\pi}
\, \frac{F^{\bar \alpha}_{\rm 4d}}{2\pi} \, \frac{F^{\bar \beta}_{\rm 4d}}{2\pi}  \ ,
\eeq
where the anomaly coefficients are given as
\begin{align} \label{c_coeffs}
c_{\bar I \bar J \bar K} &= 3 \, N^2 \, (2\pi) \, \int_{\rm SE_5} \overline \omega_{(\bar I} \, 
\iota_{\bar J} \, \overline \omega_{\bar K)} \ , \nn \\
c_{\bar I \bar J \bar \alpha} & = N^2 \, (2\pi) \, \int_{\rm SE_5}\Big[
  \overline \omega_{(\bar I} \, \iota_{\bar J)} \, \overline \omega_{\bar \alpha}
 + \overline \omega_{\bar \alpha} \, \iota_{( \bar I} \, \overline \omega_{\bar J)} 
\Big]
= 2 \, N^2 \, (2\pi) \, \int_{\rm SE_5}
\overline \omega_{\bar \alpha} \, \iota_{( \bar I} \, \overline \omega_{\bar J)}  \ , \nn \\
c_{\bar I \bar \alpha \bar \beta} & = N^2 \, (2\pi) \, \int_{\rm SE_5} 
\overline \omega_{(\bar \alpha|} \, \iota_{\bar I} \, \overline \omega_{|\bar \beta )}
= N^2 \, (2\pi) \, \int_{\rm SE_5} 
\overline \omega_{\bar \alpha} \, \iota_{\bar I} \, \overline \omega_{\bar \beta } \ .
\end{align}
In \eqref{parent_inflow} we have separated the collective index $X$
of \eqref{cone_inflow} into $(\overline I, \bar \alpha)$ and we have written explicitly the terms
associated to isometries of $\rm SE_5$ and to harmonic 3-forms
of $\rm SE_5$.
The 2-forms $F^{\bar I}_{\rm 4d}$, $F^{\bar \alpha}_{\rm 4d}$
are the 4d field strengths of the connections associated to the
symmetries of the parent 4d theory.

The result of anomaly inflow for the 2d theory can then be stated as follows.
The 2d inflow anomaly polynomial is obtained from
integration on $\Sigma_g$ of the parent 4d inflow anomaly polynomial,
\beq
I_4^{\rm inflow} = \int_{\Sigma_g} I_6^{\rm inflow} \ ,
\eeq
with the following identifications between the 4d and 2d background field
strengths,
\beq \label{4d_vs_2d}
F^{\bar I}_{\rm 4d} = F^I \, s_I{}^{\bar I} + p^{\bar I} \, V_\Sigma \ ,\qquad
F^{\bar \alpha}_{\rm 4d} = F^I \, s_I{}^{\bar \alpha} \ .
\eeq
The quantities $p^{\bar I}$ are the twist parameters introduced in \eqref{Sigma_background},
while the tensor $s_I{}^{\bar I}$ introduced in \eqref{embedding_isom} describes the embedding
of the residual isometry group after the twist inside the original
isometry group of $\rm SE_5$.
The new quantities $s_I{}^{\bar \alpha}$ in \eqref{4d_vs_2d}
are   determined by the following linear equation,
\beq \label{new_equation}
 p^{\bar K} \, c_{\bar K \bar \alpha \bar \beta} \, s_I{}^{\bar \beta}   +  s_I{}^{\bar J} \, p^{\bar K} \, 
 c_{\bar J \bar K \bar \alpha} = 0 \ .
\eeq
In general, the quantities $s_I{}^{\bar \alpha}$ are non-zero.
This means that, in uplifting the 2d curvatures $F^I$ to four dimensions,
we must also activate the vectors $F^{\bar \alpha}_{\rm 4d}$ associated 
to baryonic symmetries of the parent 4d theory.
For each fixed $I$, \eqref{new_equation} admits a unique
solution $s_I{}^{\overline \beta}$ if and only if the matrix $m_{\alpha \beta} =  p^{\bar K} \, c_{\bar K \bar \alpha \bar \beta}$ is invertible. We argue below that
this is the case for the universal supersymmetric twist.
In more general situations, invertibility of $m_{\alpha \beta}$
seems to be a consistency requirement on the choice of twist
parameters $p^{\bar K}$.

The condition \eqref{new_equation} admits an interesting interpretation.
Consider the integration of the 4d inflow anomaly polynomial
on the Riemann surface, keeping the constants
$s_I{}^{\bar \alpha}$ in \eqref{4d_vs_2d} as free parameters.
The resulting inflow anomaly polynomial in 2d  has the form $I_4^{\rm inflow} = a(s_I{}^{\bar \alpha})_{IJ} \, F^I \, F^J$,
with the anomaly coefficients $ a(s_I{}^{\bar \alpha})_{IJ} $
given as a function of the free parameters 
 $s_I{}^{\bar \alpha}$.
We have checked that imposing the condition \eqref{new_equation}
on the parameters  $s_I{}^{\bar \alpha}$ is equivalent
to extremizing simultaneously all
2d anomaly coefficients $ a(s_I{}^{\bar \alpha})_{IJ} $.

The non-trivial interplay between mesonic symmetries in 2d and baryonic
symmetries in 4d encoded in \eqref{4d_vs_2d}, \eqref{new_equation} has been observed
in
\cite{Benini:2015bwz}.

\subsection*{A comment on the universal supersymmetric twist}
By universal supersymmetric twist we mean the twist in which the vector
$p^{\bar  I}$ points exactly in the direction of the exact superconformal
R-symmetry of the parent 4d theory, as studied in \cite{Benini:2015bwz,Bobev:2017uzs}.
If the generator $R_{\cN = 1}$ of the exact superconformal R-symmetry
is given in terms of isometries of $\rm SE_5$ by
\beq
R_{\cN= 1} = s_{R_{\cN = 1}} {}^{\bar I} \, t_{\bar I} \ , 
\eeq
then the twist parameters for the universal supersymmetric twist
read 
\beq
p^{\bar I} = - \chi \,  s_{R_{\cN = 1}} {}^{\bar I} \ .
\eeq
We should stress that, as explained in \cite{Benini:2015bwz,Bobev:2017uzs}, this is a viable choice
only if the charges of all gauge-invariant operators of the 4d QFT
under $R_{\cN = 1}$ are rational. In what follows, we assume that
this condition is met.

If we choose the universal supersymmetric twist, 
the quantity $p^{\bar K} \, c_{\bar K \bar \alpha \bar \beta}$
is proportional to ${\rm Tr} (R_{\cN = 1} \, J_{\bar \alpha} \, J_{\bar \beta})$ in the SCFT,
where  $J_{\bar \alpha}$ is the generator of the $U(1)$ baryonic flavor
symmetry associated to the harmonic 3-form $\overline \omega_{\bar \alpha}$
in $\rm SE_5$. As explained in \cite{Intriligator:2003jj}, 
if we let the index $X$ label all flavor symmetries of the 4d SCFT,
the 
matrix ${\rm Tr} (R_{\cN = 1} \, J_{  X} \, J_{  Y})$ 
is negative-definite.
This implies that also the sub-matrix ${\rm Tr} (R_{\cN = 1} \, J_{\bar \alpha} \, J_{\bar \beta})$ is negative-definite.
As a result, $m_{\alpha \beta} = p^{\bar K} \, c_{\bar K \bar \alpha \bar \beta}$
is invertible, and \eqref{new_equation} admits a unique solution
for $s_I{}^{\bar \alpha}$, for each $I$.
If we consider a more general twist, in which the vector $p^{\bar I}$ deviates
from the direction of the 4d superconformal R-symmetry,
we have no general  argument to guarantee that   
$p^{\bar K} \, c_{\bar K \bar \alpha \bar \beta}$ is invertible.
We may conjecture, however, that the matrix 
$p^{\bar K} \, c_{\bar K \bar \alpha \bar \beta}$ remains non-singular
for choices of twists that do not deviate too much
from the universal supersymmetric twist.



\subsection{$\cN = 4$ SYM with half-BPS punctures} \label{sec_puncture}

In this section we consider 4d $\cN = 4$ SYM theory with gauge group
$SU(N)$, compactified on a Riemann surface with a partial topological twist
to yield a 2d $\cN = (4,4)$ theory.
This type IIB setup is the direct analog of the  M-theory setup
in which the 6d $\cN = (2,0)$ theory living on a stack of M5-branes
is compactified on a Riemann surface with a partial topological twist
to give a 4d $\cN = 2$ theory.
In this case, 
it is known how to introduce punctures
on the Riemann surface preserving $\cN = 2$ supersymmetry \cite{Gaiotto:2009we,Gaiotto:2009hg}.
In particular, we may consider a Riemann surface $\Sigma_{g,n}$ of arbitrary genus $g$
and 
 with an arbitrary number $n$ of regular punctures.

The purpose of this section is to exploit the analogy with 
the M5-brane construction to introduce punctures in the reduction
of 4d $\cN = 4$ SYM. 
We bypass a direct field-theoretic analysis
of the punctures, and instead study anomaly inflow
from the ambient space. 
In this way, we extend the M-theory anomaly inflow
approach of \cite{Bah:2018gwc,Bah:2018jrv,Bah:2019jts} to 
analogous configurations in type IIB.

In order to streamline our exposition, all derivations
for the results of this section are relegated to appendix \ref{sec_punctureapp},
together with useful background material on the
treatment of punctures along the lines of \cite{Bah:2018jrv,Bah:2019jts}.

\subsubsection{Outline of the computation} \label{sec_outline}

The computation of anomaly inflow
in the presence of (regular) punctures is based on a suitable decomposition
of the internal space $M_7$
that enters the anomaly inflow formula
\beq
I_4^{\rm inflow}= \int_{M_7} \cI_{11} \ .
\eeq
More precisely, if we consider a setup with $n$ punctures, the
space $M_7$ takes the form
\beq \label{M7_collage}
M_7 = M_7^{\rm bulk} \cup \bigcup_{\alpha = 1}^n X_7^\alpha \ ,
\eeq
where the label $\alpha$ enumerates the punctures.
The space $M_7^{\rm bulk}$ 
encodes the geometry away from the punctures and
is an $S^5$ fibration over the punctured Riemann surface,
\beq
S^5 \hookrightarrow  M_7^{\rm bulk} \rightarrow \Sigma_{g,n} \ .
\eeq
The presence of $S^5$ is due to the fact that the parent 4d theory
is $\cN = 4$ SYM. 
The fibration of $S^5$ over $\Sigma_{g,n}$ encodes the partial
topological twist. As mentioned earlier, we only consider setups that
preserve $\cN = (4,4)$ supersymmetry in 2d. In this case,
the $SO(6)$ isometry of $S^5$ (the R-symmetry of 4d $\cN = 4$ SYM)
is broken as
\beq
SO(6) \rightarrow SO(4) \times SO(2) \ ,
\eeq
and the topological twist is performed by turning a background
connection for the $SO(2)$ factor.
The residual isometry group $SO(4) \times SO(2)$ of $M_7^{\rm bulk}$
is identified with the $SU(2)^2 \times U(1)$ R-symmetry
of the 2d theory.

The spaces $X_7^\alpha$ in \eqref{M7_collage} encode the local geometry near each puncture.
Crucially, $X_7^\alpha$ is \emph{not} an $S^5$ fibration
over a 2d base space. Some aspects of the geometry of $X_7^\alpha$
are described below; a more thorough account can be found in appendix
\ref{sec_punctureapp}.

The decomposition  \eqref{M7_collage} of the internal space $M_7$ implies
a  corresponding decomposition  of the inflow anomaly
polynomial into a bulk piece, plus puncture pieces,
\beq \label{pieces_of_I4}
I_4^{\rm inflow} =  I_4^{\rm inflow} (\Sigma_{g,n})
+ \sum_{\alpha = 1}^n I_6^{\rm inflow} (P_\alpha)  \ ,
\eeq
where one has
\beq \label{various_pieces}
I_4^{\rm inflow} = \int_{M_7} \cI_{11}    \ ,
\qquad I_4^{\rm inflow} (\Sigma_{g,n}) = \int_{M_7^{\rm bulk}} \cI_{11} \ ,  
\qquad I_4^{\rm inflow} (P_\alpha) = \int_{X_7^\alpha} \cI_{11} \ .
\eeq
The task at hand is the construction of the 5-form $E_5$ for $M_7^{\rm bulk}$
and $X_7^\alpha$ and the computation of the above integrals.

\subsubsection{The bulk contribution to anomaly inflow}

The bulk anomaly inflow polynomial $I_4^{\rm inflow} (\Sigma_{g,n})$
in \eqref{various_pieces}
 can be obtained in various equivalent ways.
One can specialize the results of section \ref{sec_smoothSigma},
which are valid for any smooth Sasaki-Einstein 5-manifold,
to the case of $S^5$. Alternatively, one can take the 6-form anomaly polynomial
of 4d $\cN = 4$ SYM and integrate it on the Riemann surface.
 The result   is
\beq \label{bulk_result}
I_4^{\rm inflow} (\Sigma_{g,n}) = \frac 12 \, \int_{M_7^{\rm bulk}} E_5 \, dE_5 = 
- \frac 12 \, N^2 \, \chi(\Sigma_{g,n}) \, 
\chi_4(SO(4)) \ , 
\eeq
where we have introduced the 4-form characteristic class
\beq
\chi_4(SO(4)) =    \frac{1}{(2\pi)^2} \, \frac{1}{8} \, \epsilon_{abcd} \, 
F^{ab} \, F^{cd} \ , 
\eeq
where $F^{ab}$ is the field strength of the connection
for the $SO(4)$ isometry of $M_7^{\rm bulk}$.
The interested reader can find the  expression for the 5-form $E_5$
for the bulk of the Riemann surface in appendix \ref{sec_punctureapp},
where we also discuss non-minimal terms in $E_5$
(in the terminology of section \ref{sec_general_properties}) and how they drop
out from the anomaly inflow result.

\subsubsection{The puncture contribution to anomaly inflow}

The contribution of each puncture to anomaly inflow can be studied independently.
For this reason, let us temporarily omit the puncture label $\alpha$ to improve readability.

The salient features of the puncture geometry $X_7$ are the following.
The space $X_7$ is an $S^3_\Omega$ fibration over a 4d space $X_4$,
which is in turn a circle fibration over $\mathbb R^3$,
\beq
S^3_\Omega \hookrightarrow X_7 \rightarrow X_4 \ , \qquad
S^1_\beta \hookrightarrow X_4 \rightarrow \mathbb R^3 \ .
\eeq
The round 3-sphere $S^3_\Omega$ has $SO(4)$ isometry,
which is identified with the $SO(4)$ isometry factor
of the bulk geometry $M_7^{\rm bulk}$.
The 4d space $X_4$ has a $U(1)^2$ isometry:
one $U(1)$ factor is associated to the $S^1_\beta$ fiber,
while one $U(1)$ factor is due to the fact that
the $S^1_\beta$ fibration is axially symmetric 
in the base $\mathbb R^3$.
The latter $U(1)$ isometry is identified with the $SO(2)$ isometry factor
of $M_7^{\rm bulk}$.
The former $U(1)$ from $S^1_\beta$ does not 
yield an isometry of the total internal space $M_7$.
In fact, when the puncture geometry is glued onto the bulk geometry,
the circle $S^1_\beta$ is identified with the boundary of the small disk $D$
that is removed from the Riemann surface to 
introduce the puncture.
A more detailed description of the gluing conditions between 
bulk and puncture geometries can be found in appendix \ref{sec_punctureapp}.

The $S^1_\beta$ fibration over $\mathbb R^3$ has $p$
monopole sources, of integer positive charges $k_a$, $a = 1,\dots,p$.
All monopoles are aligned along a line in the base space $\mathbb R^3$ of $X_4$.
The positions of the monopoles are encoded in a set of parameters $\{w_a\}_{a=1}^p$.
Flux quantization implies that $\{w_a\}_{a=1}^p$ is an increasing sequence
of positive integers. The integers $\{k_a\}_{a=1}^p$, $\{w_a\}_{a=1}^p$
determine a partition of $N$,
\beq
N = \sum_{a=1}^p k_a \, w_a \ .
\eeq
This partition labels the puncture. The partition can be chosen
independently for each puncture on the Riemann surface.  As we shall see below,
the anomaly contribution of a given puncture depends on its associated 
partition of $N$.

It is worth pointing out that, at the location of the $a$-th monopole,
the 4d space $X_4$ is locally of the form
$\mathbb R^4/\mathbb Z_{k_a}$. As a result,
$X_4$ has orbifold singularities if $k_a \ge 2$.
These orbifold singularities can be resolved by blow-up 
preserving supersymmetry. The resolution introduces
additional 2-cycles in the geometry, as well as additional
harmonic 2-forms.

In the M-theory setup with wrapped M5-branes,
expansion of the $C_3$ potential onto these harmonic 2-forms
yields additional vectors. This mechanism is the origin
of flavor symmetries associated to regular punctures \cite{Gaiotto:2009gz}.
In type IIB, expansion of the $C_4$ potential onto these
harmonic 2-forms does not yield extra vectors.
As a result, the punctures in the type IIB construction
do not carry any flavor symmetry.

We are now in a position to give
the anomaly inflow polynomial 
$I_6^{\rm inflow} (P_\alpha)$ for the $\alpha$-th puncture.
It is given by
\beq \label{final_result}
I_6^{\rm inflow} (P_\alpha) =  -\chi_4(SO(4))    \,
\sum_{a=1}^{p_\alpha} \ell_{\alpha,a} \, (w_{\alpha,a}^2 - w_{\alpha,a-1}^2) \ , \qquad
\ell_{\alpha,a} = \sum_{b=a}^p k_{\alpha,b} \ .
\eeq
Since we have reintroduced the puncture label $\alpha$ on the LHS,
we have done so on the RHS too, to stress that each puncture
comes with its partition data $p_\alpha$, $k_{\alpha,a}$, $w_{\alpha,a}$.
The derivation of \eqref{final_result}
is performed in appendix \ref{sec_punctureapp},
where we also discuss in detail the 5-form $E_5$
for a puncture.



\section{Towards F-theory anomaly inflow} \label{sec_ftheory}

In this section we collect preliminary remarks on the generalization
of our anomaly inflow tools to F-theory setups.
More precisely, we want to study configurations in which the axio-dilaton
field $\tau  =C_0 + i \, e^{-\phi}$ of type IIB supergravity has a non-trivial profile over 10d spacetime
and is allowed to be multivalued,  \emph{i.e.}~to have monodromies
around singular loci.
Different values of $\tau$ at the same spacetime point
are related by the action of an element of
$SL(2,\mathbb Z)$, 
\beq \label{tau_transf}
\tau' = \frac{a \, \tau + b}{c\, \tau + d} \ , \qquad
\begin{pmatrix}
a & b \\
c & d
\end{pmatrix} \in SL(2,\mathbb Z) \ .
\eeq
A non-trivial monodromy for $\tau$ signals the presence of a 7-brane.
We refer the reader to \emph{e.g.}~\cite{Denef:2008wq,Weigand:2018rez} for  reviews on F-theory.

The $\tau$ profile in 10d spacetime is conveniently captured
by introducing an auxiliary $T^2$, or more precisely an elliptic curve 
$\mathbb E_\tau = \mathbb C/\Lambda_\tau$, 
where $\Lambda_\tau$ is the lattice in $\mathbb C$ generated
by $1$ and $\tau = \tau_1 + i \, \tau_2$, with $\tau_2>0$.
The  complex structure parameter $\tau$ of $\mathbb E_\tau$
is identified with the axio-dilaton field of type IIB supergravity.
As a result, a non-trivial axio-dilaton profile is encoded in an auxiliary 12d geometry $M_{12}$,
obtained fibering $\mathbb E_\tau$ over the physical 10d spacetime $M_{10}$,
\beq \label{M12}
\mathbb E_\tau \hookrightarrow M_{12} \xrightarrow{\pi} M_{10} \ .
\eeq
The volume of $\mathbb E_\tau$
is   constant over $M_{10}$.
The loci on the base $M_{10}$ where the fiber $\mathbb E_\tau$ degenerates
correspond to locations of 7-branes.

\subsubsection*{A new term in $\cI_{11}$}

Making use of the geometry of the auxiliary space $M_{12}$,
we can construct a new term in $\cI_{11}$, to be added to
\eqref{cI11_medium}. It takes the form
\beq \label{new_term}
\Delta \cI_{11} = - E_5 \, \pi_* X_8[M_{12}] \  , \qquad
X_8[M_{12}] = \frac{1}{192} \bigg[ p_1(TM_{12})^2
- 4 \, p_2(TM_{12}) \bigg] \ .
\eeq
The 5-form $E_5$ is the same as in \eqref{cI11_medium}.
The characteristic class $X_8$  is as in \eqref{X8def},
but it is computed not in the physical 10d spacetime,
but in the auxiliary 12d geometry \eqref{M12}.
The symbol $\pi_*$ denotes the pushforward 
of $X_8$ associated to the map $\pi$ in \eqref{M12}.\footnote{If
we were to consider a fibration $\mathbb E_\tau \hookrightarrow M_{12} \xrightarrow \pi M_{10}$
with $\mathbb E_\tau$ smooth everywhere,
$\pi_*$ would be identified
with integration along the $\mathbb E_\tau$ fibers. The latter operation
 is characterized by the property
\beq
\int_{M_{10}} \pi_* \alpha_p  \, \beta_{12-p} = \int_{M_{12}} \alpha_p \, \pi^* \beta_{12-p}   \ ,
\eeq
where $\alpha_p$ is an arbitrary compactly supported smooth $p$-form on $M_{12}$,
$\beta_{12-p}$ is an arbitrary compactly supported smooth $(12-p)$-form on the base
$M_{10}$, and $\pi^*$ is the standard pullback of differential forms.
Since the fibration \eqref{M12} is necessarily singular in the presence of 7-branes,
we need a refined notion of $\pi_*$. We can still think intuitively 
of $\pi_*$ as integration along the $\mathbb E_\tau$ fiber directions.
}
In analogy with the M-theory
anomaly inflow analysis, $\pi_* X_8[M_{12}]$ is implicitly pulled
back to $r = \epsilon$ at the location of the boundary of $M_{10}$ which
appears after we remove the sources.

As a small sanity check,
let us first verify that the new term \eqref{new_term}
is immaterial if we consider a trivial fibration,
\emph{i.e.}~a direct product $M_{12} = \mathbb E_\tau \times M_{10}$.
In this case $p_1(TM_{12}) = 0 = p_2(TM_{12})$,
and the new term vanishes.

Let us now illustrate the role of the new term \eqref{new_term} in an example
based on   the construction  of \cite{Lawrie:2018jut}.
Our discussion will be somewhat heuristic, and it would be interesting to revisit
this problem to address it in a more precise way. 

We know that if we consider a stack of $N$ D3-branes away from any singularities
we obtain a worldvolume theory which is $\cN = 4$ SYM with gauge group $SU(N)$,
together with a free $\cN = 4$ vector multiplet.
The complexified coupling constant $\tau_{\rm YM}$ of the gauge theory
is identified with the constant value of the type IIB dilaton $\tau$ throughout
10d spacetime. Morevoer, the six transverse directions to the D3-brane stack
encode the $SO(6)$ R-symmetry bundle of the 4d worldvolume theory.
Let us now consider a situation in which we turn on a non-trivial
background profile for $\tau$ along the worldvolume $W_4$ of the D3-branes.
We expect to obtain $\cN = 4$ SYM with varying complexified coupling constant
$\tau_{\rm YM}$, as studied in \cite{Lawrie:2018jut}.
We do not activate a non-trivial $\tau$ profile in the directions transverse to the D3-branes.
As a result, we can write
\begin{align}
p_1(TM_{12}) &= p_1(TW_6) + p_1(SO(6))  \ , \nn \\
p_2(TM_{12}) &= p_2(TW_6) + p_2(SO(6)) + p_1(TW_6) \, p_1(SO(6)) \ .
\end{align}
In the previous expressions, we have separated the contributions
of the $SO(6)$ vector bundle that is associated to the R-symmetry
of the worldvolume theory. The space $W_6$ encodes 
the external spacetime $W_4$ together with its non-trivial $\tau$ profile.
More precisely, we Wick rotate to Euclidean signature and take $W_4$
to be a (not necessarily compact) complex surface.
The total space $W_6$ has the form\footnote{By slight abuse of notation,
we are using $\pi$ for the projection map of $W_6$, and not of the total 12d space $M_{12}$.
This is not problematic because  $\mathbb E_\tau$ varies only over $W_4$.}
\beq \label{elliptic_fibration}
\mathbb E_\tau \hookrightarrow W_6 \xrightarrow \pi W_4  \ ,
\eeq
and is an elliptic fibration with a section,
described by a Weierstrass model.
The latter is specified by a holomorphic line bundle $\mathbb L$ on $W_4$,
together with a section $f$ of $\mathbb L^4$
and a section $g$ of $\mathbb L^6$.
The elliptic fibration is then described by the Weierstrass equation
\beq \label{Weierstrass}
y^2 = x^3+ f \, x  + g \ .
\eeq
To evaluate the new term \eqref{new_term} in this background
we need the quantity
\beq \label{piX8example}
\pi_* X_8[M_{12}] = \frac{1}{192} \, \pi_* \Big[ p_1(TW_6)^2 - 4 \, p_2(TW_6)  \Big]
- \frac{1}{96} \, p_1(SO(6)) \, \pi_* p_1(TW_6) \ ,
\eeq
where we have ignored terms with $p_1(SO(6))^2$ and $p_2(SO(6))$, because
they are 8-form on external spacetime $W_4$.
Notice that \eqref{piX8example}
does not have any legs along the directions of the $S^5$ that surrounds the
D3-brane stack. The integration over this $S^5$ is saturated by the $E_5$
factor in $\Delta \cI_{11}$, yielding a factor $N$. In summary, the new contribution
to the inflow anomaly polynomial reads
\begin{align} \label{new_inflow}
- \Delta I_6^{\rm inflow} & = \int_{S^5} E_5 \, \pi_* X_8[M_{12}]
\nn \\
& = \frac{N}{48} \, \bigg[ 
\pi_* \bigg( - p_2(TW_6) + \frac 14 \, p_1(TW_6)^2 \bigg)
- \frac 12 \, p_1(SO(6)) \, \pi_* p_1(TW_6)
\bigg] \ . 
\end{align}
This expression agrees exactly with (5.5) in \cite{Lawrie:2018jut},
which gives the anomaly polynomial for 4d $\cN  = 4$ SYM
with varying $\tau$, as described by the elliptic fibration $W_6$.

The analysis of \cite{Lawrie:2018jut} demonstrates how to perform the
pushforwards $\pi_*$ in \eqref{new_inflow}.
The result 
is written in terms of the first Chern class of the Weierstrass line bundle
$\mathbb L$. We recall some well-known facts about this
object in appendix \ref{sec_ftheoryapp}.
The pushforwards in \eqref{new_inflow}
take the form
\begin{align} \label{example}
\pi_* p_1(TW_6) & = -24 \, c_1(\mathbb L) \ , \nn \\
\pi_* \bigg( - p_2(TW_6) + \frac 14 \, p_1(TW_6)^2 \bigg)
& = 12 \, c_1(\mathbb L) \, p_1(TW_4) + \text{(non-universal terms)} \ .
\end{align}
The terms displayed explicitly on the RHSs of the previous expressions
are universal, in the sense that they only depend on the choice of
Weierstrass line bundle $\mathbb L$, but not on 
the details of the singularities of the fibration.
In contrast, the non-universal terms are indeed sensitive to these
details. We refer the reader to \cite{Lawrie:2018jut} for a thorough
analysis of this point.

\subsubsection*{A further generalization of $\cI_{11}$}

Let us conclude this section by suggesting a further generalization of $\cI_{11}$,
which combines the fluxes $F_3$, $H_3$ with a non-trivial axio-dilaton
profile. The suggested form of $\cI_{11}$
is
\beq \label{further_extension}
\cI_{11}  = \frac 12 \, E_5\, dE_5 - E_5 \, \pi_*  \, \bigg[   X_8[M_{12}] + \frac 12 \, 
\mathcal E_4^2  \bigg] \ .
\eeq
The 4-form $\mathcal E_4$ is defined on the auxiliary 12d geometry \eqref{M12}.
The object $\mathcal E_4$ combines the type IIB fluxes $\cF_3$, $\cH_3$
discussed in section \ref{cI11_in_IIB}. In the case of a trivial fibration, \emph{i.e.}~a direct
product $M_{12} = \mathbb E_\tau \times M_{10}$,
the relation between $\mathcal E_4$, $\cF_3$, $\cH_3$ is simply
\beq \label{simpleE4}
\mathcal E_4 =  \cF_3 \, dx + \cH_3 \, dy \ ,
\eeq
where $dx$, $dy$ are the 1-forms on the elliptic curve $\mathbb E_\tau$
corresponding to usual basis of A and B 1-cycles. 
The 4-form $\mathcal E_4$ is invariant under $SL(2,\mathbb Z)$ transformations
(which are simply diffeomorphisms in $M_{12}$).
It follows from   \eqref{simpleE4}  that $\cF_3$, $\cH_3$
transform as a doublet under $SL(2,\mathbb Z)$,
as expected.

In the case of a non-trivial fibration of $\mathbb E_\tau$ over $M_{10}$,
the relation \eqref{simpleE4} is only schematic, because the 1-forms $dx$ and $dy$
are no longer well-defined. To define $\mathcal E_4$ more precisely,
we need to study well-defined cycles in the elliptic fibration $M_{12}$,
and restrict to those cycles which have ``one leg along the elliptic
fiber.'' Interestingly, this condition is the same condition that
a $G_4$ flux configuration for M-theory on an elliptically fibered Calabi-Yau
four-fold has to satisfy in order to be compatible with 4d Lorentz invariance
in the F-theory dual \cite{Dasgupta:1999ss,Grimm:2011fx, Cvetic:2012xn}.
Our proposal \eqref{further_extension} makes therefore natural contact
with the subject of $G_4$ flux configurations in F-theory.
A detailed analysis of this problem goes beyond the scope of this work,
but we hope to return to it in the future.



\section{Discussion}

In this work we   studied anomaly inflow for field theories engineered on the worldvolume
of a stack of D3-branes in type IIB string theory. 
Our main proposal 
can be summarized as
\beq \label{the_summary}
I_{d+2}^{\rm inflow} = \int_{M_{9-d}} \cI_{11} \ , \qquad
\cI_{11} = \frac 12 \, E_5 \, dE_5 
 \ ,
\eeq
where $d$ is the spacetime dimension of the field theory and $I_{d+2}^{\rm inflow}$
is its inflow anomaly polynomial, equal to minus the total
anomaly of all degrees of freedom on the worldvolume theory
(including modes that decouple in the  IR).
The compact $(9-d)$-dimensional
space $M_{9-d}$   encodes the geometry of the 
directions transverse to external spacetime.
The 11-form $\cI_{11}$ is constructed in terms of the 5-form, which encodes the boundary conditions near the D3-brane stack
for
the type IIB field strengths
$F_5$.
Our approach applies both to ``mesonic'' symmetries, \emph{i.e.}~symmetries
associated to isometries of the internal space $M_{9-d}$,
and to ``baryonic'' symmetries, \emph{i.e.}~symmetries associated
to expansion of the type IIB 4-form $C_4$ onto harmonic
3-forms on $M_{9-d}$.

We have tested our proposal in the case of 4d $\cN = 1$ field theories 
engineered by 
D3-branes at the tip of a Calabi-Yau cone,
as well as 4d $\cN = 2$ field theories originating from D3-branes
probing a $\mathbb C^2/\Gamma$ singularity,
with $\Gamma$ an ADE subgroup of $SU(2)$.
In all these scenarios we get a perfect match
with the field theory results,
provided decoupling modes and accidental symmetries 
in the IR are taken into account properly.

Moreover, we have checked our formula for 2d $\cN = (0,2)$ theories
obtained from putting 
D3-branes at the tip of a Calabi-Yau cone and further wrapping their
worldvolume on a smooth genus-$g$ Riemann surface.
Our results confirm the expectation  that the inflow anomaly polynomial
$I_4^{\rm inflow}$ of the 2d $\cN = (0,2)$ theory
can be obtained by integrating the inflow anomaly
polynomial $I_6^{\rm inflow}$ of the parent 4d $\cN = 1$ theory
over the Riemann surface. 
In performing the integration, however,
one has to identify the correct relation between
2d background connections and 4d background connections.
Our geometric formalism makes it manifest that there is a non-trivial
interplay between 2d mesonic symmetries and 4d baryonic
symmetries, encoded in \eqref{4d_vs_2d} and \eqref{new_equation},
and first observed in \cite{Benini:2015bwz}.

We   applied \eqref{the_summary} to
  a class of 2d $\cN = (2,2)$ theories
obtained by compactification of 
4d $\cN = 2$ SYM theory with gauge group $SU(N)$
 on a Riemann surface with half-BPS punctures.
 The latter are 
 labelled by partitions of $N$.
Following the approach  of \cite{Bah:2018jrv,Bah:2019jts} 
for the geometry and flux configuration near the
punctures, we computed the contributions of punctures
to the 2d inflow anomaly polynomial.

We have also outlined a proposal to generalize
$\cI_{11}$
to include the contributions of the type IIB field strengths $F_3$, $H_3$, 
as well as a generalization to F-theory backgrounds.
We performed a preliminary check of the latter against the
constructions studied in \cite{Lawrie:2018jut}.

There are several future directions to explore.
Firstly, it would be desirable to have a first principle
derivation of the inflow formula \eqref{the_summary}.
Moreover, it is interesting to study the interplay between
\eqref{the_summary} and the analogous formula
in M-theory, also in connection
with the duality between F-theory and M-theory.

Our approach can be applied to holographic solutions
of type IIB supergravity supported by $F_5$ and/or
$F_3$, $H_3$ background fluxes. 
An example of regular solution with non-zero $F_5$, $F_3$, and $H_3$
is the $AdS_5$ Pilch-Warner solution \cite{Pilch:2000ej}.
Other solutions with non-zero $F_3$,  $H_3$
fluxes are known, including solutions with $F_5  = 0$,
but they are singular \cite{Gauntlett:2005ww,Couzens:2016iot}.
It would be interesting to investigate whether
they might still allow for a field theory interpretation,
and what anomaly inflow would predict for 
such field theories.

The compactification of 4d gauge theories on a Riemann
surface with punctures is an interesting problem
that is still eluding a fully systematic understanding
and is recently attracting renewed attention, see \emph{e.g.}~\cite{Bobev:2019ore}.
It would be beneficial to further
study punctures from the perspective of the
anomaly inflow formula \eqref{the_summary},
in combination with insights from holography
and  purely field theoretic analysis.

The proposed F-theoretic generalization of \eqref{the_summary}
can be further studied   in relation to the constructions analyzed in
\cite{Apruzzi:2016iac,Apruzzi:2016nfr,Lawrie:2016axq,Couzens:2017way,Couzens:2017nnr}.
A more complete understanding of anomaly inflow in F-theory
would be useful,  for instance in relation to the
vast class  of 6d $\cN =(1,0)$ SCFTs
realized in F-theory \cite{Heckman:2018jxk}.

Finally, we expect to be able to generalize
the anomaly inflow formalism based on the class $\cI_{11}$
to include also higher-form and/or discrete symmetries 
and   compute  their 't Hooft anomalies geometrically.


\section*{Acknowledgments}

We would like to thank
Nikolay Bobev,
Fri\dh rik Freyr Gautason,
Craig Lawrie,
Emily Nardoni, and
Raffaele Savelli
for interesting conversations and correspondence. 
We thank 
Sakura Sch\"afer-Nameki for 
comments on the draft.
The work of IB, FB, and PW is supported in part by NSF grant PHY-1820784. RM is supported in part by ERC Grant 787320 - QBH Structure. 
The work of PW is
supported in part by the Chateaubriand Fellowship of the Office for Science \&
Technology of the Embassy of France in the United States.
We gratefully acknowledge the Aspen Center for Physics, supported by NSF grant PHY-1607611, for hospitality during part of this work.  

\newpage


\appendix
 


\section{Type IIB on a circle and $\cI_{11}$} \label{app_circle}

In this appendix we provide indirect evidence for \eqref{cI11_medium}
by considering type IIB supergravity reduced on a circle to nine dimensions. 
The starting point is 
the 10d bosonic pseudo-action in Einstein frame,
\begin{align}
S_{10d}= \frac{1}{2\kappa_{10}^2} \, \int \bigg[ &R \, *1
- \frac 12 \, d\phi \, *d\phi
- \frac 12 \, e^{2\phi} \, F_1 \, *F_1
- \frac 12 \, e^{-\phi} \, H_3 \, *H_3
- \frac 12 \, e^\phi \, F_3 \, * F_3
\nn \\
& - \frac 14 \, F_5 \, *F_5
- \frac 12 \, C_4 \, H_3 \, F_3 \bigg] \ ,
\end{align}
where the field strengths are given in terms of the potentials 
according to
\begin{align} \label{10d_field_strengths}
H_3 &= dB_2 \ , \qquad
F_1 = dC_0 \ , \qquad
F_3 = dC_2 - C_0 \, dB_2 \ , \nn \\
F_5 & = dC_4 - \frac 12  \, C_2 \, dB_2 + \frac 12 \, B_2 \, dC_2 \ .
\end{align}
Our convention for the Hodge star of a $p$-form $\alpha_p$ is
\beq \label{Hodge_convention}
(*\alpha_p)_{M_1 \dots M_q} = \frac{1}{p!} \, \alpha^{N_1 \dots N_p} \, \epsilon_{N_1\dots
N_p M_1 \dots M_q} \ , \qquad
p+q=10 \ ,
\eeq
with $\epsilon_{0123456789} = + 1$ in an orthonormal frame.

The metric ansatz for the reduction to nine dimensions reads
\beq \label{metric_ansatz}
ds^2_{10} =  \widetilde g_{\mu\nu} \, dx^\mu \, dx^\nu
+ e^{2 \tilde \sigma} \, D\theta^2 \ , \qquad
D\theta = d\theta + \widetilde V \ , \qquad
\theta \sim \theta +  L \ ,
\eeq
where $\theta$ is the coordinate on the circle
of circumference $L$,
$\widetilde g_{\mu\nu}$ is the 9d metric,
$\widetilde V$ is the Kaluza-Klein vector, 
and $\tilde \sigma$ is the radion field.
Throughout this appendix we use a tilde to denote 9d fields.
The reduction ansatz for the $p$-forms of type IIB is
\begin{align} \label{KK_potentials}
B_2 = \widetilde B_2 + \widetilde B_1 \, D\theta  \ , \qquad
C_0 = \widetilde C_0 \ , \qquad
C_2 = \widetilde C_2 + \widetilde C_1 \, D\theta \ , \qquad
C_4 = \widetilde C_4 + \widetilde C_3 \, D\theta  \ .
\end{align}
In a similar way, the field strengths
in 10d dimensions are reduced 
as 
\begin{align}
H_3 = \widetilde H_3 + \widetilde H_2 \, D\theta  \ , \qquad
F_1 = \widetilde F_1   \ , \qquad
F_3 = \widetilde F_3 + \widetilde F_2 \, D\theta \ , \qquad
F_5 = \widetilde F_5 + \widetilde F_4 \, D\theta  \ .
\end{align}
The expressions for the 9d field strengths $\widetilde H_3$, \dots,
$\widetilde F_4$ in terms of the 9d potentials
are readily extracted from \eqref{10d_field_strengths}, \eqref{KK_potentials},
if needed.

In ten dimensions, the self-duality constraint
\beq \label{selfduality}
F_5 = *F_5
\eeq
must be imposed by hand after deriving the equations of motion.
We identify $\theta$ with the 9-th direction,
and we use the 
 orientation convention
$\epsilon_{0123456789} = \epsilon_{012345678} = 1$
in an orthonormal frame.
As a result, \eqref{selfduality} implies
\beq \label{tildeF5}
\widetilde F_5 = - e^{-\tilde \sigma} \,  \tilde * \widetilde F_4 \ ,
\eeq
where $\tilde *$ is the Hodge star 
computed with the 9d metric $\widetilde g_{\mu\nu}$,
using conventions analogous to \eqref{Hodge_convention}.
In nine dimensions we can write a proper
action, which contains  $\widetilde F_4$
but does not contain $\widetilde F_5$.
A convenient way to obtain it is as follows.
One first reduces the 10d pseudo-action on the $\theta$ circle,
and then adds a total derivative in nine dimensions
of the form $\int d\widetilde C_3 \, d\widetilde C_4$.
The coefficient of this term is selected in such a way that,
after some integration by parts, 
the 9d action depends on $\widetilde C_4$ via $\widetilde F_5$ only,
and the equation of motion for $\widetilde F_5$ 
coincides with \eqref{tildeF5}.
We may then treat $\widetilde F_5$ as an independent variable,
and integrate it out using its algebraic equation of motion.\footnote{Treating
$\widetilde F_5$ as an independent variable means that
the 9d Bianchi identity for $\widetilde F_5$ does not hold off-shell,
but one verifies that it still holds on-shell.}
The outcome of this procedure is the following 9d action,
\begin{align} \label{9daction}
S_{9d}   = \frac{L}{2\kappa_{10}^2} \, \int \bigg[ 
&e^{\tilde \sigma} \, R *1 - \frac 12 \, e^{3 \tilde \sigma} \, \widetilde W_2 \, *  \widetilde W_2
 - \frac 12 \, e^{\tilde \sigma} \, d\phi \, *d\phi
- \frac 12 \, e^{2\phi} \, e^{\tilde \sigma} \, \widetilde F_1 \, * \widetilde F_1
\nn \\
& - \frac 12 \, e^{-\phi} \, e^{\tilde \sigma} \, \widetilde H_3 \, *\widetilde H_3
- \frac 12 \, e^{-\phi} \, e^{-   \tilde \sigma} \, \widetilde H_2 \, *\widetilde H_2
- \frac 12 \, e^\phi \, e^{\tilde \sigma} \, \widetilde F_3 \, * \widetilde F_3
\nn \\
& - \frac 12 \, e^\phi \, e^{-  \tilde \sigma} \, \widetilde F_2 \, * \widetilde F_2
 - \frac 12 \, e^{-\tilde \sigma} \, \widetilde F_4 \, * \widetilde F_4
+ \widetilde \Omega_9 \bigg] \ .
\end{align}
In the above expression, $\widetilde W_2 = d\widetilde V$ is the field
strength of the Kaluza-Klein vector
and the Chern-Simons 9-form $\widetilde \Omega_9$ reads
\begin{align}
\widetilde \Omega_9 & = - \frac 14 \, \widetilde B_2 \, \widetilde F_3 \, \widetilde F_4
+ \frac 14 \, \widetilde C_2 \, \widetilde F_4 \, \widetilde H_3
+ \frac 12 \, \widetilde C_3 \, \widetilde F_3 \, \widetilde H_3
+ \frac 12  \, \widetilde C_3 \, \widetilde F_4 \, \widetilde W_2
- \frac 14 \, \widetilde B_2 \, \widetilde F_4 \, \widetilde H_3 \, C_0 \ , \nn \\
d\widetilde \Omega_9 & = \frac 12 \, \widetilde F_4 \, \widetilde F_4 \, \widetilde W_2
+ \widetilde F_4 \, \widetilde F_3 \, \widetilde H_3  \ .
\end{align}
Let us stress that \eqref{9daction} is not written in the 9d Einstein frame,
which could be reached with a Weyl rescaling of 9d the metric.

We are mainly interested in the structure of the Chern-Simons
term $\widetilde \Omega_9$.
While the term $\widetilde F_4 \, \widetilde F_3 \, \widetilde H_3$ is the
straightforward reduction of its 10d counterpart
$F_5 \, F_3 \, H_3$, the term $\widetilde F_4 \, \widetilde F_4 \, \widetilde W_2$
is generated by the self-duality of $F_5$ in ten dimensions.
The structure of $d\widetilde \Omega_9$
provides indirect support for the relative weight of the two terms in  \eqref{cI11_medium}.
To see this, we observe that
\beq
\begin{array}{rcl}
\cI_{11} &=&  \frac 12 \, E_5 \, dE_5 + E_5 \, \cF_3 \, \cH_3 \\[1mm]
E_5 &=& \widetilde F_4 \, D\theta \\[1mm]
\cF_3 &=& \widetilde F_3 + \widetilde F_2 \, D\theta \\[1mm]
\cH_3 &=& \widetilde H_3 + \widetilde H_2 \, D\theta
\end{array}
\qquad \Rightarrow \qquad
L^{-1}\int_{S^1_\theta} \cI_{11} =  \frac 12 \, \widetilde F_4 \, \widetilde F_4 \, \widetilde W_2
+ \widetilde F_4 \, \widetilde F_3 \, \widetilde H_3 \ .
\eeq
The above argument is only schematic and  we have ignored the factors
$2\pi$ and the bump function $\rho$ 
that enter the relation between $F_5$ and $E_5$, $F_3$ and $\cF_3$,
and $H_3$ and $\cH_3$. 

As a side remark, the same effective action in nine dimensions
should  be equivalently obtained by reducing M-theory on a $T^2$.
In the process, the $G_4 \, X_8$ term in eleven dimensions
generates a correction to $\widetilde \Omega_9$,
in such a way that $d\widetilde \Omega_9$
is shifted by a term $X_8 \, \widetilde W_2$.
From a type IIB perspective,
this higher-derivative coupling in nine dimensions
originates from winding modes of fundamental
strings \cite{Antoniadis:1997eg,Liu:2010gz}. As a result, while this coupling is present in nine dimensions
for any finite circumference $L$,
it does not uplift to a 10d Lorentz invariant higher-derivative
correction to the 10d type IIB effective action.
This observation is consistent with the argument in section \ref{cI11_in_IIB}
that rules out corrections to $\cI_{11}$
(for $dC_0 = 0 = d\phi$).



\section{Remarks on $E_5$} \label{sec_dF5app}

This appendix contains remarks and observation on $E_5$
that complement the discussion given in section \ref{sec_general_properties}
and provide derivations for some of the results stated there.

\subsection{The form $E_5$ and closure of $F_5$}
We consider type IIB setups with D3-brane charge only,
preserving $\cN = 1$ superconformal symmetry in 4d. 
Before turning on external connections, the only non-zero flux
in the background is $F_5$  and the internal space is a Sasaki-Einstein
manifold $\rm SE_5$. 
We assume that, even after turning on external connections,
the fluxes $F_3$ and $H_3$ and the axion remain identically zero,
and the dilaton remains constant.
This assumption is motivated by the observation that, in the 10d type IIB equations
of motions, it is consistent to set $F_3$ and $H_3$ to zero, and the axiodilaton
to a constant.

The boundary condition for $F_5$ near the D3-brane source
is parametrized in terms of the form $E_5$, in such a way that 
$F_5$ is manifestly self-dual,
\beq
F_5 = E_5 + *_{10} \, E_5 \ .
\eeq
The on-shell condition for $F_5$, in the absence of $F_3$, $H_3$,
 amounts simply to $dF_5 =0$. 
The form $E_5$ is as in \eqref{simplest_E5}, repeated here for convenience
\beq \label{ourE5}
E_5 = N \, \bigg( V_5^{\rm g} + \frac{F^I}{2\pi} \, \omega_I^{\rm g}
+ \frac{F^\alpha}{2\pi} \, \omega_\alpha^{\rm g} \bigg ) \ .
\eeq
Recall that the superscript ``g'' signals the gauging of
internal forms, defined in \eqref{gauging_def}. The 5-form $V_5$ is the volume form
on $\rm SE_5$, normalized to integrate to 1. The 3-forms $\omega_\alpha$
are a basis of harmonic 3-forms on $\rm SE_5$.
In this appendix, we regard 
$\omega_I$ as unspecified 3-forms on $\rm SE_5$.
The importance of $\omega_I$ for achieving $dF_5=0$ 
will be clear momentarily.
All terms in $E_5$ contain  at least three internal gauged legs;
terms with fewer internal gauged legs in $F_5$ originate from $*_{10} \, E_5$.
We do not include terms in $E_5$ with four internal gauged legs,
because there are no harmonic 4-forms on $\rm SE_5$.

Let us now impose $dF_5 = 0$. Our analysis is similar to the one in \cite{Benvenuti:2006xg}.
We can compute $dF_5$ with the help of \eqref{isom_identity}
and the Bianchi identity for $F^I$. The result reads\footnote{Our conventions for the Hodge star are such that
$*_{10} [ \alpha_{{\rm ext},p}  \,   (\beta_{{\rm int},q})^{\rm g}] = (-)^{(5-p)q} \, (* \alpha_{{\rm ext},p}) \, (*  \beta_{{\rm int},q})^{\rm g}$, where $\alpha_{{\rm ext},p}$ is a $p$-form
in the external 5d spacetime, and $\beta_{{\rm int},q}$ is a $q$-form on $\rm SE_5$.}
\begin{align} \label{total_dF5}
dF_5 & = N \, F^I \,  \bigg( \iota_I V_5 + \frac{d\omega_I}{2\pi} \bigg)^{\rm g} 
\nn \\
& + N \, dF^\alpha \, \frac{\omega_\alpha^{\rm g}}{2\pi}
+ N \, (*F^I) \,  \frac{(d* \omega_I)^{\rm g}}{2\pi}
\nn \\
& 
+ N \,    F^I \, F^J \, \frac{(\iota_I \omega_J)^{\rm g}}{2\pi} 
+ N \, F^I \, F^\alpha \,  \frac{(\iota_I \omega_\alpha)^{\rm g}}{2\pi} 
  - N \, (D * F^I) \, \frac{(* \omega_I)^{\rm g}}{2\pi}
  - N \, (d * F^\alpha) \, \frac{(* \omega_\alpha)^{\rm g}}{2\pi}
\nn \\
& 
+ N \, (*F^I) \, F^J \, \frac{(\iota_J * \omega_I)^{\rm g}}{2\pi}
+ N \, (*F^\alpha) \, F^J \, \frac{(\iota_J * \omega_\alpha)^{\rm g}}{2\pi} \ .
\end{align}
The Hodge star is understood to be computed with the external 5d metric
if it acts on an external forms, and to be computed with the metric on $\rm SE_5$
if it acts on an internal form.
The symbol $D$ denotes exterior covariant differentiation with respect to the isometries
of $\rm SE_5$, and is defined by the LHS of identity \eqref{isom_identity}.
For the sake of argument, we have not yet imposed the Bianchi identity for $F^\alpha$.
Each line in the expression \eqref{total_dF5} for $dF_5$
has a different number of external legs and gauged internal legs.
Hence, each line must vanish separately.

The first line of \eqref{total_dF5} implies that the 3-forms $\omega_I$ must be chosen
in such a way that  \eqref{domegaI} holds, an anticipated in the main text.
As explained there, the existence of $\omega_I$ with the desired property
is guaranteed by the absence of harmonic 4-forms on $\rm SE_5$.

On the second line of \eqref{total_dF5}, the first term contains an internal harmonic 3-form,
while the second contains an internal exact 3-form. Such terms must vanish
independently, from which we recover the expected Bianchi identity for $F^\alpha$,
as well as co-closure of $\omega_I$,
\beq
dF^\alpha = 0 \ , \qquad d* \omega_I = 0 \ .
\eeq
On a Sasaki-Einstein  manifold, \eqref{domegaI} can be solved explicitly
by $\omega_I \propto * dk_I$, where $k_I$ are the 1-forms
dual to the Killing vectors. Co-closure of $\omega_I$ is then automatically
satisfied.

The third and fourth lines of \eqref{total_dF5} contain terms that are zero
by virtue of the 5d equations of motion in the 5d supergravity theory
obtained from reduction of type IIB supergravity on $\rm SE_5$.\footnote{The relevant
5d equations of motion are those of the vector modes, but also
of their scalar superpartners, which are implicitly frozen to zero
in our discussion.}
These terms in $dF_5$ do not impose new constraints on the form of $E_5$.
Therefore, they are not directly  relevant  for anomaly inflow, and will not be discussed
further.

\subsection{Non-minimal terms in $E_5$}

In this subsection, we make use of the collective notation
introduced in \eqref{collective_notation}.
Let us   add   terms to $E_5$ in \eqref{simplest_E5}
built using external 4-forms,
\beq
E_5' = E_5 + \Delta E_5 \ ,\qquad
\Delta E_5 = F^X \, F^Y \, \lambda_{XY}^{\rm g} + p_1(TW_4) \, \lambda^{\rm g} \ , \qquad
\lambda_{XY} = \begin{pmatrix}
\lambda_{IJ} & \lambda_{I\beta} \\
\lambda_{J\alpha} & \lambda_{\alpha \beta}
\end{pmatrix} \ ,
\eeq
where $p_1(TW_4)$ is the first Pontryagin class
of the tangent bundle to external spacetime
and $\lambda_{XY}$ are 1-forms on $\rm SE_5$.
The form $E_5'$ 
is the most general polynomial in $F^X$, $p_1(TW_4)$
with coefficients given by gauged internal forms on $\rm SE_5$.
In order for $E_5'$ to be invariant under gauge transformations
of the connections $A^I$, we must demand
\beq
\pounds_I \lambda_{J_1 J_2} = f_{IJ_1}{}^K \, \lambda_{K J_2}
+ f_{I J_2}{}^K \, \lambda_{J_1 K} \ , \qquad
\pounds_I \lambda_{I\alpha} = f_{IJ}{}^K \, \lambda_{K\alpha} \ , \qquad
\pounds_I \lambda_{\alpha \beta} = 0 = \pounds_I \lambda \ .
\eeq
The 1-forms $\lambda_{XY}$ are otherwise arbitrary.

The claim we want to verify is
\beq \label{qed}
\int_{\rm SE_5} E_5' \, dE_5' = \int_{\rm SE_5} E_5 \, dE_5 \ .
\eeq
As a first step, we compute
\begin{align} \label{dE5prime}
dE_5' & = F^X \, F^Y \, \bigg(d\lambda_{XY} +\frac{N}{2\pi} \, \iota_X \omega_Y \bigg)^{\rm g}
+ p_1(TW_4) \, (d\lambda)^{\rm g}
\nn \\
&+ F^X \, F^Y \, F^Z \, \iota_X \lambda_{YZ}
+ p_1(TW_4) \, F^X \, \iota_X \lambda \ .
\end{align}
We can now collect all terms in $E_5' \, dE_5'$ that give a non-zero
result upon integration on $\rm SE_5$,
\begin{align}
\int_{\rm SE_5} E_5' \, dE_5' & = F^X \, F^Y \, F^Z \, \int_{\rm SE_5} \bigg[
\frac{N^2}{(2\pi)^2} \, \omega_X \, \iota_Y \omega_Z
+ N\, V_5 \, \iota_X \lambda_{YZ}
+\frac{N}{2\pi} \, \omega_X \, d\lambda_{YZ}
\bigg]
\nn \\
& + F^X \, p_1(TW_4) \, \int_{\rm SE_5} \bigg[
N\, V_5 \, \iota_X \lambda + \frac{N}{2\pi} \, \omega_X \, d\lambda
\bigg] \ .
\end{align}
The integrals over $\rm SE_5$ can be manipulated
by adding total derivatives $d(\dots)$ and total
interior products $\iota_X(\dots)$ without changing the result.
We then see that, by virtue of the condition \eqref{domegaX},
all dependence on $\lambda_{XY}$ and $\lambda$ drops away.
We thus establish \eqref{qed}.

\subsection{Obstruction to horizontality of $dE_5$}

Let us inspect $dE_5'$ in \eqref{dE5prime}.
In order to achieve horizontality of $dE_5'$
we must eliminate all terms in the first line of \eqref{dE5prime}.
Setting $d\lambda = 0$ eliminates the term with $p_1(TW_4)$.
In order to eliminate the remaining term, we would need
\beq \label{what_we_want}
N \, \iota_{(X} \omega_{Y)} + 2\pi \, d\lambda_{XY} = 0 \ .
\eeq
The 2-form $\iota_{(X} \omega_{Y)}$ is closed for any $X$, $Y$,
\begin{align}
d \iota_{(X} \omega_{Y)} & = \pounds_{(X} \omega_{Y)} - \iota_{(X } d\omega_{Y)}
= f_{(XY)}{}^K \, \omega_K + (2\pi)^{-1} \, \iota_{(X} \iota_{Y)} V_5 = 0 \ .
\end{align}
In the collective notation, $\pounds_\alpha := 0$, and the only non-zero
components of $f_{XY}{}^K$ are the Lie algebra structure
constants $f_{IJ}{}^K$, antisymmetric in $IJ$.

If $\rm SE_5$ admits harmonic 3-forms, it also admits harmonic 2-forms
and therefore there is no guarantee that $\iota_{(X} \omega_{Y)}$
is exact and that $\lambda_{XY}$ solving \eqref{what_we_want} exists.
The obstruction to exactness of $\iota_{(X} \omega_{Y)}$ is measured by
the integrals\footnote{To check the equality in \eqref{obstruction}, use $\iota_\alpha = 0$
and the symmetry property $\int_{\rm SE_5} \omega_X \iota_Y \omega_X = \int_{\rm SE_5} \omega_Z \iota_Y \omega_X$, which follows from integrating $0=\iota_Y(\omega_X \omega_Z)$.}
\beq \label{obstruction}
2\, \int_{\rm SE_5}\omega_\alpha \, \iota_{(X} \omega_{Y)} = 3 \, \int_{\rm SE_5}
\omega_{(\alpha} \, \iota_{X} \omega_{Y)} \ .
\eeq
The quantity on the RHS is proportional to the 't Hooft anomaly
coefficient $c_{\alpha XY}$ in the term $c_{\alpha XY} \, F^\alpha \, F^X \, F^Y$
in the inflow anomaly polynomial, see \eqref{cone_inflow}.
We conclude that, as soon as the anomaly polynomial contains
any term with $F^\alpha$, we have an obstruction to horizontality of $dE_5'$.

\subsection{Shifts of $\omega_I$}
The 3-forms $\omega_I$ are not uniquely determined by the relation \eqref{domegaI}.
In fact, we can shift $\omega_I$ with a closed 3-form, which we may parametrize as an
exact part, plus a linear combination of the harmonic 3-forms $\omega_\alpha$,
\beq \label{omegaI_shift}
\widehat \omega_I := \omega_I + d \Omega^2_I + \cC_I{}^\alpha \, \omega_\alpha \ .
\eeq
We use the symbol $\widehat E_5$ to denote $E_5$ as in \eqref{ourE5}
with $\omega_I$ replaced by $\widehat \omega_I$.
Gauge invariance of $\widehat E_5$ requires that
the 2-forms $\Omega^2_I$
and the constants $\cC_I{}^\alpha$   satisfy\footnote{Notice that, since $\cC_I{}^\alpha$ are constants, $\pounds_I \cC_J{}^\alpha = 0$,
and therefore the   condition on $\cC_I{}^\alpha$ translates to the requirement that
$\cC_I{}^\alpha$ be an invariant tensor of the Lie algebra of isometries of $\rm SE_5$.
As a result, $\cC_I{}^\alpha$ can only be non-zero if the index $I$ is associated
to a  generator of an Abelian subgroup of the isometry group.}
\beq
d\pounds_I \Omega^2_J = f_{IJ}{}^K \, d\Omega^2_K \ , \qquad
\pounds_I \cC_J{}^\alpha = f_{IJ}{}^K \, \cC_K{}^\alpha \ .
\eeq
By shifting $\Omega_I^2$ by a closed 2-form if necessary,
we can achieve
\beq
\pounds_I \Omega^2_J = f_{IJ}{}^K \, \Omega^2_K \ .
\eeq
As a result, the following 4-form is   gauge invariant,
\beq
\Omega_4 = -N \, \frac{F^I}{2\pi} \, (\Omega^2_I)^{\rm g} \ .
\eeq 
On the one hand, making use of $d\widehat \omega_I + 2 \pi \, \iota_I V_5  = 0$,
we verify that
$\int_{\rm SE_5} (\widehat E_5 + d\Omega_4) \, d(\widehat E_5 + d\Omega_4) =  \int_{\rm SE_5} \widehat E_5 \, d\widehat E_5$.
On the other hand, we compute
\begin{align} \label{newhatE5}
\widehat E_5 + d\Omega_4 = N \, \bigg( V_5^{\rm g} + \frac{F^I}{2\pi} \, \omega_I^{\rm g}
+ \frac{F^\alpha + F^I \, \cC_I{}^\alpha}{2\pi} \, \omega_\alpha^{\rm g} \bigg ) 
- \frac{N}{2\pi} \, F^I \, F^J \, (\iota_I \Omega^2_J)^{\rm g} \ .
\end{align}
The quantity on the RHS differs from $E_5$ in \eqref{ourE5} in two respects:
the  non-minimal term quadratic in $F$, and the fact that
$F^\alpha$ in \eqref{ourE5} is replaced by $F^\alpha + F^I \, \cC_I{}^\alpha$ in \eqref{newhatE5}.
We have already argued that non-minimal terms can be safely ignored
for the purposes of anomaly inflow.
The fact that $F^\alpha$ is replaced by 
$F^\alpha + F^I \, \cC_I{}^\alpha$
can be undone by a redefinition of the external connections,
of the form
$F^\alpha + F^I \, \cC_I{}^\alpha = F^\alpha_{\rm new}$.

In conclusion, if we shift from $\omega_I$ to $\widehat \omega_I$
as in \eqref{omegaI_shift}, the inflow anomaly polynomial is not affected,
up to a redefinition of the external connections $A^\alpha$.
The latter is merely a change of basis and does not change the
physics of the system.



\section{Inflow derivation for D3-branes probing $\mathbb C^2/\Gamma$} \label{app_Gamma}

In this appendix we use $E_5$ in \eqref{GammaE5} to compute the
inflow anomaly polynomial for a stack of D3-branes
probing a $\mathbb C^2/\Gamma$ singularity.
First of all, let us record the explicit expression of $e_5^{S^5}$ in \eqref{GammaE5}.
It is given by
\begin{align} \label{quotient_F5}
e_5^{S^5} & = (V_5)^{\rm g} + F^{AB} \, (\omega_{AB})^{\rm g}
 + F^{AB} \, F^{CD} \, (\lambda_{AB,CD})^{\rm g} \ , \\
  (V_5)^{\rm g} & = \frac{1}{\pi^3} \cdot \frac{1}{5!} \, \epsilon_{ABCDEF} \,
  y^A \, Dy^B \,  Dy^C \,  Dy^D \,  Dy^E \,  Dy^F  \ , \\
  (\omega_{AB})^{\rm g} & = \frac{1}{\pi^3} \, \cdot \frac{-1}{48} \, 
  \epsilon_{ABCDEF} \,
   y^C \,  Dy^D \,  Dy^E \,  Dy^F \ , \qquad \qquad
   Dy^A = dy^A - A^{AB} \, y_B \ . \label{omegaAB_expr}
\end{align}
The indices $A,\dots,F= 1,\dots 6$ are vector indices of $SO(6)$,
and $y^A$ are constrained coordinates on $S^5$.
The above expression is manifestly $SO(6)$ covariant.
It is understood, however, that the background field strength
$F^{AB}$ is only non-zero
along the generators of the
subgroup $G_L \times SU(2)_R \times U(1)_\phi \subset SO(6)$.
The 3-forms $\omega_{AB}$ are such that\footnote{Compared
with \eqref{domegaX}, the normalization of $\omega_{AB}$ differs
from that of $\omega_X$ by a factor $2\pi$. While the latter is convenient
in comparing our results with \cite{Benvenuti:2006xg}, in this section
we prefer not to  include this $2\pi$ factor.}
\beq \label{good_omegaAB}
\iota_{AB} V_5 + d\omega_{AB} = 0 \ .
\eeq
The 1-forms $\lambda_{AB,CD}$ can be left arbitrary,
since we verify below that the anomaly does not depend
on them. If we make the choice
\beq
(\lambda_{AB,CD})^{\rm g} = \frac{1}{\pi^3} \cdot \frac{1}{64} \,
  \epsilon_{ABCDEF} \,
   y^E \,  Dy^F  \ ,
\eeq
the 5-form $e_5^{S^5}$ reduces exactly to the global angular form of $SO(6)$,
as stated in the main text.
In this situation,   the 6-form $de_5^{S^5}$ is purely external (or horizontal),
\beq
de_5^{S^5} = \frac{1}{(2\pi)^3} \, \frac{1}{48} \, \epsilon_{ABCDEF} \, 
F^{AB} \, F^{CD} \, F^{EF} 
=: -  \chi_6(SO(6))   \ .
\eeq

We can now make use of \eqref{GammaE5}, \eqref{quotient_F5},
and \eqref{gauged_omega_alpha}
and compute
\begin{align}
\int_{S^5/\Gamma} E_5 \, dE_5
& =N^2 \, |\Gamma|^2 \, F^{AB} \, F^{CD} \, F^{EF} \, \int_{S^5/\Gamma} \bigg[
\omega_{AB} \, \iota_{CD} \omega_{EF}
+V_5 \, \iota_{AB} \lambda_{CD,EF}
+ \omega_{AB} \, d\lambda_{CD,EF}
\bigg] \nn \\
& + N \, |\Gamma | \, \frac{F^\alpha}{2\pi} \, F^{AB} \, F^{CD} \, \int_{S^5/\Gamma} \bigg[
 V_5 \, \iota_{AB} \lambda_{CD \alpha}
+ \omega_{AB} \, d\lambda_{CD\alpha}
\bigg]
\nn \\
& + \frac{F^\alpha}{2\pi} \, F^{AB} \, F^{CD} \, \int_{S^5/\Gamma} \bigg[
 \widetilde \omega_\alpha \, \frac{d\phi}{2\pi} \, \iota_{AB} \omega_{CD}
 + \widetilde \omega_\alpha \, \frac{d\phi}{2\pi} \, d\lambda_{AB,CD}
\bigg]
\nn \\
& - N \, |\Gamma | \, \frac{F^\alpha}{2\pi} \, \frac{F_\phi}{2\pi} \, F^{AB} \, 
\int_{S^5/\Gamma} \omega_{AB} \, \widetilde \omega_\alpha
- \frac{F^\alpha}{2\pi} \, \frac{F^\beta}{2\pi} \, \frac{F_\phi}{2\pi} \, \int_{S^5/\Gamma}
\widetilde \omega_\alpha \, \widetilde \omega_\beta \, \frac{d\phi}{2\pi}
\nn \\
& + \frac{F^\alpha}{2\pi} \, \frac{F^\beta}{2\pi} \, F^{AB} \, \int_{S^5/\Gamma}
\widetilde \omega_\alpha \, \frac{d\phi}{2\pi} \, d\lambda_{AB\alpha} \ .
\end{align}
Making use of \eqref{good_omegaAB}
and of the fact that $\widetilde \omega_\alpha \, d\phi$ is closed,
we see that all dependence on $\lambda_{AB,CD}$ and $\lambda_{AB\alpha}$
drops away, as anticipated.
Moreover, we have
\beq
\int_{S^5/\Gamma} \omega_{AB} \, \widetilde \omega_\alpha = 0 \ , \qquad
\int_{S^5/\Gamma}  \widetilde \omega_\alpha \, \frac{d\phi}{2\pi} \, \iota_{AB} \omega_{CD} = 0 \ .
\eeq
These relations follow from the fact that $\widetilde \omega_\alpha$
is supported on the locus $y_1 = \dots = y_4 = 0$.
Using \eqref{omegaAB_expr}, we see that both $\omega_{AB}$
and $d\phi \, \iota_{AB} \omega_{CD}$ are zero on this locus.
To proceed, we use the relation
\begin{align}
\int_{S^5/\Gamma} \omega_{AB} \, \iota_{CD} \omega_{EF}
=\frac{1}{|\Gamma|} \int_{S^5} \omega_{AB} \, \iota_{CD} \omega_{EF}
= \frac{1}{(2\pi)^3} \, \frac{1}{48 \, |\Gamma|} \, 
\epsilon_{ABCDEF} \ .
\end{align}
We also need the integral
\beq
\int_{S^5/\Gamma} \widetilde \omega_\alpha \, \widetilde \omega_\beta \, \frac{d\phi}{2\pi} = \int_{\mathbb C^2/\Gamma} \, \widetilde \omega_\alpha \, \widetilde \omega_\beta
= - \cC_{\alpha \beta} \ ,
\eeq
where we recalled \eqref{Cartan_integral}.

In summary, the integral of $E_5 \, dE_5$ yields
\begin{align} \label{almost_final}
\int_{S^5/\Gamma} E_5 \, dE_5 = - \frac{1}{(2\pi)^3} \, N^2 \, |\Gamma| \, 
\chi_{6} (SO(6))
+ \cC_{\alpha \beta} \,  \frac{F^\alpha}{2\pi} \, \frac{F^\beta}{2\pi} \, \frac{F_\phi}{2\pi}   \ .
\end{align}
Since only a subgroup of $SO(6)$ is a symmetry of the system,
we decompose $\chi_6(SO(6))$ as\footnote{Following \cite{Nakahara:2003nw},
we define the Euler classes of $SO(6)$ and $SO(4)$ vector bundles as
\beq
\chi_6(SO(6)) =  - \frac{1}{(2\pi)^3} \, \frac{1}{48} \, \epsilon_{ABCDEF} \, 
F^{AB} \, F^{CD} \, F^{EF}  \ , \qquad
\chi_4(SO(4)) =  + \frac{1}{(2\pi)^2} \, \frac{1}{8} \, \epsilon_{ABCD} \, 
F^{AB} \, F^{CD}   \ , \nn
\eeq
where $A$, $B$, \dots, are vector indices of $SO(6)$, $SO(4)$
respectively.
}
\beq \label{chiSO6}
\chi_6(SO(6)) = - \chi_4(SO(4)) \, \frac{F^{56}}{2\pi}
=  \chi_4(SO(4)) \, \frac{F_\phi}{2\pi}
=  \Big[ c_2(G_L) - c_2(SU(2)_R)\Big] \, \frac{  F_\phi}{2\pi} \ .
\eeq
We have used the notation $c_2(G_L)$ defined in \eqref{c2GL}.
The final result \eqref{Gamma_inflow} quoted in the main text
is obtained from \eqref{almost_final} using \eqref{chiSO6}
and recalling the identifications \eqref{4d_identifications}.



\section{Inflow derivation for smooth $\rm SE_5$ fibrations over $\Sigma_g$}
\label{app_no_punctures}

In this appendix we compute  the inflow anomaly polynomial
for the 2d theories considered in section \ref{sec_smoothSigma}.
Recall that the relevant 7d internal space is
\beq
{\rm SE}_5 \hookrightarrow M_7 \rightarrow \Sigma_g \ ,
\eeq
and that we use a bar to distinguish quantities and indices
relative to the $\rm SE_5$ fiber. The fibration is specified
by the background flux
\eqref{Sigma_background}.

\subsection*{The form $V_5$}
Because of the fact that the fiber $\rm SE_5$ is non-trivially twisted over the base
$\Sigma_g$, $p$-forms on $\rm SE_5$ are generically no longer
well-defined on the total space $M_7$.
We must instead consider their twisted counterparts,
denoted with a superscript `t'. 
Twisting here  means gauging with the background connections. 
For example, the volume form $V_5$
on $\rm SE_5$ is promoted to its twisted version
$\overline V_5^{\rm t}$, which is no longer closed,
\beq
d(\overline V_5^{\rm t}) = F_\Sigma^{\bar I} \, (\iota_{\bar I} \, \overline V_5)^{\rm t}
= V_\Sigma \, p^{\bar I} \, (\iota_{\bar I} \, \overline V_5)^{\rm t} \ .
\eeq
Even though $\overline V_5^{\rm t}$ is not closed,
we can restore closure by adding terms
linear in $F_\Sigma^{\bar I}$. More precisely,
we define the quantity
\beq  \label{V5def}
V_5 = \overline V_5^{\rm t} + p^{\bar I} \, \frac{V_\Sigma}{2\pi} \,  \, \overline \omega_{\bar I}^{\rm t} \ ,
\eeq
which is well-defined on $M_7$ and closed, thanks to \eqref{iotaV5bar}
and $V_\Sigma \, V_\Sigma = 0$.

\subsection*{The 3-forms $\omega_I$}

In order to implement anomaly inflow, for each generator $t_I$
of the preserved isometry group of $\rm SE_5$ we must find
a 3-form $\omega_I$ on $M_7$ such that
\beq \label{omega_eq}
d\omega_I  + 2\pi \, \iota_I V_5 = 0 \ ,
\eeq
with $V_5$ given by \eqref{V5def}. 
While it is always true that $d(\iota_I V_5) = 0$,
the space $M_7$ generically has non-trivial harmonic 4-forms.
It follows that the existence of a globally well-defined 3-form $\omega_I$
such that \eqref{omega_eq} holds is not guaranteed a priori,
and should be rather considered to be a restriction
on the allowed choices of twist.
This point is addressed in greater detail later.

Assuming that a solution for $\omega_I$ in \eqref{omega_eq} exists,
it can be written in the form
\beq \label{omega_def}
\omega_I = s_I{}^{\bar I} \, \overline \omega_{\bar I}^{\rm t}
+ s_I{}^{\bar \alpha} \, \overline \omega_{\bar \alpha}^{\rm t}
+ V_\Sigma \, \overline \Lambda_I^{\rm t} \ .
\eeq
Recall that the 3-forms $\overline \omega_{\bar I}$ on $\rm SE_5$
satisfy \eqref{iotaV5bar}, while $\overline \omega_{\bar \alpha}$ is a basis
of harmonic 3-forms on $\rm SE_5$. The quantities $\overline \Lambda_I$
are 1-forms on $\rm SE_5$ and must be such that
\beq \label{dLambda_eq}
d\Lambda_I + s_I{}^{\bar \alpha} \, p^{\bar K} \, \iota_{\bar K} \overline \omega_{\bar \alpha}
+ 2 \, s_I{}^{\bar J} \, p^{\bar K} \, \iota_{(\bar K} \overline \omega_{\bar J)} = 0\ .
\eeq
Finally, the constants $s_I{}^{\bar \alpha}$ are determined by the condition
\beq \label{salpha_eq}
 p^{\bar K} \, c_{\bar K \bar \alpha \bar \beta} \, s_I{}^{\bar \beta}  =  - s_I{}^{\bar J} \, p^{\bar K} \, 
 c_{\bar J \bar K \bar \alpha} \ .
\eeq
The interpretation of the above statements is the following.
The equation \eqref{omega_eq}
sets a closed 4-form on $M_4$ to zero.
Its harmonic and exact parts have to vanish separately.
The equation \eqref{salpha_eq} for the constants $s_I{}^{\bar \alpha}$
ensures that the harmonic part vanishes,
while the condition \eqref{dLambda_eq} on $\overline \Lambda_I$
takes care of the exact piece.
In reference to the last statement,
it should be noted that the 2-form
$s_I{}^{\bar \alpha} \, p^{\bar K} \, \iota_{\bar K} \overline \omega_{\bar \alpha}
+ 2 \, s_I{}^{\bar J} \, p^{\bar K} \, \iota_{(\bar K} \overline \omega_{\bar J)}$
is not only closed, but also exact.
Indeed, it cannot have any harmonic part, because 
its pairing with any harmonic 3-form on $\rm SE_5$ is zero,
\beq
2\pi \int_{\rm SE_5} \overline \omega_{\bar \beta} \, \Big[
s_I{}^{\bar \alpha} \, p^{\bar K} \, \iota_{\bar K} \overline \omega_{\bar \alpha}
+ 2 \, s_I{}^{\bar J} \, p^{\bar K} \, \iota_{(\bar K} \overline \omega_{\bar J)}
\Big] 
= \frac{1}{N^2 } \Big[ s_I{}^{\bar \alpha} \, p^{\bar K} \, c_{\bar K \bar \alpha \bar \beta}
+  s_I{}^{\bar J} \, p^{\bar K} \, c_{\bar J \bar K \bar \alpha}
\Big] = 0
   \ ,
\eeq
where we recalled the expressions \eqref{c_coeffs} for the $c$ coefficients
and we used \eqref{salpha_eq}.
As a result, the existence of $\Lambda_I$ solving \eqref{dLambda_eq}
is guaranteed.

\subsection*{The 5-form $E_5$ and inflow anomaly polynomial}

In the previous subsections we have determined $V_5$ and $\omega_I$.
This data is all we need to perform anomaly inflow
for symmetries related to the isometries of the fiber ${\rm SE_5}$ of $M_7$.
Let us stress that there are additional sources of symmetries for the 2d theory,
including: additional isometries of $M_7$ originating from isometries of the Riemann surface,
when the latter is a 2-sphere; harmonic 3-forms  on $M_7$.
We do not investigate these symmetries of the 2d theory in this work.
With this caveat in mind, the 5-form $E_5$ is given by
\beq
E_5 = N \, V_5 ^{\rm g} + N \, \frac{F^I}{2\pi} \, \omega_I^{\rm g} + F^I \, F^J \, \lambda_{IJ}^{\rm g} 
+ p_1(TW_2) \, \lambda^{\rm g}\ .
\eeq
The superscript `g' stands for gauged, and refers to gauging with the 2d external
connections $F^I$. The quantities $\lambda$, $\lambda_{IJ}$ are arbitrary 1-forms
on $M_7$. Indeed, we find
\beq
\int_{M_7} E_5 \, dE_5 = \frac{N^2}{2\pi} \, F^I \, F^J \, \int_{M_7} V_5 \, \iota_I \omega_J \ ,
\eeq
with the 1-forms $\lambda$, $\lambda_{IJ}$ dropping out by virtue of
\eqref{omega_eq}.  
Making use of \eqref{V5def}, \eqref{iotaV5bar}, \eqref{omega_def},
and \eqref{dLambda_eq} we compute
\begin{align} \label{result}
\int_{M_7} E_5 \, dE_5 = \frac{ F^I}{2\pi} \, \frac{ F^J }{2\pi} \,\Big[   s_I{}^{\bar I} \, 
s_J{}^{\bar J} \, p^{\bar K} \, c_{\bar I \bar J \bar K}
+   s_I{}^{\bar I} \, 
s_J{}^{\bar \alpha} \, p^{\bar K} \, c_{\bar I \bar K \bar \alpha}
\Big] \ ,
\end{align}
with a $2\pi$ factor being generated  from the integral of $V_\Sigma$
over $\Sigma_g$.
The result \eqref{result} can be cast in a more suggestive form,
\begin{align} \label{resultBIS}
I_4^{\rm inflow} = \frac 12 \, \int_{M_7} E_5 \, dE_5  =  (2\pi)^{-2}  \, p^{\bar K} \, \Big[
 &\frac 12\, c_{\bar K \bar I \bar J} \, (F^I \, s_I{}^{\bar I}) \, (F^J \, s_J{}^{\bar J})
 +   c_{\bar K \bar I \bar \alpha} \, (F^I \, s_I{}^{\bar I}) \, (F^J \, s_J{}^{\bar \alpha})
 \nn \\
 & +  \frac 12 \, c_{\bar K \bar \alpha \bar \beta} \, (F^I \, s_I{}^{\bar \alpha}) \,(F^J \, s_J{}^{\bar \beta})
 \Big] \ .
\end{align}
The equivalence  of  \eqref{result} and  \eqref{resultBIS}
relies on the condition \eqref{salpha_eq} on the $s_I{}^{\bar \alpha}$ coefficients.
The form \eqref{resultBIS} makes it easy to see that
$I_4^{\rm inflow}$ is obtained from the integral of the 4d anomaly polynomial
\beq \label{parent_inflow2}
I_6^{\rm inflow} = \frac 16 \, c_{\bar I \bar J \bar K}  \, \frac{F^{\bar I}_{\rm 4d}}{2\pi}
\, \frac{F^{\bar J}_{\rm 4d}}{2\pi} \, \frac{F^{\bar K}_{\rm 4d}}{2\pi}
+ \frac 12 \, c_{\bar I \bar J \bar \alpha}  \, \frac{F^{\bar I}_{\rm 4d}}{2\pi}
\, \frac{F^{\bar J}_{\rm 4d}}{2\pi} \, \frac{F^{\bar \alpha}_{\rm 4d}}{2\pi}
+ \frac 12 \, c_{\bar I \bar \alpha \bar \beta}  \, \frac{F^{\bar I}_{\rm 4d}}{2\pi}
\, \frac{F^{\bar \alpha}_{\rm 4d}}{2\pi} \, \frac{F^{\bar \beta}_{\rm 4d}}{2\pi}  \ ,
\eeq
with the identifications
\beq  
F^{\bar I}_{\rm 4d} = F^I \, s_I{}^{\bar I} + p^{\bar I} \, V_\Sigma \ ,\qquad
F^{\bar \alpha}_{\rm 4d} = F^I \, s_I{}^{\bar \alpha} \ .
\eeq
We have thus verified the claim made in the main text.



\section{Punctures in 4d $\cN = 4$ SYM} \label{sec_punctureapp}

This appendix collects further details and derivations
about the setup studied in section \ref{sec_puncture}.
We begin collecting useful background material
for the discussion of punctures.

\subsection{Inclusion of punctures: generalities} \label{sec_generalities}

The strategy of \cite{Bah:2018jrv,Bah:2019jts} for the study of regular punctures in 4d $\cN = 2$ class $\cS$
theories from M-theory can be directly generalized to study a class of punctures
in 4d $\cN = 4$ SYM. 

Our starting point is the internal space $M_7^{n=0}$ 
for 4d $\cN = 4$ SYM compactified on a genus-$g$ Riemann surface without punctures
$\Sigma_{g,0}$. The 7d space $M_7^{n=0}$ is of the form
\beq \label{before_punctures}
S^5 \hookrightarrow M_7^{n=0} \rightarrow \Sigma_{g,0} \ .
\eeq
The topology of this $S^5$ fibration over $\Sigma_{g,0}$ depends
on the choice of topological twist. In this work, we consider the Maldacena-Nu\~nez  twist \cite{Maldacena:2000mw},
which we describe in more detail below.
Let us now select $n$ distinct points on 
$\Sigma_{g,0}$, labeled by the index $\alpha = 1, \dots , n$.
Let $D_\alpha$ denote a small disk on $\Sigma_{g,0}$ centered
at the $\alpha$-th point.  
The space $M_7^{n=0}$ can be presented as
\beq \label{M7_decomp}
M_7^{n=0} = M_7^{\rm bulk} \cup  \bigcup_{\alpha=1}^n (D_\alpha \times S^5) \ ,
\eeq
where $M_7^{\rm bulk}$ is the space obtained
from $M_7^{n=0}$ by removing the small disks $D_\alpha$ and the $S^5$ fibers
on top of them. 
The 7d space that is relevant for a configuration with punctures
is obtained from \eqref{M7_decomp}
by replacing each $D_\alpha \times S^5$ term with a 
puncture geometry $X_7^\alpha$,
\beq  
M_7 = M_7^{\rm bulk} \cup \bigcup_{\alpha = 1}^n X_7^\alpha \ .
\eeq
This decomposition of the internal space $M_7$ implies
an analogous decomposition  of the inflow anomaly
polynomial into a bulk piece plus puncture pieces,
as stated in \eqref{pieces_of_I4}.
The task at hand is the description of the topology and isometries
of the bulk geometry $M_7^{\rm bulk}$ and the puncture geometries $X_7^\alpha$,
and the construction of the 5-form $E_5$ for $M_7^{\rm bulk}$
and $X_7^\alpha$.

\subsection{The bulk of the Riemann surface}

The topology of the $S^5$ fibration \eqref{before_punctures}
is chosen in such a way that 
the isometry group $SO(6)$ of $S^5$ is broken as
\beq \label{SO6_split}
SO(6) \rightarrow SO(4) \times SO(2) \ ,
\eeq
and the twist is performed by turning on a background field strength
for the $SO(2)$ connection.

To describe the setup more precisely we need some additional notation.
Let us describe $S^5$ as the locus $Y^A \, Y_A = 1$,
where $Y^A$, $A = 1,\dots, 6$ are Cartesian coordinates on $\mathbb R^6$,
and the $A$ index is raised/lowered with $\delta$.
With reference to \eqref{SO6_split},
it is convenient to 
parametrize the coordinates $Y^A$ subject to $Y^A \, Y_A = 1$ as
\beq \label{Y_param}
Y^{a} = \mu \, y^{a} \ , \quad a  = 1,2,3,4 \ ,   \qquad Y^5 = \sqrt{1-\mu^2} \, \cos \phi \ ,
\qquad Y^6 = \sqrt{1-\mu^2} \, \sin \phi \ ,
\eeq
where the four quantities $y^a$ obey the constrain $y^a \, y_a = 1$,
with the $a$ index raised/lowered with $\delta$.
The coordinate $\mu$ has range $[0,1]$, and the angle $\phi$
has periodicity $2\pi$. The parametrization \eqref{SO6_split}
shows that we can regard $S^5$ as an $S^1_\phi \times S^3_\Omega$ fibration
over the $\mu$-interval, where $S^1_\phi$ is the circle
parametrized by $\phi$ and $S^3_\Omega$ is the round 3-sphere
described by $y^a \, y_a = 1$.
The $SO(4)$ factor in \eqref{SO6_split} is identified
with the isometry group of $S^3_\Omega$,
while the $SO(2)$ factor is the isometry group of $S^1_\phi$.
We also see from \eqref{Y_param} that $S^1_\phi$ shrinks
at $\mu = 1$, while $S^3_\Omega$ shrinks at $\mu = 0$.

The total $SO(2)$ connection contains an internal contribution with legs on the 
Riemann surface, corresponding to the topological twist,
and an external contribution, corresponding to 
gauging the $SO(2)$ isometry. We then write
\beq \label{Dphi_def}
D\phi = d\phi - \cA \ , \qquad \cA = A^\phi + \cA_\Sigma \ , \qquad
\cF = d \cA =  p^\phi \, V_\Sigma + F^\phi   \ ,
\eeq
where $V_\Sigma$ is the volume form on the Riemann surface,
normalized as in \eqref{Sigma_background}.
The twist parameter $p^\phi$ is fixed by supersymmetry,
\beq
p^\phi = - \chi(\Sigma_{g,n}) \ , \qquad
\chi(\Sigma_{g,n}) = - 2 \, (g-1) -n  \ .
\eeq
In contrast, the $SO(4)$ connection is purely external.
In our conventions, the constrained coordinates $y^a$ on $S^3_\Omega$
couple to the $SO(4)$ background connection $A^{ab}$ according to
\beq \label{Dy_def}
D y^a = dy^a - A^{ab} \, y_b \ .
\eeq

\subsection{The form $E_5$ in the bulk of the Riemann surface}
As a warm-up exercise for the discussion of $E_5$ for a puncture,
we first reconsider $E_5$ for the bulk of the Riemann surface.
Instead of applying the recipe of section \ref{sec_smoothSigma} and appendix \ref{app_no_punctures},
we proceed by writing down the most general ansatz for
$E_5$ compatible with the topology and isometries of 
the bulk geometry.
Next, we impose that each term in $dE_5$  has at most two legs
along the internal space.
The outcome of this analysis is the following $E_5$,
\begin{align} \label{general_bulk_E5}
E_5 & = N \, \bigg[ d\gamma \, \frac{D\phi}{2\pi}
-  \gamma \, \frac{\cF}{2\pi} \bigg] \, e_3^{SO(4)} \nn \\
& + \bigg[ du_1 \, \frac{D\phi}{2\pi}
-  u_1 \, \frac{\cF}{2\pi} \bigg] \, \frac{ \epsilon_{abcd} \, F^{ab} \, y^c \, Dy^d }{ (2\pi)^2 }
- u_1 \, \frac{D\phi}{2\pi} \,  \frac{ \epsilon_{abcd} \, F^{ab} \, Dy^c \, Dy^d }{ (2\pi)^2 }
\nn \\
& + u_2 \, \frac{F^\phi}{2\pi} \, \frac{ \epsilon_{abcd} \, F^{ab} \, y^c \, Dy^d }{ (2\pi)^2 }
+ u_3 \, \frac{D\phi}{2\pi}  \, \frac{ \epsilon_{abcd} \, F^{ab} \, F^{cd} }{ (2\pi)^2 } \ .
\end{align}
In the above expression, we recalled $\cF = F^\phi + p^\phi \, V_\Sigma = - dD\phi$,
we used the global angular form of $SO(4)$ given in \eqref{global_ang_SO4},
and we introduced the quantities $\gamma$, $u_1$, $u_2$, $u_3$,
which are functions of $\mu$ only.
The function $\gamma$ satisfies
\beq \label{gamma_values}
\gamma(0) = 0 \ , \qquad \gamma(1) = 1 \ .
\eeq
Indeed, $\gamma$ must vanish at $\mu = 0$ to have a regular $E_5$,
since $S^3_\Omega$ shrinks at $\mu = 0$.
The difference $\gamma(1)  - \gamma(0)$ is   fixed to be 1
from the flux quantization condition
\beq
N = \int_{S^5} E_5 \ .
\eeq
The function $\gamma$ in the interior of the $\mu$ interval
is smooth, but otherwise arbitrary.
By a similar token, the functions $u_1$, $u_2$, $u_3$
are smooth and arbitrary, up to the requirements
\beq \label{u_values}
u_1(0) = u_1(1) = 0 \ , \qquad
u_2(0) = 0 \ , \qquad
u_3(1) = 0 \ ,
\eeq
which follow from regularity of $E_5$.
(Recall that $S^1_\phi$ shrinks at $\mu = 1$.)

Recall that the 5-form $E_5$ for $\cN = 4$ SYM
is the global angular form of $SO(6)$, 
given in \eqref{quotient_F5}.
If we take the global angular form of $SO(6)$,
and we only activate the background connections $A^{AB}$
along the generators of the subgroup $SO(4) \times SO(2)$,
we get a 5-form that is of the form \eqref{general_bulk_E5}.
In this special case, the functions $\gamma$, $u_1$, $u_2$, $u_3$
are given by
\beq
\gamma = \mu^4 \ , \qquad
u_1 = - \frac 12 \, N \, \mu^2 \, (1- \mu^2) \ ,  \qquad
u_2 = 0 \ , \qquad
u_3 = - \frac 18 \, N \, (1-\mu^2) \ .
\eeq

Next, let us evaluate the integral of $E_5 \, dE_5$ over the internal space.
The integration over $S^3_\Omega$ is conveniently performed
using the identity
\beq \label{S3_identity}
\int_{S^3_\Omega} y^a \, Dy^b \, Dy^c \, Dy^d = \frac{\pi^2}{2} \, \epsilon^{abcd} \ .
\eeq
Moreover, we recall that $\phi$ has period $2\pi$, that $\int_{\Sigma_{g,n}} \cF = - 2\pi \, \chi(\Sigma_{g,n})$, and we choose a convention that gives positive orientation
to $d\mu \, d\phi \, {\rm vol}_{S^3_\Omega}$.
We then obtain
\beq
\int_{M_7^{\rm bulk}} E_5 \, dE_5 = \frac{ \epsilon_{abcd} \, F^{ab} \, F^{cd} }{(2\pi)^2} \, 
\chi(\Sigma_{g,n}) \, \bigg[
- \frac 18 \, N^2 \, \gamma^2  + \frac 14 \, N \, \gamma\, u_1 + N \, \gamma \, u_3
\bigg]_{\mu = 0}^{\mu = 1} \ .
\eeq
As we can see, the arbitrary function $u_2$ completely drops from the
result.
Moreover,   $u_1$ and $u_3$ drop out as well,
thanks to the regularity conditions \eqref{gamma_values}, \eqref{u_values}.
In conclusion,
\beq
\int_{M_7^{\rm bulk}} E_5 \, dE_5 = - \frac 18 \, N^2 \, \chi(\Sigma_{g,n}) \, \frac{ \epsilon_{abcd} \, F^{ab} \, F^{cd} }{(2\pi)^2}   \ .
\eeq
Since the result is independent of $u_1$, $u_3$, $u_2$,
a viable choice of $E_5$ is given simply by the first line of 
\eqref{general_bulk_E5}, which exhibits a simple factorized structure
and is the direct analog of the 4-form $E_4$ in the bulk of the Riemann
surface in the M-theory analysis of \cite{Bah:2018jrv,Bah:2019jts}.

\subsection{The puncture geometry}

Let us now turn to a description of the puncture geometry $X_7^\alpha$.
Since each puncture can be analyzed in isolation,
for the sake of brevity we omit the puncture label $\alpha$
for the remainder of this section.
The analogous problem in M-theory
has been studied in \cite{Bah:2018jrv,Bah:2019jts}.
The arguments presented there
can be repeated with minimal modifications in the present context.
The only difference is that the 2-sphere $S^2_\Omega$ of the M-theory
analysis is replaced by the 3-sphere $S^3_\Omega$ in our type IIB setup.
For this reason, we 
proceed with a description of the puncture geometry without
derivations.

Before discussing the puncture geometry $X_7$,
we need to introduce an auxiliary 4d space $X_4$.
The latter can be described as a circle fibration
over  $\mathbb R^3$,
\beq
S^1_\beta \hookrightarrow X_4 \rightarrow \mathbb R^3 \ .
\eeq
Let us introduce cylindrical coordinates $(\rho, \chi, \eta)$
in $\mathbb R^3$, where $\eta \in \mathbb R$ is the coordinate along the 
cylindrical axis of symmetry, $\rho\ge 0$ is the distance from the axis,
and $\chi$ is the azimuthal angle around the axis,
with periodicity $2\pi$.
Axial symmetry restricts the 
fibration of the $\beta$ circle, which is described by 
\beq \label{naked_Dbeta}
D\beta = d\beta - L \, d\chi \ ,
\eeq
where $L$ is a function of $\rho$ and $\eta$, independent of $\chi$.
The function $L$ encodes the fact that 
the $S^1_\beta$ fibration has $p$ monopole sources.
The latter are   
located along the positive $\eta$ semiaxis at $\rho = 0$ at positions
$\eta_a$, $a = 1, \dots, p$ (ordered as $0<\eta_1 < \eta_2 < \dots < \eta_p$).
The function $L$ is piecewise constant along the $\eta$ axis,
with jumps at the location of each monopole.
The value of $L$ in the interval $(\eta_{a-1}, \eta_a)$ is an integer,
which we denote $\ell_a$,
\beq
L(0,\eta) = \ell_a \qquad \text{for $\eta_{a-1} < \eta < \eta_{a}$} \ , \qquad a = 1, \dots, p \ ,
\eeq
with the convention $\eta_0 = 0$. 
The value of $L$ on the $\eta$ axis past the last monopole is zero,
\beq \label{L_is_zero}
L(0,\eta) = 0 \qquad \text{for $\eta > \eta_p$} \ .
\eeq
The charge $k_a$ of the monopole at $\eta = \eta_a$ is a positive integer and is
measured
by the discontinuity in $L$ across $\eta = \eta_a$,
\beq  \label{ell_vs_k}
k_a = \ell_a - \ell_{a+1} \ ,
\eeq
which holds for all $a = 1, \dots, p$ with the understanding that $\ell_{p+1} = 0$.
Notice that the circle $S^1_\beta$ shrinks at the location of the monopoles.

Having described the salient features of the space $X_4$,
we can now describe the puncture geometry $X_7$.
It is obtained by fibering
$S^3_\Omega$ over $X_4$,
\beq \label{X7_fibration}
S^3_\Omega \hookrightarrow X_7 \rightarrow X_4 \ .
\eeq
The 3-sphere $S^3_\Omega$ shrinks at $\eta = 0$.
This ensures that the total space $X_7$ caps off smoothly
at $\eta = 0$, and therefore we only consider the
half space in $\mathbb R^3$ with $\eta \ge 0$. 
In the fibration \eqref{X7_fibration},
we do not turn on any $SO(4)$ background field strength
with legs along $X_4$.

\subsection{Compatibility between puncture and bulk}
\begin{figure}
\centering
\includegraphics[width = 12.85cm]{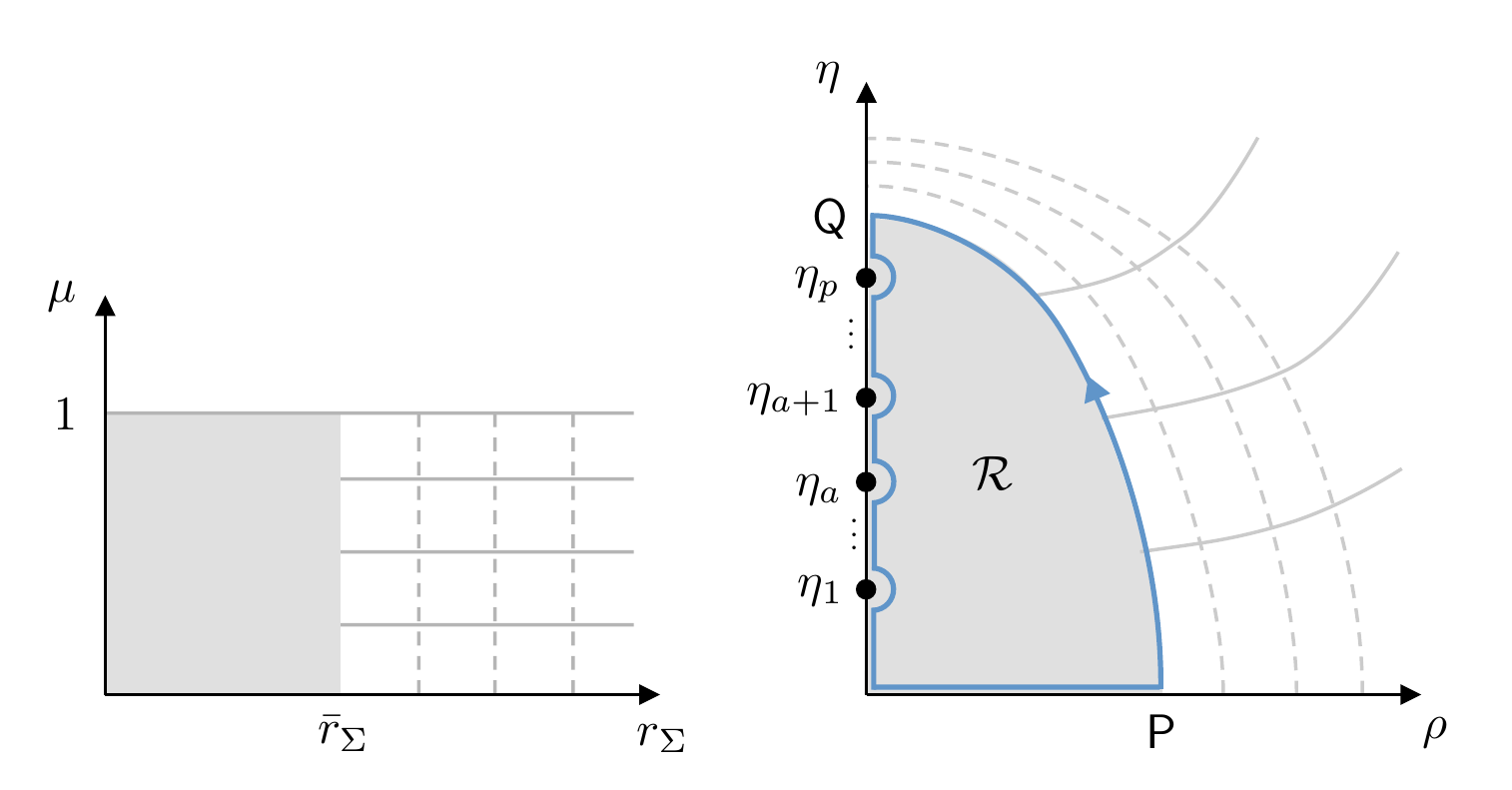}
\caption{
On the left  we depict the $(r_\Sigma,\mu)$ plane.
The relevant region is the strip $r_\Sigma \ge 0$,
$0 \le \mu \le 1$. The shaded area corresponds
to the portion $D\times S^5$ that is excised
to make room for the puncture. The value $\bar r_\Sigma$
is the radius of the disk $D$.
We also include lines of constant $r_\Sigma$ and lines of constant $\mu$.
On the right we depict the $(\rho, \eta)$ plane.
The region $\cR$
corresponds to the relevant portion of the puncture geometry $X_7$.
The portion of the $(\rho,\eta)$ plane outside
the region $\cR$ corresponds to the bulk of the Riemann surface.
We depict the qualitative behavior of lines of constant
$r_\Sigma$ and $\mu$ as they appear in $(\rho,\eta)$ coordinates.
}
\label{new_fig}
\end{figure}

According to the general strategy outlined in section \ref{sec_generalities},
inserting a puncture means replacing $D \times S^5$
with a new geometry. The latter is a portion of the full space
$X_7$ described in the previous section.
More precisely, the relevant portion of $X_7$
is the one that is obtained by restricting the coordinates
$(\rho,\eta)$ to lie in the shaded region $\cR$
depicted in figure \ref{new_fig} on the right.
The gluing of the puncture geometry to the bulk
is performed along the $\mathsf{PQ}$ arc.

To discuss this more precisely, let
us introduce polar coordinates $(r_\Sigma, \beta)$
on the small disk $D$ on the Riemann surface.
As our notation anticipates, the polar angle $\beta$ on the disk $D$
is identified with the angle $\beta$ in the puncture geometry.
The relation between the bulk coordinates $(r_\Sigma,\mu)$
and the puncture coordinates $(\rho,\eta)$ is more involved.
Figure \ref{new_fig} includes a schematic depiction
of lines of constant $r_\Sigma$ and $\mu$
in the $(\rho,\eta)$ plane.
In particular, in the gluing we identify the
vertical line at $r_\Sigma = \bar r_\Sigma$ on the left
with the $\mathsf{PQ}$ arc on the right.

In performing the gluing of puncture geometry and bulk geometry,
the angular coordinate $\chi$
in the puncture geometry is 
given in terms of bulk coordinates by 
\beq
\chi  = \phi + \beta \ .
\eeq
In particular, this relation implies that the angle $\chi$ is gauged
by the external connection for the angle $\phi$,
\beq
D\chi = d\chi  - A^\phi \ , \qquad F^\phi = dA^\phi \ ,
\eeq
where $A^\phi$, $F^\phi$ are the same as in \eqref{Dphi_def}.
As soon as the external connection $A^\phi$ is activated,
the 1-form $D\beta$ in \eqref{naked_Dbeta} has to be improved to
\beq
\widetilde { D\beta} = d\beta - L \, D\chi \ .
\eeq

For later applications, we also need to point out
that the internal part of the $\phi$ connection
on the disk $D$ on the Riemann surface is conveniently
parametrized as
\beq \label{U_def}
D\phi = d\phi - A^\phi - \cA_\Sigma \ , \qquad 
\cA_\Sigma = U(r_\Sigma) \, d\beta \  ,
\eeq
where the function $U$ vanishes at $r_\Sigma  = 0$
in order to ensure regularity of $A_\Sigma$.
Since $U$ is a function of $r_\Sigma$ only,
it is constant on the locus $r_\Sigma = \bar r_\Sigma$.
Let us therefore write $\overline U = U(\bar r_\Sigma)$.
Recall that the gluing is implicitly performed in the limit
of small disk,
$\bar r_\Sigma \rightarrow 0$.
In this limit, we have $\overline U \rightarrow 0$.

\subsection*{The form $E_5$ for the puncture geometry}
Our next task is to write down the most general $E_5$ compatible
with the topology and isometries of the puncture
geometry, and impose that each term in $dE_5$ has at most
two internal legs.
The most general allowed $E_5$ is found to be 
\begin{align} \label{puncture_E5}
E_5 & = \bigg[ d \bigg( Y \, \frac{D\chi}{2\pi} - W \, \frac{\widetilde {D\beta}}{2\pi} \bigg) 
+ \Lambda \, d\rho \, d\eta \bigg] \, e_3^{SO(4)}
\nn \\
& + \bigg[  \sigma_1 \, \frac{D\chi}{2\pi}   +  \sigma_2 \, \frac{ \widetilde {D\beta } }{2\pi}
+ \lambda_1  \bigg] \,  \frac{ \epsilon_{abcd} \, F^{ab} \, F^{cd}}{(2\pi)^2}  
\nn \\
& + \bigg[ \sigma_3 \, \frac{D\chi}{2\pi}  +  \sigma_4 \, \frac{ \widetilde {D\beta}   }{2\pi}
+ \lambda_2 
\bigg] \, \frac{ \epsilon_{abcd} \, F^{ab} \, D y^c \, D y^d  }{(2\pi)^2}   \nn \\
&  + \bigg[   \sigma_0 \,  \frac{ F^\phi}{2\pi}
   - d \bigg( \sigma_3    \, \frac{ D\chi  }{2\pi} \bigg)
   - d  \bigg(\sigma_4   \, \frac{ \widetilde {D\beta}  }{2\pi}  \bigg)
- d \lambda_2  
\Big] \, \frac{  \epsilon_{abcd} \, F^{ab} \, y^c \, Dy^d }{(2\pi)^2}   \ .
\end{align}
In the above expression,
the 3-form $e_3^{SO(4)}$ is the global angular form of $SO(4)$,
\beq \label{global_ang_SO4}
 e_3^{SO(4)} = \frac{1}{2\pi^2} \, \bigg[
 \frac{1}{3!} \, \epsilon_{abcd} \, y^a \, Dy^b \, Dy^c \, Dy^d
 - \frac 14 \, \epsilon_{abcd} \, F^{ab} \, y^c \, Dy^d
 \bigg] \ .
\eeq
It satisfies
\beq
\int_{S^3_\Omega}  e_3^{SO(4)}  = 1 \ , \qquad
d  e_3^{SO(4)}  = - \frac 18 \, \frac{\epsilon_{abcd} \, F^{ab} \, F^{cd}}{(2\pi)^2}
= - \chi_4(SO(4)) \ .
\eeq
The quantities 
 $Y$, $W$, $\sigma_{0,1,2,3}$, $\Lambda$ are functions of 
$\rho$, $\eta$, while $\lambda_{1,2}$ are 1-forms in the $(\rho,\eta)$ plane.
These objects are not uniquely determined,
but are constrained by regularity of $E_5$ and flux
quantization.\footnote{While $E_5$ is not closed,
it does yield a closed 5-form $\overline E_5$
if we turn off
all external connections.
It is therefore meaningful to impose integrality of the periods of the 5-form
$\overline E_5$ over 5-cycles in $X_7$.
}

Let us first focus on the functions $Y$, $W$. They enter $E_5$ via the closed 2-form
\beq
\cE_2 = d\bigg[ Y \, \frac{D\chi}{2\pi} - W \, \frac{\widetilde{D\beta}}{2\pi} \bigg]
= (dY + W \, dL) \, \frac{D\chi}{2\pi}
- dW \, \frac{\widetilde{D\beta}}{2\pi}
- (Y + W \,L) \, \frac{F^\phi}{2\pi} \ .
\eeq
This 2-form   is exactly the same
as the one that appears in the M-theory setup of \cite{Bah:2018jrv,Bah:2019jts}.
This means that we can repeat the flux quantization analysis of \cite{Bah:2018jrv,Bah:2019jts}
almost  verbatim, keeping in mind that 
the role of $S^2_\Omega$ in M-theory is now played by $S^3_\Omega$.
It follows that the conditions on $Y$, $W$
that were derived in \cite{Bah:2018jrv,Bah:2019jts} are also true in the present
context.
They can be summarized as follows:
\begin{itemize}
\item The function $W = W(\rho,\eta)$ is smooth for   $\rho \ge 0$, $\eta \ge 0$,
and vanishes for $\eta = 0$
for any $\rho$,
\beq
W(\rho,0) = 0 \ .
\eeq
The values of $W$  at the locations
of the monopoles along the $\eta$ axis at $\rho = 0$ satisfy
\beq
W(0,\eta_a) = w_a \ , 
\eeq
where  $\{ w_a\}_{a=1}^p$ is   an increasing sequence of positive integers.

\item The function $Y = Y(\rho,\eta)$ is smooth 
away from the $\eta$ axis at $\rho = 0$, 
and vanishes at $\eta = 0$ for any $\rho$,
\beq
Y(\rho,0) = 0 \ .
\eeq
Moreover, $Y$   is piecewise constant
(hence discontinuous) along the $\eta$ axis,
\begin{align}
Y(0,\eta) &= y_a \phantom{ {}:=N } \qquad \text{for $\eta_a < \eta < \eta_{a+1}$} \ , \qquad
a =1, \dots, p-1 \ ,  \nn \\
Y(0,\eta) &= y_p :=N \qquad \text{for $\eta> \eta_p$}   \ .
\end{align}
The quantities $y_a$ are all positive integers.

\item Even though $L$ and $Y$ are both discontinuous
along the $\eta$ axis at $\rho = 0$, the form $\cE_2$ is free
from discontinuities, thanks to the ``sum rule''
\beq
y_a = \sum_{b=1}^a w_b \, k_b \ .
\eeq
In particular, selecting $a  = p$ and using $y_p = N$, we get the relation 
\beq
N = \sum_{a=1}^p w_a k_a \ ,
\eeq
which defines a partition of $N$.
\end{itemize}
In direct analogy with the M-theory analysis,
we observe that regularity and flux quantization
of $E_5$ imply that the class of punctures we are studying
are labelled by partitions of $N$.
It would be interesting to have a purely field-theoretic
understanding of this feature of punctures in 4d $\cN = 4$ SYM theory.

When the puncture geometry is glued to the bulk
geometry, the functions $Y$, $W$ are related to the function $\gamma$
in \eqref{general_bulk_E5} and $U$ in \eqref{U_def}.
The analysis of \cite{Bah:2019jts} shows that the gluing condition is
\beq \label{YW_gluing}
Y + W \, L = N \, \gamma \ , \qquad
W = N \, \gamma \, (1 + \overline U) \qquad
\text{along the $\mathsf{PQ}$ arc} \ .
\eeq
We have recalled that the $\mathsf {PQ}$ arc 
sits at $r_\Sigma = \bar r_\Sigma$, hence $U = \overline U$ constant
along the $\mathsf{PQ}$ arc.

Finally, 
let us collect some conditions on the functions $\sigma_{1,2,3,4}$
that stem from regularity of $E_5$ and smooth gluing onto the bulk
geometry.
The $\chi$ circle in $\mathbb R^3$ shrinks along the $\eta$ axis.
This implies the regularity conditions
\beq \label{sigma13_reg}
\sigma_1 \Big|_{\rho = 0} = \sigma_3 \Big|_{\rho = 0} = 0 \ .
\eeq
We also know that the circle $S^1_\beta$ shrinks
at the location of the monopoles. This gives the regularity conditions
\beq \label{monopole_reg}
\sigma_2(0, \eta_a) = \sigma_4(0,\eta_a) = 0 \ , \qquad a = 1, \dots, p \ .
\eeq
Next, let us compare the terms with $\epsilon_{abcd} \, F^{ab} \, F^{cd}$
in the expressions \eqref{general_bulk_E5} and \eqref{puncture_E5} for $E_5$ in the bulk
and for a puncture.
In \eqref{general_bulk_E5} the prefactor of $\epsilon_{abcd} \, F^{ab} \, F^{cd}$
has only legs along $D\phi$, while in \eqref{puncture_E5}
it is a combination of $D\chi$ and $\widetilde{D\beta}$.
These different prefactors must agree along the $\mathsf{PQ}$ arc.
In particular, on this arc there should be no $d\beta$ term
in the prefactor of $\epsilon_{abcd} \, F^{ab} \, F^{cd}$ in \eqref{puncture_E5}. 
This implies
\beq \label{first_arc_rel}
\sigma_1 + \sigma_2 - L \, \sigma_2 = 0 \qquad \text{along the $\mathsf{PQ}$ arc} \ .
\eeq
In a similar way, matching terms with $\epsilon_{abcd} \, F^{ab} \, Dy^d \, Dy^d$
in \eqref{general_bulk_E5} and \eqref{puncture_E5} leads to the condition
\beq  \label{second_arc_rel}
\sigma_3 + \sigma_4 - L \, \sigma_4 = 0 \qquad \text{along the $\mathsf{PQ}$ arc} \ .
\eeq

Let us point out that, by arguments similar to those of the previous paragraphs,
one can also argue that $\lambda_{1,2}$ and $\Lambda$ should be zero
in order to ensure a smooth gluing between puncture and bulk $E_5$ forms.
We will not make direct use of this observation, however, because
the anomaly inflow result turns out to be automatically independent of
$\lambda_{1,2}$,  $\Lambda$.

\subsection{The integral of $E_5 \, dE_5$ in the puncture geometry}

We may use again \eqref{S3_identity} for the integration over $S^3_\Omega$.
Both the $\chi$ and the $\beta$ circles have periodicity $2\pi$.
The orientation convention that fits with the orientation of the bulk
is the one that assigns a positive orientation
to $d\rho \, d\eta\, d\chi \, d\beta \, {\rm vol}_{S^3_\Omega}$.
One finds
\beq
\int_{X_7} E_5 \, dE_5 = \frac{\epsilon_{abcd} \, F^{ab} \, F^{cd}}{ (2\pi)^2 } \, \int_{\cR_2} \, \cS_2 \ ,
\eeq
where $\cR_2$ is the region in the $(\rho, \eta)$ plane depicted in figure \ref{new_fig},
and $\cS_2$ is the following 2-form in the $(\rho, \eta)$ plane,
\begin{align}
\cS_2 & = - \frac 14 \, d(Y + W \, L) \, dW \nn \\
& - \bigg[ dW \, d(\sigma_1 + \sigma_2 - L \, \sigma_2)
+ d(Y + W \, L) \, d\sigma_2 - dW \, d\sigma_2 \bigg]
\nn \\
& + \frac 14 \,
 \bigg[ dW \, d(\sigma_3 + \sigma_4 - L \, \sigma_4)
+ d(Y + W \, L) \, d\sigma_4 - dW \, d\sigma_4 \bigg]   \ .
\end{align}
As we can see, the result seems to depend on the unspecified
functions $\sigma_{1,2,3,4}$. We now demonstrate, however,
that all dependence on $\sigma_{1,2,3,4}$ drops away
after integrating on the region $\cR_2$.
To this end, it is convenient to write
\begin{align}
\cS_2 = d\cS_1 \ , \qquad 
\cS_1 & =   \frac 14 \, W \, d(Y + W \, L)  \nn \\
& - \bigg[ W \, d(\sigma_1 + \sigma_2 - L \, \sigma_2)
+ (Y + W \, L) \, d\sigma_2 - W \, d\sigma_2 \bigg]
\nn \\
& + \frac 14 \,
 \bigg[ W \, d(\sigma_3 + \sigma_4 - L \, \sigma_4)
+ (Y + W \, L) \, d\sigma_4 - W \, d\sigma_4 \bigg]  \ .
\end{align}
By Stokes' theorem,
\beq \label{stokes}
\int_{\cR_2} \cS_2 = \int_{\partial \cR_2} \cS_1     \ .
\eeq
The boundary $\partial \cR_2$ consists of a horizontal segment
along the $\rho$ axis,   the $\mathsf{PQ}$ arc,
and a collection of intervals along the $\eta$ axis, connected
by small semicircles around the monopole sources,
as shown in figure \ref{new_fig}.
We discuss these boundary components in turn.

\paragraph{The horizontal segment along the $\rho$ axis.}
We know that $W$ and $Y$ vanish along the $\rho$ axis at $\eta = 0$.
It follows that $\cS_1$ is zero along the horizontal segment
of $\partial \cR_2$.

\paragraph{The $\mathsf {PQ}$ arc.}
The integral of the term $\frac 14 \, W \, d(Y + W \, L)$ in $\cS_1$ along the 
$\mathsf{PQ}$
arc
is non-zero.
As in appendix B of \cite{Bah:2019jts}, this term is interpreted as a bulk contribution,
rather than as a puncture contribution.\footnote{More precisely,
we can imagine to perform the integral of $E_5 \, dE_5$ in the bulk
geometry using the Euler characteristic $\chi(\Sigma_{g,0})$ of the unpunctured Riemann surface. The contribution of $\cS_1 \supset \frac 14 \, W \, d(Y + W \, L)$
from the $\mathsf{PQ}$ arc
is computed using the gluing conditions \eqref{YW_gluing} between puncture and bulk,
and is found to be independent on the details of the puncture.
The net effect of these terms is to shift the Euler characteristic
from $\chi(\Sigma_{g,0})$ to the correct value  $\chi(\Sigma_{g,n})$
for the punctured Riemann surface.
}
Next, we argue that all  terms in $\cS_1$ with $\sigma_{1,2,3,4}$
integrate to zero along the 
$\mathsf{PQ}$ arc.
To see this, we use \eqref{first_arc_rel} and \eqref{YW_gluing} to write
\begin{align}
& W \, d(\sigma_1 + \sigma_2 - L \, \sigma_2)
+ (Y + W \, L) \, d\sigma_2 - W \, d\sigma_2  = \nn \\
& = - N \, \overline U \, \gamma \, d\sigma_2 \qquad
\text{along the $\mathsf{PQ}$ arc} \ .
\end{align}
The gluing is performed in the limit of small disk radius,
which implies $\overline U \rightarrow 0$. As a result,
the terms in $\cS_1$ with $\sigma_{1,2}$ do not yield
any contribution from integration along the 
$\mathsf{PQ}$ arc. The terms with $\sigma_{3,4}$
are treated in a completely analogous way,
making use of  \eqref{second_arc_rel}.

\paragraph{The intervals  along the $\eta$ axis.}
We consider each interval $(\eta_{a-1}, \eta_a)$,
$a = 1, \dots, p$,
together with the interval that connects the last monopole at $\eta  = \eta_p$
with the point $\mathsf{Q}$, which we denote schematically as $(\eta_p, \mathsf Q)$.
First of all, we compute
\beq \label{nonzero}
\int_{(\eta_{a-1}, \eta_a)} \frac 14 \, W \, d(Y + W\,L)
= \int_{(\eta_{a-1}, \eta_a)} d\bigg[ \frac 18 \, \ell_a \, W^2 \bigg] 
= \frac 18 \, \ell_a \, (w_a^2 - w_{a-1}^{2}) \ ,
\eeq
where we used the fact that  $Y = y_{a-1}$ constant and $L = \ell_a$   constant
in the interval $(\eta_{a-1}, \eta_a)$. 
If we consider the last interval $(\eta_p,\mathsf Q)$, we have $L = 0$
and therefore we get no contribution.

Next, we argue that the terms with $\sigma_{1,2,3,4}$ in $\cS_1$
drop away
from all integrals over $(\eta_{a-1},\eta_a)$ and $(\eta_p, \mathsf Q)$.
If we consider the interval $(\eta_{a-1},\eta_a)$, we can 
use $L = \ell_a$, $Y = y_{a-1}$, and the regularity condition \eqref{sigma13_reg}
on $\sigma_1$ to observe that 
\begin{align}
 & W \, d(\sigma_1 + \sigma_2 - L \, \sigma_2)
+ (Y + W \, L) \, d\sigma_2 - W \, d\sigma_2  =   \\
& =  y_{a-1} \, d\sigma_2 \qquad
\text{along the interval $(\eta_{a-1}, \eta_a)$} \ . \nn
\end{align}
When this 1-form is integrated on  $(\eta_{a-1}, \eta_a)$,
the result is proportional to
the difference $\sigma_2(0,\eta_a) - \sigma_{2}(0,\eta_{a-1})$,
which is zero thanks to the regularity condition \eqref{monopole_reg}.
In a similar way, if we consider the last interval $(\eta_p, \mathsf Q)$,
we can use $L =0$, $Y = N$, and get
\begin{align}
 & W \, d(\sigma_1 + \sigma_2 - L \, \sigma_2)
+ (Y + W \, L) \, d\sigma_2 - W \, d\sigma_2  =   \\
& =  N \, d\sigma_2 \qquad
\text{along the interval $(\eta_{p}, \mathsf Q)$} \ . \nn
\end{align}
To show that this integrates to zero we must argue that
$\sigma_2$ vanishes at point $\mathsf Q$.
This is indeed the case, because $\mathsf Q$
lies at the intersection of the $\eta$ axis with the $\mathsf {PQ}$ arc,
and therefore we can combine \eqref{sigma13_reg} and \eqref{first_arc_rel} and infer
that $\sigma_2$ is zero at $\mathsf Q$.
 
The fact that all terms in $\cS_1$ with $\sigma_{1,3}$ do not contribute
to integrals over $(\eta_{a-1},\eta_a)$ and $(\eta_p, \mathsf Q)$ is shown
in a completely analogous way.

\paragraph{Small semicircles around the monopole sources.}
The small semicircles do not give any non-zero contribution
in the limit in which their radius goes to zero. 
To see this, let us introduce coordinates $(R_a,\tau_a)$ in the vicinity of the $a$-th monopole,
as
\beq
\eta = \eta_a + R_a \, \tau_a \ , \qquad \rho = R_a \, \sqrt{1 - \tau_1^a} \ ,
\eeq
with the range of $\tau_a$ being $[-1,1]$. The small semicircle is described by
$R_a = \text{const} \rightarrow 0$.
To argue that the term $W \, d(Y + W \,L)$ in $\cS_1$ does not contribute
when integrated on the small semicircle around the $a$-th monopole,
we recall that both $W$ and the combination $Y + W \, L$ are continuous
along the $\eta$ axis (while $Y$ and $L$ separately are piecewise constant).
As a result, for small constant $R_a$, we have
\beq
\int_{\text{semicircle}} W \, d(Y + W \, L) \approx w_a \,\int_{-1}^1 d\tau_a \, \partial_{\tau_a}(Y + W \, L) 
= w_a \, \Big[ Y + W\, L \Big]_{\eta = \eta_a - R_a}^{\eta = \eta_a + R_a}  \rightarrow 0\ .
\eeq 
In the first step we used the fact that, to leading order as $R_a \rightarrow 0$,
$W$ is approximated by its value $w_a$ at $(\rho,\eta) =(0, \eta_a)$
because it is continuous near that point.
In the last step we get zero because $Y+W \, L$ tends to the same value
as we approach $\eta_a$ from below or above.
All other terms in $\cS_1$ are treated in a similar way. We need to recall that
$\sigma_{1}$ and $\sigma_3$ vanish along the $\eta$ axis,
and that $\sigma_2$ and $\sigma_4$ vanish at the location of the monopoles.

\paragraph{Summary.}
There is only one non-zero contribution to the puncture anomaly,
given by summing terms of the form  \eqref{nonzero}. Notice that the boundary $\partial \cR_2$ must be
traversed in counterclockwise orientation, which means that each
interval on the $\eta$ axis is considered with a negative orientation.
As a result, we arrive at
\beq
\int_{X_7} E_5 \, dE_5 = - \frac 18 \,  \frac{\epsilon_{abcd} \, F^{ab} \, F^{cd}}{ (2\pi)^2 }  \,
\sum_{a=1}^p \ell_a \, (w_a^2 - w_{a-1}^2)\ .
\eeq



\section{Remarks on $c_1(\mathbb L)$} \label{sec_ftheoryapp}

In this appendix we recall some well-known facts about the
Weierstrass line
bundle $\mathbb L$ introduced in section \ref{sec_ftheory}.
These remarks   are useful in 
elucidating the physical interpretation of the new term \eqref{new_term}.
We follow the exposition of \cite{Weigand:2018rez,Bianchi:2011qh}.

Classical type IIB supergravity has a rigid $SL(2,\mathbb R)$ symmetry.
In the quantum theory, 
this
is broken by non-perturbative effects.
A discrete $SL(2,\mathbb Z)$ subgroup is preserved,
and is a local symmetry of the theory.\footnote{More precisely,
the quantum symmetry group is the metaplectic group $Mp(2,\mathbb Z)$,
which is the unique non-trivial $\mathbb Z_2$ central extension of $SL(2,\mathbb Z)$ \cite{Pantev:2016nze}.}
In F-theory constructions, we imagine to cover spacetime with
overlapping patches and we allow non-trivial
$SL(2,\mathbb Z)$ transformations in the transition functions.
Let $\cU$, $\cU'$ be a generic pair of overlapping patches.
The local expressions $\tau$ and $\tau'$ for the axio-dilaton
on $\cU$, $\cU'$ are related on $\cU \cap \cU'$ by \eqref{tau_transf}
for some $\left(\begin{smallmatrix} a & b \\ c & d\end{smallmatrix}
\right) \in SL(2, \mathbb Z)$.
Using the $\tau$ profile and the same 
transition matrix $\left(\begin{smallmatrix} a & b \\ c & d\end{smallmatrix}
\right) \in SL(2, \mathbb Z)$, 
we can define a complex line bundle
by the following gluing condition on $\cU \cap \cU'$, 
\beq \label{U1bundle}
s' = e^{i\theta} \, s \ , \qquad
e^{i\theta} := \frac{c\, \tau + d}{|c \, \tau + d|} \ .
\eeq
In the previous expression,
$s$, $s'$ are local trivializations of a   section of the complex line bundle
on $\cU$, $\cU'$ respectively.
There is a simple local expression for 
a connection $Q$ on this bundle.
It is given by
\beq \label{Q_expr}
Q = - \frac{1}{2 \, \tau_2} \, d\tau_1 \ .
\eeq
Indeed, if $\tau$ and $\tau'$ are related by \eqref{tau_transf},
the expression \eqref{Q_expr} implies 
\beq
Q' = Q - d\theta \ ,
\eeq
which is   the expected gluing condition for a connection
on the bundle satisfying \eqref{U1bundle}.
The field strength of $Q$ reads
\beq \label{FD_expr}
F_D = dQ = \frac{d\tau \, d\bar \tau}{4 \, i \,\tau_2^2} \ .
\eeq

In a setup described by a Weierstrass model
\eqref{Weierstrass}, $\tau$ varies holomorphically
over $W_4$ and the field strength $F_D$ is of $(1,1)$ type.
In this situation, there is a canonical way to turn the
complex line bundle defined by \eqref{U1bundle} into
a holomorphic line bundle, defined by the gluing condition
\beq \label{holo_bundle}
\hat s' = (c\tau + d) \, \hat s \ ,
\eeq
where $\hat s$, $\hat s'$ are local trivializations on $\cU$,
$\cU'$ of a section of the holomorphic line bundle.
The relation between $\hat s$ and $s$ in each patch is
\beq
\hat s  = (\tau_2)^{-1/2} \, s \ .
\eeq
In fact, \eqref{tau_transf} and \eqref{U1bundle} imply \eqref{holo_bundle}.
But the gluing condition \eqref{holo_bundle} is exactly the one that corresponds to
the Weierstrass line bundle $\mathbb L$.\footnote{Indeed,
as explained for instance in \cite{Weigand:2018rez},
the transformation properties of $f$ and $g$ 
under \eqref{tau_transf} are
\beq
f' = (c\, \tau + d)^4 \, f \ , \qquad
g' = (c \, \tau + d)^6 \, g \ ,
\nn
\eeq
and $f$ (resp.~$g$) is a section of $\mathbb L^4$ (resp.~$\mathbb L^6$).
}
As a result, we may identify the 
first Chern class of $\mathbb L$
with the field strength $F_D$,
\beq
c_1(\mathbb L) = \frac{F_D}{2\pi} \ .
\eeq
The non-triviality of $c_1(\mathbb L)$ is thus a precise
measure of a non-zero gradient for the axio-dilaton.
This fits with our intuition of the new term \eqref{new_term}
as being built with derivatives of $\tau$.

It should be stressed that the expression for $F_D$
in terms of $d\tau$, $d\bar \tau$ 
must be taken with a grain of salt.
In the presence of 7-branes, $\tau$ is multivalued and $d\tau$
is not a good 1-form.
In particular, despite what the form \eqref{FD_expr} suggests,
we have in general $F_D^2 \neq 0$.
Indeed, in many examples the non-universal terms in \eqref{example}
contain $c_1(\mathbb L)^2$ and $c_1(\mathbb L)^3$ terms \cite{Lawrie:2018jut}. 
By a similar token, higher powers of $c_1(\mathbb L)$
are encountered in the analysis of discrete
anomalies in supergravities of \cite{Minasian:2016hoh}.


\bibliographystyle{./ytphys}
\bibliography{./refs}

\end{document}